\documentclass[fleqn,usenatbib,useAMS]{mnras}
\usepackage{newtxtext,newtxmath}
\usepackage[T1]{fontenc}
\let\VANthebibliography\thebibliography
\def\thebibliography{\DeclareRobustCommand{\VAN}[3]{##3}\VANthebibliography}

\usepackage{graphicx}	% Including figure files
\usepackage{tabularx}
\usepackage{authblk}
\usepackage{orcidlink}
\usepackage{color}
\usepackage{xcolor}

\usepackage{natbib}
\usepackage{booktabs}
\usepackage{subcaption}
\usetikzlibrary{shapes,arrows,fit}
\usepackage{tikz}
\usepackage{arydshln}

\defcitealias{sdc3a1}{SDC3aI}

\title[SKA Science Data Challenge 3b]{Square Kilometre Array Science Data Challenge 3b: EoR inference}
\author{ O.  Bait \orcidlink{0000-0003-2722-8841}$^{1,2,3}$, 
 A.  Bonaldi \orcidlink{0000-0003-4787-2888}$^{1}$\thanks{E-mail: anna.bonaldi@skao.int}, 
 M.  Bianco \orcidlink{0000-0002-6766-0017}$^{4}$, 
 S. K.  Giri \orcidlink{0000-0002-2560-536X}$^{5}$, 
 G.  Mellema \orcidlink{0000-0002-2512-6748}$^{5}$, 
 P.  Hartley$^{1}$, 
 S.  Purser \orcidlink{0000-0002-6358-5152}$^{1}$, 
 R.  Braun$^{1}$, 
 A.  Acharya \orcidlink{0000-0003-3401-4884}$^{6,7}$, 
 D.  Banik$^{8}$
% ...
}

\begin{document}
\date{Accepted XXX. Received XXX; in original form XXX}

%\pagerange{\pageref{firstpage}--\pageref{lastpage}} \pubyear{2017}

\maketitle

%\clearpage
\begin{abstract}
We present and analyse the results of Science Data Challenge 3b (SDC3b; \url{https://sdc3.skao.int/challenges/inference}), an Epoch of Reionization (EoR) inference exercise organised by the Square Kilometre Array Observatory (SKAO) using simulated SKA data. The nine-month challenge, running from October 2024 to May 2025, represents the first community-wide comparison of EoR inference pipelines on common simulated datasets. We consider multiple datasets designed to assess performance on both foreground-free and foreground-contaminated cylindrical power spectra and imaging products, adopting different reionization scenarios and EoR simulation models. We include results from 16 teams, which employ a range of methodologies to infer the posterior distribution of the hydrogen neutral fraction across three redshift intervals spanning $z=8.41$ to $z=6.25$. The results show significant inter-team variability of the astrophysical constraints even in the simpler case where only instrumental noise is added to the cosmological signal. This underscores the need for further investigation into the relative contributions of signal modelling, the treatment of instrumental effects, and inference methods to the robustness of the resulting conclusions. Although individual methods can produce biased parameter estimates, the combined average of their estimates remains close to the input values, suggesting that independent approaches can offset method-specific biases.

\end{abstract}
%\twocolumn 

\section{Introduction}
One of the main science goals for the Square Kilometre Array (SKA) is to map the distribution of the 21-cm signal from the eras referred to as the Cosmic Dawn and Epoch of Reionization (CD/EoR), roughly corresponding to the first billion years in the history of the Universe \citep[see][for an overview]{skao2026overview}. During these times, as well as during the preceding Dark Ages, the hydrogen gas in the Intergalactic Medium (IGM) was mostly atomic and neutral. Non-ionizing radiation produced by the first generations of stars and accreting black holes during the CD/EoR could, through the so-called Wouthuysen-Field effect, decouple the HI quantum state from the background photons of the Cosmic Microwave Background. This led to the generation of an observable 21-cm signal \citep[see e.g.][for a review on the 21-cm signal from the CD/EoR]{2012RPPh...75h6901P}. During the reionization process, H-ionizing photons produced by stars and/or accreting black holes turned intergalactic HI into HII, ultimately resulting in a fully reionized Universe in which most HI resides inside galaxies.

By measuring the evolution of the redshifted 21-cm signal over a range of frequencies with the SKA-Low telescope, it should be possible to derive information about the state of the IGM at different times, and through this, about the radiation fields produced by the galaxies. However, these measurements are far from trivial. The radio sky at the relevant frequencies (50--250\,MHz) is filled with strong foreground signals both from the Milky Way and from extragalactic radio sources. Separating the redshifted 21-cm signal from these foregrounds is challenging as they are several orders of magnitude brighter and therefore require excellent calibration of instrumental effects. Once a 21-cm signal has been extracted, deriving information is not straightforward either but requires inference based on simulations of the signal. Recent examples of the application of inference to data from 21-cm experiments can be found in \citet{2026ApJ...998...33A} and \citet{2025A&A...699A.109G}. We refer interested readers to \citet{skao2026inference} for more discussion.

To test the ability of existing frameworks to study the redshifted 21-cm signal from the CD/EoR, the SKA Observatory initiated an SKA Data Challenge, the third of such challenges after SDC1 \citep{2021MNRAS.500.3821B} and SDC2 \citep{2023MNRAS.523.1967H}. It was decided to split SDC3 into two parts. SDC3a tested the ability to extract the 21-cm signal from simulated SKA-Low data including foreground and telescope effects. Its results were published in \cite{sdc3a} and a description of how the data was created in \citet{sdc3a1}, henceforth SDC3a1. The second part, SDC3b, tests the ability to derive information about the state of the IGM from an already extracted 21-cm signal. This paper presents the results from SDC3b.

The specific aim of the data challenge was to provide posteriors on the value of the mean neutral fraction from an extracted 21-cm power spectrum, as well as from tomographic image cubes. The provided 21-cm data was produced blindly and the correct answers were not provided until after the participating teams had submitted their results. The choice was made to focus on the mean neutral fraction as this is a single parameter that measures the progress of reionization in the IGM. It is also a quantity which is well defined and does not depend on the details of the reionization model used.

The structure of this paper is as follows. Section 2 describes the data sets which were made available to the participating teams. Section 3 defines the goals of the challenge. The participating teams describe their inference tools in Section 4. The results are presented and analysed in Section 5 and Section 6 discusses and summarizes the conclusions reached.

\section{Simulated data}\label{sec:datasim}
The data for SDC3b use a full simulation of the SKA-Low instrumental response to the sky signal through the generation of visibilities, image cubes and, finally, power spectra. The simulation pipeline is mostly the same as the one described in \citetalias{sdc3a1}. Here, we summarise the aspects already covered there and describe the differences introduced in this simulation. Multiple datasets were generated for SDC3b, as listed in Sec. \ref{sec:products}, enabling tests for different aspects related to inference, and also including an increasing range of corrupting effects.  

The SKA-Low telescope response was simulated with the {\tt OSKAR}\footnote{\url{https://github.com/OxfordSKA/OSKAR/releases}} package \citep{oskar}, which generates visibilities using a sky model, a telescope model and an error model. A four hour observation was simulated, directed at (RA, Dec) = (00h, -30deg) using a 10 second integration time and a 100\,kHz frequency sampling between 151 and 196\,MHz. The resulting instrumental noise was afterwards rescaled in amplitude to correspond to a 100\,h or a 1000\,h observation, depending on the dataset (see Sec. \ref{sec:products}). The basic telescope model makes use of the SKA-Low configuration of 512 stations \citep{ska_low_conf}, also known as ``AA4'', with the so-called “Vogel” layout \citep{vogel} for the stations, each consisting of 256 antennas.

The full sky model consists of the following components:
\begin{itemize}
\item Epoch of reionization (EoR) 21-cm emission modelled as detailed in Sec. \ref{sec:eor} over an $8 \times 8$ deg field of view (FoV) centred on the pointing centre;
\item Galactic and extragalactic foreground emission modelled as described in \citetalias{sdc3a1} over an $8 \times 8$ deg field of view (FoV) centred on the pointing centre;
\item Strong extragalactic out-of-field sources, modelled over the full 2$\pi$\,sr above the horizon at any time, as described in \citetalias{sdc3a1} (see "outer sky model"). 
\end{itemize}
Calibration errors, both direction-dependent (DD) and direction-independent (DI) have been introduced into the visibilities. The amplitude of these errors has been reduced with respect to the full expected effects to represent the final outcome after a successful DD and DI calibration process (see \citetalias{sdc3a1} for more details). 

The foreground contamination in the simulation has also been strongly reduced to represent foreground residuals after a successful foreground mitigation process; this step is described in detail in Sec.~\ref{sec:pca}. 

The {\tt wsclean}\footnote{\url{https://wsclean.readthedocs.io/}} task \citep{2014MNRAS.444..606O} was used to produce images from the visibilities and the corresponding synthesised beam, with “natural” weighting and no deconvolution. 

The public package \texttt{ps\_eor}\footnote{\url{https://gitlab.com/flomertens/ps\_eor}} has been used to compute all power spectra. 
Given the 3D power spectrum $P(\mathbf{k})=V|\tilde{T}(\mathbf{x})|^2$, where $V$ is the comoving volume and $\tilde{T}_{\rm b}(\mathbf{x})$ is the Fourier transform of the brightness temperature at a physical coordinate $\mathbf{x}$,

\begin{itemize}
\item the cylindrical (2D) power spectrum $P(k_\parallel,k_\bot)$, where $k_\parallel$ and $k_\bot$ denote modes parallel and perpendicular to the line of sight, is obtained by averaging $P(\mathbf{k})$ over cylindrical shells with radius $k_\bot = |\mathbf{k}_\bot|$; 
\item the spherical (1D) power spectrum $P(k)$ is obtained by averaging $P(\mathbf{k})$ in spherical shells with radius $k=|\mathbf{k}|=\sqrt{k_\parallel^2+k_\bot^2}$;
\item the dimensionless power spectrum is $\Delta^{2}(k) = \dfrac{k^{3}}{2\pi^{2}}P(k)$.
\end{itemize}
The challenge adopts the cylindrical (2D) power spectrum $P(k_{\parallel},k_{\bot})$, as done in SDC3a, because it allows easier separation from foreground components, which are intrinsically smooth in $k_\parallel$ and present structures in $k_\bot$.

\subsection{The cosmological 21-cm signal}\label{sec:eor}
Here we present the simulation of the 21-cm signal.

\subsubsection{Cosmological evolution}\label{sec:cosmo_evol} 
The set of cosmological parameters adopted are referred to as {Planck18}: $\Omega_{\rm m}=0.30964$, $\Omega_{\rm b} = 0.04897$, $\Omega_\Lambda=0.69036$ H$_0=67.66$~km~s$^{-1}$~Mpc$^{-1}$, $\sigma_8=0.8102$, $n_s=0.9665$ \citep{2020A&A...641A...6P}.
The public code {\tt 21cmFAST\_v4} \citep{davies25} was used to model the cosmological evolution of the matter distribution and the formation of large-scale structures. 
Specifically, Gaussian initial conditions were generated in a box of side length 700 cMpc sampled by 1050 pixels.  The matter density field was evolved to Eulerian space using second order Lagrangian Perturbation theory (2LPT), smoothed to a grid of 350 pixels per dimension, implying a resolution of 2 comoving Mpc. The same set-up was used for all datasets, but with a different random seed for each (see Sec.~\ref{sec:products}).

The sources of ionizing photons form in dark matter (DM) haloes \citep[e.g.,][]{1978MNRAS.183..341W,2001PhR...349..125B,2018PhR...780....1D}. Two different algorithms were used for the cosmological radiative transfer (RT) of ionizing photons, labelled Simulator 1 and Simulator 2, and described in Sec. \ref{sec:21cm_Tb}. Both simulators derive the halo population from the conditional halo mass function, but use it in different ways: Simulator 1 uses it to generate a catalogue of discrete haloes, whereas Simulator 2 integrates it over halo mass to obtain the collapsed fraction in each cell.

\subsubsection{The 21-cm differential brightness temperature}\label{sec:21cm_Tb}

The 21-cm signal was assumed to be in the saturated spin temperature regime, with $T_{\rm S} \gg T_{\rm CMB}$ (where $T_{\rm S}$ is the spin temperature and $T_{\rm CMB}(z)$=2.7255$(1+z)$\,K is the temperature of the Cosmic Microwave Background), for the redshifts ($z$) of interest. Under this assumption, the differential brightness temperature at any position $\pmb{r}$ can be given as
\begin{equation}
    \delta T_\mathrm{b}(\pmb{r},z) \approx 27 x_\mathrm{HI}(\pmb{r},z)\left[1+\delta(\pmb{r},z)\right]\left(\frac{1+z}{10}\frac{0.15}{\Omega_\mathrm{m} h^2}\right)^\frac{1}{2}\left(\frac{\Omega_\mathrm{b}h^2}{0.023}\right) \mathrm{mK} \ ,
    \label{eq:dTb}
\end{equation}
where $x_\mathrm{HI}$ and $\delta$ are the neutral fraction and matter overdensity, respectively. The neutral fraction field depends on the emission and propagation of ionising photons in the IGM \citep[see e.g.,][]{iliev2006cosmological,mellema2006simulating}.

Both Simulator 1 and Simulator 2 use the comoving IGM density fields and galaxy emissivity fields described above as their source model, but generated with different initial seeds and astrophysical parameter choices.
In both simulators, the ionizing sources are described by the stellar-to-halo mass ratio, $f_\star$, and the escape fraction of ionizing photons, $f_\mathrm{esc}$, which depend on the mass of the host DM halo, $M_\mathrm{h}$, through a power-law of the following form
\begin{equation}
    f_i = f_{i,10} \left(\frac{M_\mathrm{h}}{10^{10} M_\odot}\right)^{\alpha_i} \ ,
\end{equation}
with $i=\{\star,\mathrm{esc}\}$. $f_{i,10}$ and $\alpha_i$ are free parameters. See \citet{park2019inferring} and \citet{giri202421} for details of this source model.

\paragraph{Simulator 1}\label{sec:C2R}
The first simulator is {\tt pyC$^2$Ray} \citep{Hirling:2023cjx}, which models the emission of ionizing photons from sources residing in DM haloes by solving the three-dimensional radiative transfer equations using the Conservative Causal Ray Tracing (C$^2$Ray) method \citep{Mellema:2005ht}. Haloes with mass $\gtrsim 3 \times 10^{11}$ $M_\odot$ were identified from the Lagrangian initial conditions using the \cite{sheth_tormen2002} ellipsoidal collapse barrier \citep[e.g.][]{1996Natur.380..603B,2007ApJ...669..663M}. Smaller mass haloes were sampled from the \cite{sheth_tormen2002} conditional mass function within each Lagrangian cell at the lowest redshift, with progenitor populations at higher redshifts being sampled from each descendant halo’s conditional mass function \citep{davies25} down to the halo sampler limit, $M_\mathrm{min}$ (see Table \ref{tab:21cmfast_parameters}). All haloes were then moved from Lagrangian to Eulerian space using 2LPT. These haloes and the matter distribution were generated using {\tt 21cmFAST\_v4}, as described in Sec.~\ref{sec:cosmo_evol} and subsequently post-processed with {\tt pyC$^2$Ray}. The haloes are assumed to grow exponentially, giving a mass accretion rate $\dot{M}_\mathrm{h}=M_\mathrm{h}/\left[(1+z)^{-1}H(z)^{-1}t_\star\right]$, where $t_\star$ is a free parameter \citep{giri202421}. Each halo then emits ionizing photons at a rate
\begin{equation}
    \dot{N}_\gamma \propto f_\star(M_\mathrm{h})\,f_\mathrm{esc}(M_\mathrm{h})\,\dot{M}_\mathrm{h}\ .
\end{equation}
These photons are propagated through the IGM with a three-dimensional photon-conserving ray-tracing method. The ray-tracing algorithm computes the neutral hydrogen column density along the line of sight from each source and uses it to calculate the corresponding optical depth. This quantity is used to determine the photoionization rate, which describes the rate of photoionizations in each cell. From the photo-ionization rate and the density field, a second algorithm finds the solution to the chemistry differential equation for neutral hydrogen using an iterative procedure that enforces photon conservation. In this work, recombinations are computed on the base density grid without accounting for sub-structure. The simulations adopted a fixed mean free path of 50 comoving Mpc using a Heaviside filter.

\paragraph{Simulator 2}\label{sec:21cmf}
The second simulator implements Fast Fourier Transform Radiative Transfer \citep[FFRT, e.g.][]{mesinger201121cmfast,2011MNRAS.414..727Z}, as implemented in the public {\tt 21cmFAST} code, which is the numerical implementation of the excursion-set algorithm of \cite{2004ApJ...613....1F}. Instead of resolving haloes, this simulator integrates the conditional halo mass function, $\mathrm{d}n/\mathrm{d}M_\mathrm{h}(M_\mathrm{h}\,|\,\delta_R,R)$, over halo mass within a spherical region of radius $R$ and mean overdensity $\delta_R$ around each cell, to obtain the number of ionizing photons per baryon emitted into the IGM \citep{park2019inferring},
\begin{equation}
    \begin{array}{r@{\;}l}
    n_\mathrm{ion}(\delta_R,R) \propto & \displaystyle\int \mathrm{d}M_\mathrm{h}\, \frac{\mathrm{d}n}{\mathrm{d}M_\mathrm{h}}(M_\mathrm{h}\,|\,\delta_R,R) \\[2ex]
    & \times\, f_\mathrm{duty}(M_\mathrm{h})\, f_\star(M_\mathrm{h})\, f_\mathrm{esc}(M_\mathrm{h})\, M_\mathrm{h}\ ,
    \end{array}
    \label{eq:nion}
\end{equation}
where the duty cycle, $f_\mathrm{duty} = \exp(-M_\mathrm{turn}/M_\mathrm{h})$, suppresses star formation in haloes below the turnover mass $M_\mathrm{turn}$.
% CHECK: confirm Eq. (nion) matches the 21cmFAST_v4 settings used (no-halo-field mode), and add where t_star enters Simulator 2 (in park2019, t_star sets the star formation rate, M_star/(t_star H^{-1}), which does not enter n_ion in the saturated spin-temperature limit).
% Each cell is determined to be ionized if $n_\mathrm{ion}$ exceeds that of the number of baryons plus (sub-grid) recombinations in surrounding spheres of decreasing radii \citep[e.g.][]{2014MNRAS.440.1662S}. The redshift-dependent mean free path of ionizing photons through the ionized IGM is implemented using the exponential filter from \cite{2022MNRAS.514.1302D}.
Each cell is tested for ionization by smoothing the field of ionizing photon production with spheres of decreasing radius. This radius runs from $R_{\rm max} = 50$ cMpc down to the cell size. The smoothing uses the exponential filter of \citet{2022MNRAS.514.1302D}. This filter weights each source within the sphere by $\exp(-r/\lambda_\mathrm{mfp})$, where $r$ is the source's distance from the cell. The weighting accounts for photons absorbed by residual neutral gas in the ionized IGM before they reach the cell. We adopt a constant mean free path $\lambda_\mathrm{mfp} = 25.48\,h^{-1}$\,cMpc. This is the value at z = 6 of the fit $\lambda_\mathrm{mfp} = 112\,h^{-1} [(1+z)/5]^{-4.4}$\,cMpc to the measurements of \citet{songaila2010evolution}. A cell is ionized when, at any $R$, the number of photons reaching it exceeds the number of baryons in the cell plus its cumulative sub-grid recombinations. The recombinations are computed per cell following \citet{2014MNRAS.440.1662S}.

Table \ref{tab:21cmfast_parameters} lists the \citet{park2019inferring} parametrization adopted as a common parameter framework across both simulators. The implementation of this parametrization in {\tt pyC$^2$-Ray} is presented in \citet{giri202421}. Some differences remain in how the parameters are used internally to each code, most notably in the halo mass modelling, as described above. In particular, $M_\mathrm{min}$ is used only by Simulator 1 and $M_\mathrm{turn}$ only by Simulator 2 (see Table~\ref{tab:eorsummary}).

\subsubsection{lightcone generation}
Both simulators produce a series of coeval cubes of the ionization fraction at different redshifts, $x_{\rm H}(z)$. From this quantity and the density field, the corresponding differential brightness coeval cubes, $\delta T_\mathrm{b}$, are calculated with the {\tt tools21cm}\footnote{\url{https://tools21cm.readthedocs.io/} } python package \citep{giri2020tools21cm}. This package was also used by several of the teams. 
%Lightcone generation
Lightcones of $\delta T_\mathrm{b}$  and the ionization fraction, $x_{\rm HII}$, were generated from the coeval cubes at different redshifts with the {\tt make\_lightcone} function from {\tt tools21cm}. For details of the algorithm implemented by this function, see \citet{mellema2006simulating} and \citet{datta2012light}.

%t2c.make_lightcone(files,z_low=6, z_high=10, file_redshifts=redshift,los_axis=2, interpolation='linear', reading_function=np.load, box_length_mpc=700). 

%Lightcones output from this step are in units of (cMpc, cMpc, z). Those have been further converted to observational quantities (deg, deg, Hz)  by using the tools21cm physical_lightcone_to_observational function.

The frequency resolution (100\,kHz) and the comoving spatial resolution (2\,cMpc) are comparable for the considered redshifts and cosmology (100\,kHz increments correspond to increments in comoving distance of 1.7–1.5 cMpc). Therefore, the transformation along the LoS essentially maps every comoving bin to the nearest frequency bin. With this nearest-neighbour interpolation, some comoving bins are repeated.

The transformation perpendicular to the line of sight involves regridding each frequency map onto an $8\times8$ degree field sampled by $512\times512$ pixels of 0.94 arcmin. The $8\times8$ FoV corresponds to a comoving transverse size of 1009--1357\,cMpc at 196--151\,MHz, which is only partially filled by the 700\,cMpc simulated cube.  For each frequency slice, the edges of the FoV were filled assuming periodic boundary conditions perpendicular to the line of sight. 

\subsection{Foreground mitigation} \label{sec:pca}
In the analysis of future SKA data, foreground mitigation would be performed prior to EoR inference. This process has been already tested in the preceding part of the SDC3 data challenge \citep{sdc3a} and is not tested here. Instead, we assume a successful foreground removal step has already taken place. As detailed in Sec. \ref{sec:products}, for two of the four simulated datasets, perfect foreground removal is assumed -- this in practice means that foreground components, either inside or outside the FoV, are simply not included in the {\tt OSKAR} sky model. 

For the two remaining datasets a more realistic foreground removal step has been modelled. The aim was to mimic the level of cleaning achieved by the best-performing teams in the foreground-removal challenge \citep[SDC3a,][]{sdc3a}. Overall, those best results were achieved with several stages of cleaning -- such as an iterative detection and subtraction of strong point sources before diffuse foreground removal, or an explicit deconvolution/convolution step to disentangle instrumental effects from the sky signal \citep[see][for more details]{sdc3a}. Such a detailed analysis was considered beyond the scope of this work and was not replicated here. Instead, we performed a simpler one-step foreground cleaning, described below, and finally scaled the resulting foreground residuals by $3\times 10^{-4}$ to achieve the desired contamination. 

Our simple foreground-estimation pipeline used a commonly employed technique called principal component analysis (PCA). We used the PCA implementation of the FOREGROUNDS-FRIENDS \footnote{\url{https://github.com/espsrc/FOREGROUNDS-FRIENDS}} team from the SDC3a foreground-removal challenge \citep{sdc3a} for our PCA analysis. PCA was applied to the entire $8\times 8$ deg FoV imaging data containing foreground components (both Galactic and extragalactic, including out-of-field) to obtain a frequency-dependent foreground model. After trial and error, two components have been assumed in the PCA, to capture the foregrounds and the chromatic effects. The resulting foreground model was subtracted from a noiseless version of the data to obtain a noiseless foreground-residual cube. This cleaned cube was further smoothed using a Gaussian kernel with a standard deviation of 5 pixels or 4.7 arcmin. 

\begin{table}
\renewcommand{\arraystretch}{1.25}
\centering
\caption{Description  of the \citet{park2019inferring} astrophysical parameters used for the simulations; these parameters follow the most recent {\tt 21cmFAST} parametrization and were adopted as a common parameter framework for both simulators. The three parameters at the bottom affect X-heating and were not used while creating the challenge dataset.}
\label{tab:21cmfast_parameters}
\begin{tabularx}{\linewidth}{@{}
  >{\raggedright\arraybackslash}m{1.2cm}
  >{\raggedright\arraybackslash}X
@{}}

\toprule
\textbf{Parameter} &
\multicolumn{1}{c}{\textbf{Description}}\\
\midrule
$f_{\star,10}$ & Normalized ratio of galactic stellar mass to baryon gas \\
$\alpha_{\star}$ & Power-law index of stellar-to-gas mass ratio \\
$f_{\mathrm{esc}, 10}$ & Normalized escape fraction of ionizing
photons of star-forming galaxies \\
$\alpha_\mathrm{esc}$ & Power-law index of escape fraction \\
$M_\mathrm{turn}$ & Turnover halo mass due to the galactic duty cycle,
[$M_\odot$] \\
$t_{\star}$ & Star formation timescale relative to the Hubble time \\
$M_\mathrm{min}$&Minimum halo mass for the selection of haloes, [$M_\odot$] \\
\hdashline
$L_\mathrm{X}/\mathrm{SFR}$ & Soft X-ray luminosity per unit star
formation rate, [$\mathrm{erg}\ \mathrm{s}^{-1}\
\mathrm{M}_\odot^{-1}\ \mathrm{yr}$] \\
$E_{0}$ & Minimum energy of X-ray photons for self-absorption, [eV] \\
$\alpha_\mathrm{X}$ & X-ray spectral energy index \\
\bottomrule
\end{tabularx}
\end{table}

\subsection{Data products} \label{sec:products}

The first three SDC3b datasets, labelled PS1, PS2 and PS3, are three sets of cylindrical (2D) EoR power spectra from an SKA-Low simulated observation. A fourth dataset, labelled IM1, consists of the EoR imaging product corresponding to PS3. 
Details of all the products are given below:

\begin{itemize}
\item {\bf PS1, PS2} -- foreground-free power spectrum datasets. They contain the EoR signal, instrumental noise for a 100\,h observation and calibration errors, but no foreground residuals. PS1 and PS2 adopt a different simulation code (Simulator 1 vs Simulator 2, see Sec. \ref{sec:eor} for more details). 
\item {\bf PS3} -- foreground-contaminated power spectrum dataset. This 
dataset contains EoR signal generated with Simulator 2, noise corresponding to 1000\,h observation, calibration errors and residual foreground contamination. 
\item {\bf IM1} -- foreground-contaminated imaging dataset -- is the imaging product from which PS3 has been generated.
\end{itemize}
Table \ref{tab:datasummary} summarises the different simulation components included in the datasets. 

All EoR components generated for PS1, PS2 and PS3/IM1 adopted a different set of astrophysical parameters to describe the sources of reionization, as well as different random seeds, therefore making the inference exercises independent. The values of the astrophysical parameters (defined in Table \ref{tab:21cmfast_parameters}) adopted for the datasets are reported in Table \ref{tab:eorsummary}. 

\begin{figure}
\includegraphics[width=9cm]{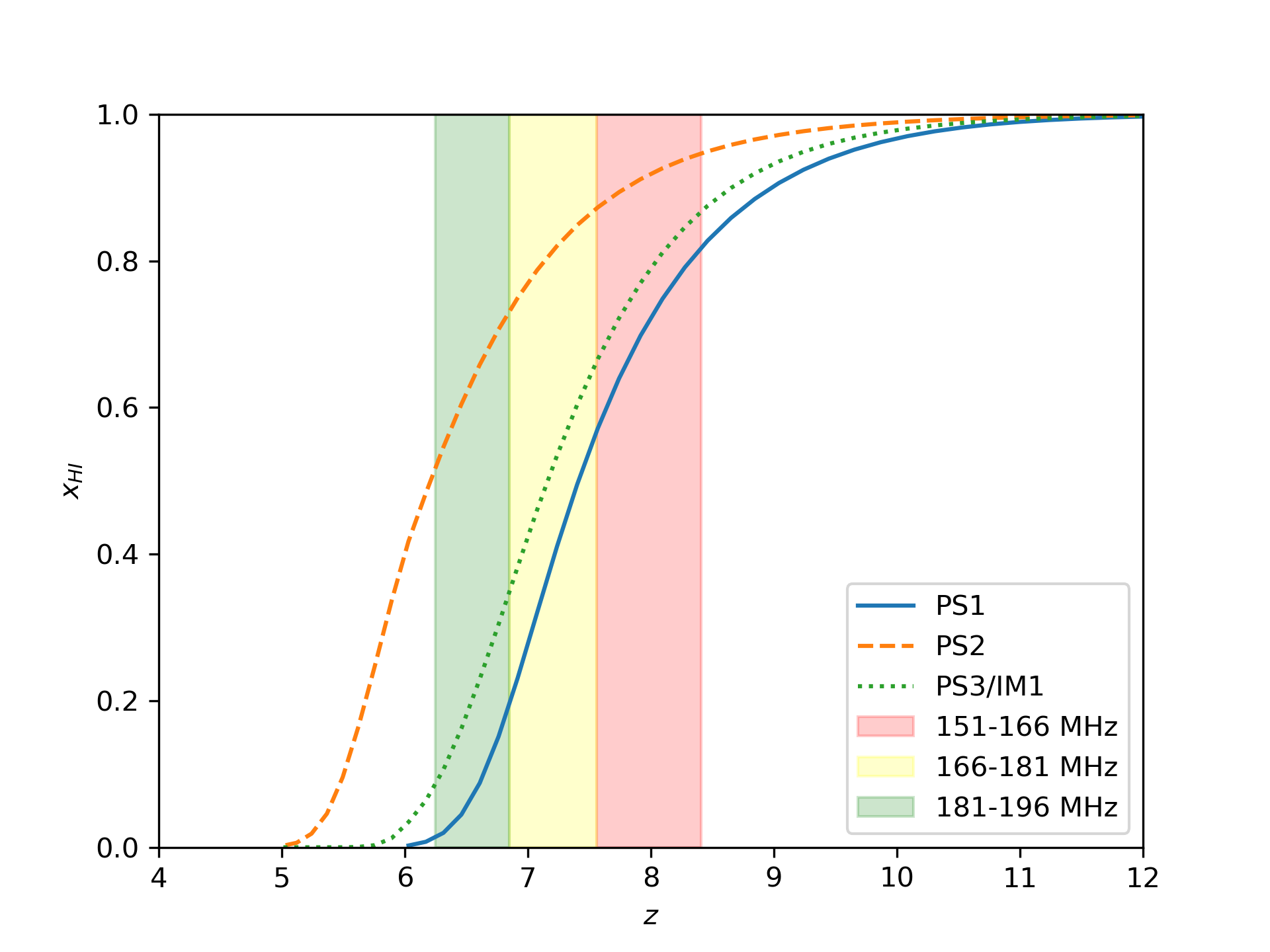}

\caption{Neutral fraction as a function of redshift for the three datasets. The frequency ranges delivered are also displayed. \label{fig:modelcomp}} 
\end{figure}

The values for the astrophysical parameters were picked within those allowed by the inference performed on LOFAR data \citep{2021MNRAS.501....1G}. 
Models inside the parameter space were further selected for having diverse reionization scenarios $x_{\rm HI}(z)$, as shown in Fig. \ref{fig:modelcomp}. 

The IM1 input consists of an observed (deg, deg, MHz) cube in Jy/beam units, covering a field of view of 8$\times$8 deg and a frequency range of 151--196\,MHz.

The power spectrum inputs PS1, PS2 and PS3 each consist of 3 power spectra $P(k)$, in units K$^2 h^{-3}$ cMpc$^3$, evaluated over a 15\,MHz frequency range from observed cubes of the same kind of IM1:
\begin{itemize}
\item 151-165.9 MHz, corresponding to a redshift range of 8.41-7.56
\item 166-180.9 MHz, corresponding to a redshift range of 7.56-6.85
\item 181-195.9 MHz, corresponding to a redshift range of 6.85-6.25
\end{itemize}
The FoV used for the power spectrum computation is the inner $4\times4$ degrees of the $8\times8$ field, to reduce noise contamination. We stress that the power spectra have been calculated on portions of lightcones, therefore they contain evolution within the considered 15\,MHz frequency interval. 

\begin{table}
\renewcommand{\arraystretch}{1.25}
\caption{Simulation components included in generating all the SDC3b datasets. \label{tab:datasummary}}
\begin{tabular}{llllll}

\hline
\textbf{Simulation component}&\textbf{PS1}&\textbf{PS2}&\textbf{PS3}&\textbf{IM1}\\
\hline
EoR (simulator 1)&\checkmark&&&\\
EoR (simulator 2)&&\checkmark&\checkmark&\checkmark\\
Foreground residuals&&&\checkmark&\checkmark\\
Calibration errors&\checkmark&\checkmark&\checkmark&\checkmark\\
Instrumental noise&100\,h&100\,h&1000\,h&1000\,h\\
Power spectrum estimation&\checkmark&\checkmark&\checkmark&\\
\hline
\end{tabular}
\end{table}

\begin{table}
\renewcommand{\arraystretch}{1.25}
\caption{EoR astrophysical parameters used for the datasets. The description of the parameters is in Table \ref{tab:21cmfast_parameters}. \label{tab:eorsummary}}
\begin{tabular}{llllll}
\hline
\textbf{Astrophysical parameter}&\textbf{PS1}&\textbf{PS2}&\textbf{PS3/IM1}\\
\hline
$f_{\star,10}$&0.0820635&0.0421095&0.0754762\\
$\alpha_\star$&0.469372&0.474432&0.607645\\
$t_\star$&0.339628&0.330636&0.265415\\
$\log\left(M_\mathrm{turn} \right)$&-&8.1824&8.45306\\
$f_\mathrm{esc,10}$&0.224732&0.064288&0.0754032\\
$\alpha_\mathrm{esc}$&-0.239605&0.217863&0.168536\\
$\log\left(M_\mathrm{min} \right)$&$10$&$-$&$-$\\
$R_\mathrm{max}$&$50$&$-$&$-$\\
\hline
\end{tabular}
\end{table}

The challenge also provided ancillary data, which are listed below:
\begin{itemize}
    \item Noise-only power spectra, generated in the same way as the true noise but with different seeds, as well as an average noise power spectrum between all realizations. This defines the noise bias in the data.
    \item Sample variance. The finite number of samples used for power spectrum estimation introduces a further source of error. The 1\,$\sigma$ sample variance error was made available. 
\end{itemize}
These data are useful for quantifying the observational uncertainties in the power spectra.

\section{Data challenge description}
This section presents the goals (Sec.~\ref{sec:goals}) and details (Sec.~\ref{sec:defined}) of the data challenge. In Sec.~\ref{sec:hpcs}, we describe the computing resources provided to the participating teams.

\subsection{Challenge goals}\label{sec:goals}
The inference of EoR properties is affected by multiple sources of uncertainty, two of which we aimed to investigate in this challenge. A first effect is related to the choice of prescription used by inference methods to model the ionization fraction field, $x_{\rm HII}$, given cosmological and astrophysical parameters. Multiple recipes are present in the literature for deriving this quantity, and the lack of strong observational constraints prevents us from clearly favouring one prescription over another. Systematic departures between different prescriptions could lead to inconsistent inference results by independent approaches. This is often referred to as simulation bias, where subtle methodological differences between forward models imprint characteristic differences in the 2D power spectra. 

The first and main goal of SDC3b, addressed by the PS1 and PS2 datasets, was to test the level of agreement between the wide range of EoR simulators used by the community. To isolate this modelling error component, no foreground contamination was added to those datasets. Two different simulators were adopted for PS1 and PS2 (as explained in Sec. \ref{sec:products}) to avoid favouring one specific approach during inference.  
The overall spread in results between different teams and for the two datasets can give us important insights on the level of agreement between different EoR simulation codes, and hopefully stimulate collaborations that will advance the field.

The second goal of the challenge was to assess the impact of residual foreground contamination on the inference; this aspect was addressed by the datasets PS3 and IM1. The modelling error discussed previously is still unavoidably present in those datasets, but the comparison between results on PS1/PS2 and PS3/IM1 was meant to assess the relative magnitude of modelling-induced and foreground-induced errors, thus giving a way to prioritise further work needed in the field ahead of the advent of the SKA data.

\subsection{The challenge defined}\label{sec:defined}
Teams were asked to analyse PS1 and PS2 and, optionally, PS3/IM1 and recover the mean neutral fraction $x_{\rm HI}=1-x_{\rm HII}$, which traces the reionization history. Most reionization models use a parametrisation that characterises the sources of ionizing radiation, and the resulting reionization history is a derived quantity. However, for this challenge it was still decided to use the mean neutral fraction as the large diversity of source parametrizations adopted in the literature made it impossible to identify a consistent parameter set that all teams could adopt. Previous inference tests on simulated data also suggested that the neutral fraction is well constrained by the power spectra \citep{2015MNRAS.449.4246G}. 

The required submission consisted of the full posterior for $x_{\rm HI}$ for the three redshift ranges under consideration (see Sec.~\ref{sec:products}) $P(x_{\rm HI}; z_1,z_2,z_3)$. This quantity encodes both the inferred neutral fractions and their uncertainties, with their full correlation information. 

Submissions were scored only at the end of the challenge and the input values were not revealed until then, ensuring that the challenge was blind. 

\subsection{Computational facility partners} \label{sec:hpcs}
Due to the computational complexity of the challenge, an effort was made to facilitate participation by offering dedicated computational resources to interested teams. This was done through a network of computational facility partners that kindly made some resources available to SDC participants. The network included various computing facilities, which generally had different hardware and software infrastructures. Globally, they were able to offer a diverse set of resources, including CPU and GPU computation, and different specifications regarding virtual and physical memory. 

No attempt was made to deliver a uniform resource allocation across different teams. Instead, teams specified their needs and were awarded the allocation that best matched their needs, subject to availability. 
The list of facility partners in alphabetical order, and a brief description of the provision they offered, is given in Appendix \ref{sec:partners}. 

\section{SDC3b teams} \label{sec:teams}
This section describes the analyses performed by the teams participating in SDC3b. Table \ref{tab:methods} provides a summary of the approaches employed by the teams, with detailed descriptions provided in a separate subsection for each team. At the heart of each analysis is a simulation of the 21-cm signal. Different simulators are used in this step. Several teams adopted the {\tt 21cmFAST} code; other approaches use a dark matter (DM) N-body simulation paired with a reionization prescription, emulators, or existing simulation sets. Instrumental effects and other possible corrupting effects are either incorporated in the data simulation step or modelled at the power spectrum level. 

Inference is performed in a variety of ways, but we can distinguish between Bayesian inference with explicit likelihood (analytic or emulated), Bayesian inference with implicit likelihood (i.e. Simulation-Based Inference, SBI), or inference where a neural network is trained to learn the relationship between parameters and data (e.g. Artificial Neural Network, ANN). The choice of free astrophysical parameters, $\theta_{\rm astro}$, depends on the chosen simulator, but even when the same one is used, teams choose to estimate a generally different number of free parameters (from 1 to 9, but typically 4). As pointed out above in Sec.~\ref{sec:defined}, for most teams the target parameter evaluated in this challenge, the neutral fraction $x_{\rm HI}$, is not directly sampled and is a parameter derived from the simulation results.

\begin{table*}
\renewcommand{\arraystretch}{1.25}
\centering
\caption{Summary of approaches used by all teams. Two independent {\tt 21cmFAST} parametrisations were used: \citet{park2019inferring} (called 21cmFAST in this Table) and \citet{mesinger201121cmfast} (called 21cmFAST (M)). The acronyms are: Markov Chain Monte Carlo (MCMC); Neural Network (NN); Artificial Neural Network (ANN); Power Spectrum (PS); Invertible Neural Network (INN); Conditional Invertible Neural Network (cINN); Convolutional Neural Network (CNN); Simulation-Based Inference (SBI); Multilayer Perceptron (MLP); Neural Ratio Estimation (NRE). $\theta_{\rm astro}$ identifies parameters used to describe the reionization scenario. $x_{\rm HI}$, the neutral fraction, is considered separately. The arrow $\rightarrow$ indicates when this parameter is derived separately from the other astrophysical parameters; $x_{\rm HI}(\theta_1,\theta_2)$ means this quantity is modelled in terms of additional parameters; $\theta_{\rm MU}$ identifies parameters to describe the model uncertainties. A full list of parameters for each approach is provided in the referenced table, where present; the approach is described in the referenced section. \label{tab:methods}}

\begin{tabularx}{\linewidth}{@{}
    >{\raggedright\arraybackslash}X
    >{\raggedright\arraybackslash}X
    >{\raggedright\arraybackslash}X
    >{\raggedright\arraybackslash}X
    >{\raggedright\arraybackslash}X
    >{\raggedright\arraybackslash}m{1.5cm}
@{}}

\toprule
\textbf{Team} &\textbf{21-cm simulation} &\textbf{Instrumental effects} &\textbf{Posterior sampling} &\textbf{Parameters}&\textbf{Section (Table)}\\
\midrule
21 To Infinities&DM PS $\rightarrow$ 21-cm PS&Noise correction applied&MCMC&1 $\theta_{\rm astro}$, $x_{\rm HI}$&\ref{sec:41}\\
Akashganga&21cmFAST&Noise correction/ Noise bias removal&ANN+MCMC&5 $\theta_{\rm astro}\rightarrow x_{\rm HI}$&\ref{sec:42} \,\,(\ref{tab:parameter_ranges_Akashganga})\\ 

Cantabrigians&21cmSPACEreion&Full mock observation&Emulator&4 $\theta_{\rm astro}\rightarrow x_{\rm HI}$&\ref{sec:43} \,\,(\ref{tab:cantabrigians_params})\\
COTSS-21&N-body sim. + Grizzly reionization scheme&&MCMC +PS grid&2 $\theta_{\rm astro}$ + $x_{\rm HI}(\theta_1,\theta_2)$&\ref{sec:44} \,\,(\ref{tab:cotss_params})\\
EoR-PIE&21cmFAST&Noise modelling&SBI using NPE with cINN&8 $\theta_{\rm astro}\rightarrow x_{\rm HI}$&\ref{sec:45}\,\,(\ref{tab:eorpie_params})\\
EoR-PIE-MC&21cmFAST (M)&Noise modelling&MCMC&3 $\theta_{\rm
astro}\rightarrow x_{\rm HI}$&\ref{sec:45} \,\,(\ref{tab:eorpiemc_params})\\
HIMALAYA&21cmFAST (M)&Noise modelling&cINN&3 $\theta_{\rm astro}\rightarrow x_{\rm HI}$&\ref{sec:46} \,\,(\ref{tab:himalaya_params})\\
HIstorian&21cmFAST&Full mock observation&Emulator+ Bayesian inference&8 $\theta_{\rm astro}\rightarrow x_{\rm HI}$&\ref{sec:47} \,\,(\ref{tab:historians_params})\\
Imperial-Nottingham&21cmFAST (M)&Noise modelling&SBI using NPE&3 $\theta_{\rm astro}\rightarrow x_{\rm HI}$&\ref{sec:48} \,\,(\ref{tab:imperial_params})\\
KUSANAGI&21cmFAST&&CNN&9 $\theta_{\rm astro}\rightarrow x_{\rm HI}$&\ref{sec:49}  \,\,(\ref{tab:21cmfast_parameters_kusanagi})\\
Loreli-A&Licorice simulations database&&Emulator+MCMC&3 $\theta_{\rm astro}\rightarrow x_{\rm HI}$&\ref{sec:410}\\
Loreli-B&Licorice simulations database&&SBI using NPE&$x_{\rm HI}$&\ref{sec:410}\\
ReionYuga PS1&N-body sim. + ionization emulator&&MCMC&1 $\theta_{\rm astro}$&\ref{sec:411}\\
ReionYuga PS2&21cmFAST (M)&&MCMC&1 $\theta_{\rm astro}$&\ref{sec:411}\\
RISE&SCRIPT &Noise modelling&MCMC&4 $\theta_{\rm astro}\rightarrow x_{\rm HI}$&\ref{sec:412} \,\,(\ref{tab:rise_params})\\
Modern SEarCH&21cmFAST&Noise modelling&MLP-Mixer&$x_{\rm HI}$&\ref{sec:413} \,\,(\ref{tab:search_params})\\
Traditional SEarCH&21cmFAST&Noise modelling&SBI with NN&$x_{\rm HI}$&\ref{sec:413}  \,\,(\ref{tab:search_params})\\
SegU-Net SEarCH&21cmFAST&Noise modelling&Feature identification with U-Net&$x_{\rm HI}$&\ref{sec:413}\\
Shuimu-Tianlai&21cmFAST (M)&Full mock observation&Neural posterior estimation&2 $\theta_{\rm astro}+ \theta_{\rm MU}\rightarrow x_{\rm HI}$&\ref{sec:414} \,\,(\ref{tab:tianlai_params})\\
ToSKA- Model selection&21cmFAST and C2Ray&&Emulators+Multinest& 4 $\theta_{\rm astro}\rightarrow x_{\rm HI}$&\ref{sec:415} \,\,(\ref{tab:toska_params})\\
ToSKA- Explicit likelihood&21cmFAST and C2Ray&&Emulators+Multinest& 4 $\theta_{\rm astro}\rightarrow x_{\rm HI}$&\ref{sec:415} \,\,(\ref{tab:toska_params})\\
ToSKA- SBI general&21cmFAST and C2Ray&&NRE&4 $\theta_{\rm astro}\rightarrow x_{\rm HI}$&\ref{sec:415} \,\,(\ref{tab:toska_params})\\
YEYE&21cmEMU&&MCMC&9 $\theta_{\rm astro}\rightarrow x_{\rm HI}$&\ref{sec:416} \,\,(\ref{tab:yeye_params})\\

\bottomrule
\end{tabularx}
\end{table*}

\subsection{21 To Infinities}\label{sec:41}
Team `21 To Infinities' considered a cosmologically motivated and computationally efficient approach for inferring the ionization fraction of the Universe for the given redshift ranges. 
The averaged noise power spectra provided as SDC3b ancillary data were used to correct for the contributions from instrumental systematics for both the PS1 and PS2 datasets. The sample variance files provided for both datasets were used as the 1\,$\sigma$ error bars. A Cold Dark Matter (CDM) power spectrum was generated by simulating the CDM density field using the code {\tt $N$-Body Tools}\footnote{\url{https://github.com/rajeshmondal18/N-body}} \citep{Mondal2015} within a co-moving volume of $[700\,{\rm cMpc}]^3$ with a grid-dimension of 1050. The post-Reionization dimensionless 21-cm power spectrum \citep{10.1111/j.1365-2966.2004.08604.x,2009MNRAS.397.1926W} was modelled as follows
\begin{equation}
\Delta_{21}^{2}(k)=\bar{T}_{\rm b}^{2}(z)[b_{\rm HI}\bar{x}_{\rm HI}]^{2}\Delta_{\rm m}^{2}(k).
\end{equation}

Here,
\begin{enumerate}

  \item $\Delta_{21}^{2}(k)$ and $\Delta^{2}_{\rm m}(k)$ are the dimensionless neutral hydrogen and CDM power spectra; 
    \item $\bar{T}_{b}(z) = 23.8\left(\dfrac{1+z}{10}\right)^{1/2}\left(\dfrac{\Omega_{\rm b}h^{2}}{0.02}\right)\left(\dfrac{0.15}{\Omega_{\rm m}h^{2}}\right)^{1/2}\text{ mK}$ is the mean 21-cm brightness temperature at the redshift $z$;
    \item $b_{\rm HI}$ is the neutral hydrogen bias factor; 
    \item $\bar{x}_{\rm HI}$ is the volume-averaged neutral hydrogen fraction;
   % \item $\Delta^{2}_{m}(k) = \dfrac{k^{3}}{2\pi^{2}}P_{CDM}(k)$ is the dimensionless CDM power spectrum, where $P_{\rm CDM}(k)$ is the 1D CDM power spectrum.
\end{enumerate}

The parameter $b_{\rm HI}$ was fixed to 0.2 as a simple phenomenological assumption, based on the fact that we consider, to a first approximation, a uniformly ionized IGM, i.e. there are no modulations by the ionizing background. The $\bar{x}_{\rm HI}$ was estimated within the range $[0.05,0.7]$ based on results from the previous measurements such as \cite{10.1093/mnras/stw3026},  \cite{10.1093/mnras/stac825} and \cite{refId0}.
The MCMC used the {\tt emcee}\footnote{\textcolor{blue}{\url{https://emcee.readthedocs.io/en/stable/}}} library with 32 walkers for 5000 steps. This yielded the volume-averaged neutral hydrogen fraction $\bar{x}_{\rm HI}$ together with its uncertainty for the given redshift range. 

This method is computationally cheap and a consumer-grade desktop computer with 32 GB of DDR5 RAM, AMD Ryzen 9 7900 CPU with 12 Cores and 24 Threads, Nvidia RTX 3060 GPU with 3584 CUDA Cores and 12 GB of GDDR6 VRAM was used.\\

\subsection{Akashganga}\label{sec:42}

Team `Akashganga' used an emulator-based pipeline to infer the posteriors of the volume-averaged neutral fraction $\bar{x}_{\rm HI}$. The inference pipeline consisted of three major stages: (1) simulating the training dataset, (2) building an emulator, (3) using the emulator for the Bayesian inference. These stages are discussed in more detail below.

Training the emulator requires a large number of mock EoR 21-cm signal 2D power spectra and the corresponding IGM parameters. The publicly available semi-numerical code {\tt 21cmFAST} v4 was used to generate a training dataset comprising 2000 different astrophysical models of EoR 21-cm lightcones within the frequency ranges provided in the challenge. The 2D power spectrum and the volume-averaged neutral fraction were computed from these boxes on the fly and stored for later use. 
Most of the parameters in the code were kept at their default values and only $5$ of them were varied to construct the training set. 
Prior ranges for these parameters 
are reported in Table \ref{tab:parameter_ranges_Akashganga} and 
were selected by running an MCMC inference on the data provided directly with {\tt 21cmFAST} low-resolution runs. As already mentioned, a parameter set of $2000$ Latin-Hypercube samples was generated and 21-cm brightness temperature lightcones were simulated following the simulation specifications mentioned in the data challenge documentation. The generation of the data set took roughly a couple of days. Finally, the volume-averaged neutral fractions $\bar{x}_{\rm HI}$ at the central redshifts and the 2D power spectra for all the $2000$ samples were computed. 
The neural networks were trained on 90 per cent of this dataset, with the remaining 10 per cent used for testing. Validation of the neural networks was performed on a handful of other cases within the range of priors.

\begin{table}
% table OK
\renewcommand{\arraystretch}{1.25}
%\centering
\caption{{\tt 21cmFAST} \citet{park2019inferring} astrophysical parameters (described in Table \ref{tab:21cmfast_parameters}) used by Akashganga and their adopted priors.}
\label{tab:parameter_ranges_Akashganga}

\begin{tabularx}{\linewidth}{@{} >{\raggedright\arraybackslash}m{4cm}
 >{\raggedright\arraybackslash}m{4cm}
@{}}
\toprule
\textbf{Parameter} &\textbf{Prior} \\
\midrule

$f_{\star,10}$ &Log$(10^{-2.5},\;10^{-1})$ \\

$\alpha_{\star}$ &Flat$(-0.5,\;1.0$) \\

$f_{\rm esc,10}$ &Log$(10^{-1.8},\;10^{-0.5})$ \\

$\alpha_{\rm esc}$ &Flat$(-1.0,\;1)$ \\

$M_{\rm turn}$ &Log$(10^{8},10^{10})$ \\

\bottomrule
\end{tabularx}
\end{table}

A combination of two Artificial Neural Network (ANN)-based models was used to obtain the final inference on $\bar{x}_{\rm HI}$. The first ANN model connects the five astrophysical parameters to the resulting 2D power spectrum on the $10 \times 10\,(k_\perp, k_\parallel)$ grid matching the SDC3b data product. The team designed a deep ANN (6 dense layers) which 
used the {\tt Adam} optimiser \citep{kingma2017} during training and minimised the mean squared error (MSE) metric at the output power spectra. The training process was regularised by using the $L2$ norm (weight decay) to reduce overfitting. The networks were trained for up to $10,000$ epochs and a good training was ensured by looking at the MSE cost function. Since the simulated EoR 21-cm power spectra span several orders of magnitude, a custom normalization was applied to bring the power spectra within a range appropriate for the ANN modules. The normalization was reverted later. With network architecture and its parameters  roughly the same, three different networks were trained for the lightcones at the given three bandwidths separately.

To infer $\bar{x}_{\rm HI}$, another (less deep, 4 dense layers) ANN network was built and trained, which takes the astrophysical parameters as input and outputs three $\bar{x}_{\rm HI}$ corresponding to the three bandwidths. This network is built to map the sampled astrophysical parameter chain to $\bar{x}_{\rm HI}$ without additional forward simulations. The same optimiser, cost function, regularisation scheme, and total number of epochs as before were used. However, it was possible to train a single ANN that provides the three $\bar{x}_{\rm HI}$ together. The goodness of training was again verified using the convergence of the cost function with training epochs. The network was built using the publicly available ANN package {\tt Pytorch} \citep{NEURIPS2019_9015}, which was run on the GPUs for faster training. Training each network took approximately 10 minutes. Finally, the trained networks were saved to be used in the final Bayesian inference step. 

The {\tt emcee} package \citep{2013PASP..125..306F} was used to build an MCMC pipeline that is dependent on the first ANN that connects the aforementioned astrophysical parameters and the cylindrical power spectra. A joint analysis combined the likelihoods at the three redshift bands provided here. The emulated power spectra are compared with the provided mock data corresponding to their frequency bands. The analysis also includes the emulation errors added in quadrature to the provided error estimates. A chain of astrophysical parameters is generated and stored and subsequently passed through the second ANN to generate a chain of $\bar{x}_{\rm HI}$ for the three frequency bands. The chain of the three $\bar{x}_{\rm HI}$ values is finally used to generate its 3D posterior normalized by the total length of the chain.

As an improvement of the current method, adding different simulations in the training dataset would help overcome simulation bias, as demonstrated and studied by \citet{Zhou_2022} and \citet{Solt_2026}. A difference in performance of this approach between the PS1 and PS2 datasets (see Sec. \ref{sec:results}) suggests that simulation bias could have impacted these results.

%%%%%%%%%%%%%%%%%%%%%%%%%%%%%%%%%%%%%%%%%%%

%%%%%%%%%%%%%%%%%%%%%%%%%%%%%%%%%%%%
%%%%%%%%%%%%%%%%%%%%%%%%%%%%%%%%%%%%%

\subsection{Cantabrigians}\label{sec:43}
For this data challenge, the Cantabrigians collaboration has developed a workflow to perform Bayesian inference on the ionization history of the Universe. The workflow can be divided into three phases. The first phase simulates the brightness temperature and neutral fraction ($x_{\rm HI}$) of the 21-cm signal. The second phase focuses on mock observations of the EoR signal and signal processing and involves instrumental modelling, image synthesis and power spectrum estimation. In the final phase, the team utilizes the generated dataset from Phase 2 to train emulators that predict EoR quantities, enabling neural-network-based Bayesian inference. 

The team started with generating 1000 realizations of the 21-cm signal using \texttt{21cmSPACEreion}. This simulation code is a new, simplified version of \texttt{21cmSPACE} \citep{2012Natur.487...70V,2012MNRAS.424.1335F,2013MNRAS.432.2909F,2014MNRAS.437L..36F} that focused exclusively on the physics of reionization. As a result, the code omits astrophysical processes that are only significant during early stages of the cosmic history and that are expected to be irrelevant at the redshifts of this data challenge, such as Lyman-$\alpha$ coupling (saturated coupling assumed), Lyman-Werner feedback, and X-ray heating (saturated heating assumed). DM haloes were modelled through a sub-grid conditional halo mass
function \citep{2004ApJ...609..474B,2011MNRAS.418..906T}, as in \texttt{21cmSPACE}. Reionization was modelled using the excursion set formalism \citep{2004ApJ...613....1F}, with 1000 simulations generated by sampling the four-dimensional astrophysical parameter space, listed in Table~\ref{tab:cantabrigians_params}. Please note that this model uses a minimum circular velocity $V_\mathrm{c}$ instead of a minimum halo mass. For each of the resulting differential brightness temperature cubes, a lightcone was constructed with the \texttt{physical\_lightcone\_to\_observational} function from \texttt{tools21cm}.

To construct a robust training dataset, the team performed mock observations of the 21-cm signal by using the \texttt{OSKAR} interferometric simulator. The simulation followed the procedures outlined by \citet{2026MNRAS.549ag878L}, which were adapted from \citet{2024MNRAS.533.2876O} and \citet{sdc3a}. This allowed the team to account for the beam response of SKA-Low and generated the more realistic data products. Although \texttt{OSKAR} supported distinct beam patterns for individual antenna elements, the team modelled each antenna response analytically as a resonant half-wavelength dipole. The lightcone was transformed into local observable coordinates, with the center of the spatial plane positioned at the zenith at LST = 0h. Visibilities were generated for an observation across the bandwidth and channel width as described in Section~\ref{sec:datasim}, however with a single 10 second integration. These visibilities were subsequently imaged using \texttt{wsclean}. The imaging used natural weighting without deconvolution. These image cubes were then Fourier transformed and azimuthally averaged using the \texttt{ps\_eor} package to produce the cylindrical 2D power spectra.

\begin{table}
%TABLE OK

\renewcommand{\arraystretch}{1.25}
\centering
\caption{21cmSPACEreion EoR parameters used by Cantabrigians to construct the training dataset of the 21-cm signal and described in \citet{2012Natur.487...70V,2012MNRAS.424.1335F,2013MNRAS.432.2909F,2014MNRAS.437L..36F}.}

\label{tab:cantabrigians_params}
\begin{tabularx}{\linewidth}{@{} >{\raggedright\arraybackslash}m{1.0cm} >{\raggedright\arraybackslash}X >{\centering\arraybackslash}m{2.7cm} @{}}
\toprule
\textbf{Parameter} &
\multicolumn{1}{c}{\textbf{Description}} &
\textbf{Prior} \\
\midrule
$\zeta_\mathrm{eff}$ &
Effective ionizing efficiency of collapsed gas &
$\mathrm{Flat}(10,100)$ \\
$\zeta_\mathrm{exp}$ &
Power-law index of ionizing efficiency &
$\mathrm{Flat}(0.5,2)$ \\
$R_\mathrm{mfp}$ & Maximum root mean free path of ionizing photons [cMpc]& $\mathrm{Flat}(10,100)$ \\
$V_\mathrm{c}$ & Minimum circular velocity of haloes that can host star formation [km\,s$^{-1}$]& $\mathrm{Log}(4.2,100)$ \\
\bottomrule
\end{tabularx}
\end{table}

\tikzstyle{start} = [diamond, draw, text centered,  rounded corners, minimum size=1cm, align=center]
\tikzstyle{analysis} = [rectangle, draw, text centered, rounded corners, minimum size=1cm, align=center]
\tikzstyle{connector} = [draw, -latex']

\tikzstyle{start} = [oval, draw, text centered,  rounded corners, minimum size=1cm, align=center]
\tikzstyle{process} = [rectangle, draw, text centered, rounded corners, minimum size=1cm, align=center]
\tikzstyle{decision} = [diamond, draw, text centered, rounded corners, minimum size=1cm, align=center]
\tikzstyle{data} = [trapezium, trapezium left angle=70, trapezium right angle=110, draw, text centered, rounded corners, minimum size=1cm, align=center]

\tikzstyle{connector} = [draw, -latex', thick]

\tikzstyle{arrow} = [draw, -latex', thick, dashed]

\begin{figure}
    \centering

    \begin{tikzpicture}
        \node[process, scale=1, fill=orange!60] at (-1, 0) (1) {Parameter \\ prior};

        \node[process, scale=1, fill=purple!50] at (1.5, 0) (2) {21-cm power \\ spectrum emulator};

        \node[data, scale=1, fill=yellow!60] at (4.5, 0) (3) {Observed \\ SKA power};

        \node[process, scale=1, fill=orange!60] at (3, -3) (4) {Likelihood};

        \node[decision, scale=1, fill=teal!60] at (0.25, -3) (5) {Termination?};

        \path [connector] (1) -- (2);
        \path [connector] (2) -- (4);
        \path [connector] (3) -- (4);
        \path [connector] (4) -- (5);

        \node[process, scale=1, fill=orange!60] at (-1, -4.8) (7) {Parameter \\ posterior};

        \node[process, scale=1, fill=orange!60] at (5, -2.2) (10) {Updated \\ samples};

        \path [connector] (5) -| (7);
        \path [arrow] (5) |- (5, -1.5) -- (10) |- (4.east);

        \node[process, scale=1, fill=purple!50] at (1.8, -4.8) (8) {xHI emulator};

        \node[process, scale=1, fill=orange!60] at (4.8, -4.8) (9) {xHI posterior};

        \path [connector] (7) -- (8);
        \path [connector] (8) -- (9);
        
    \end{tikzpicture}
     \caption{Flowchart that illustrates the inference procedures that the Cantabrigians team adopted during SDC3b. Geometric shapes denote the type of steps. Rectangles represent processes, parallelograms denote data inputs, and diamonds indicate decisions. In addition, the colours further categorize each type. Orange corresponds to processes related to nested sampling, purple highlights processes that predict astrophysical quantities using emulators, and teal represents the nested sampling termination. \label{fig:cantab_flowchart}}
\end{figure}
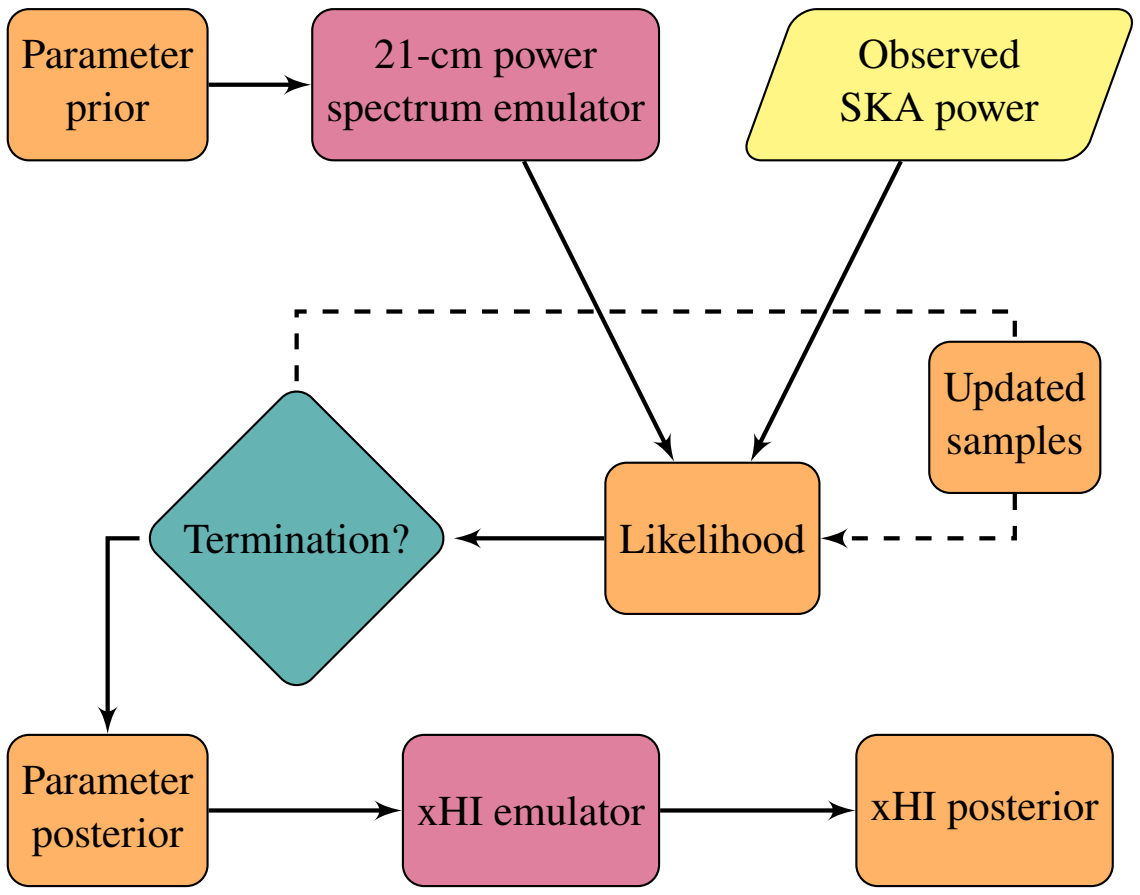

The inference framework utilized two least-squares regression neural networks - specifically multilayer perceptrons (MLPs). These MLP emulators took $[-1,1]$ min-max normalized astrophysical parameters and data parameters, 
\begin{equation}
    \boldsymbol{\theta} = (\mathbf{d}, \zeta_\mathrm{eff}, \zeta_\mathrm{exp}, R_\mathrm{mfp}, V_c)
\end{equation}
where $\mathbf{d} = (z, k_\perp, k_\parallel)$ for the power spectrum emulator, and $\mathbf{d}=(z)$ for the neutral fraction emulator, and predicted the 21-cm power spectra or neutral fraction. Emulator training was inspired by \cite{pochinda_inference_2024} and employed stochastic gradient descent and the Adam optimizer \citep{kingma2017} with a learning rate of $10^{-3}$ to fit network parameters. To prevent overfitting regularization techniques were used, including weight decay with a coefficient of $10^{-4}$ and early stopping.
As illustrated in Fig.~\ref{fig:cantab_flowchart}, the Bayesian inference started by drawing astrophysical parameters from the prior and feeding them to the 21-cm power spectrum emulator to predict the 21-cm power. To determine the likelihood of each sample, the predicted power was compared against the observed SKA power in a Gaussian likelihood, 
\begin{equation}
\begin{split}
    \log \left(\mathcal{L}(\boldsymbol{\theta})\right) = \sum_{i}^{N_\mathrm{d}} 
    \left(
    -\frac{1}{2} \log\left(2\pi\left(\sigma_\mathrm{noise}^2 + \sigma_\mathrm{emu}^2\right)\right)
    \right. \\
    \left.
    -\frac{1}{2} 
    \frac{(P(k_\perp, k_\parallel,z) - P_\text{avg err} (k_\perp, k_\parallel,z) - P(\boldsymbol{\theta}))^2}{(\sigma_\mathrm{noise}^2+\sigma_\mathrm{emu}^2)}
    \right).
\end{split}
\end{equation}
Here $N_\mathrm{d}$ is the number of data points, $P(k_\perp, k_\parallel,z)$ is the observed power spectrum, $P_\text{avg err} (k_\perp, k_\parallel,z)$ is the average noise power spectrum provided for the challenge, and $P(\boldsymbol{\theta})$ is the emulated power spectrum for parameters $\boldsymbol{\theta}$. Lastly, $\sigma_\mathrm{noise}$ is the aforementioned fitted residual noise, and $\sigma_\mathrm{emu}=0.1\,P(\boldsymbol{\theta})$ denotes an assumed ten percent uncertainty in emulator predictions.
For this inference, we used the nested sampling algorithm implemented in PolyChord \citep{2015MNRAS.450L..61H,2015MNRAS.453.4384H}. The sampling generated likelihood-constrained samples and iteratively accumulated the Bayesian evidence. Upon completion, the loop concluded and output the posterior distribution on the astrophysical parameters. We refer interested readers to \citet{10.1214/06-BA127,2022NRvMP...2...39A} for more detailed nested sampling procedures. The posterior samples of astrophysical paramters were propagated through the $x_\mathrm{HI}$ emulator to yield the $x_\mathrm{HI}$ posterior distribution.

\subsection{COTSS-21 - COnstraining The SKA 21-cm Signal}\label{sec:44}

\begin{table}
% Table OK 
\renewcommand{\arraystretch}{1.25}
\centering
\caption{Parameters used by COTSS-21 (see Sec. \ref{sec:sourcemodel} for more details) and their  priors.}
\label{tab:cotss_params}
\begin{tabularx}{\linewidth}{@{} >{\raggedright\arraybackslash}m{1.0cm} >{\raggedright\arraybackslash}X >{\centering\arraybackslash}m{2.7cm} @{}}
\toprule
\textbf{Parameter} &
\multicolumn{1}{c}{\textbf{Description}} &
\textbf{Prior} \\
\midrule
$M_\mathrm{min}$ &Minimum halo mass of ionising sources [M$_\odot$]&$\mathrm{Log}(10^{9},10^{12.5})$ \\
$\alpha$ &emissivity-mass scaling index&$\mathrm{Flat}(0.1,3)$ \\
$z_\mathrm{mid}$ & Midpoint of reionization& $\mathrm{Flat}(5.5,8)$ \\
$\Delta z$ & Reionization duration & $\mathrm{Flat}(0.1,3)$ \\
\bottomrule
\end{tabularx}
\end{table}

The \textsc{COTSS-21} team used a semi-numerical forward-modelling framework to interpret the PS1 and PS2 21-cm power-spectrum data in terms of the underlying astrophysical parameters driving the EoR. 
The methodology combines an $N$-body cosmological simulation providing the DM and halo distributions with a photon-conservation reionization scheme to produce 21-cm signal lightcones. 
Four key reionization parameters were used, listed in Table \ref{tab:cotss_params} and described in Sec. \ref{sec:sourcemodel}. 
A likelihood-based comparison was then performed between the mock 2D power spectra, $P(k_\perp,k_\parallel)$, and the PS1 and PS2 data to constrain the ionization fractions at three specific redshifts. 
This approach enables a physically interpretable yet computationally efficient link between 21-cm observables and the astrophysics of early galaxies and the intergalactic medium (IGM). A brief description of the framework is as follows.

\subsubsection{N-body simulation}
The analysis adopts the $\Lambda \rm CDM$ cosmology with cosmological parameters $\Omega_{\mathrm{m}}=0.27$, $\Omega_\Lambda=0.73$, $\Omega_{\mathrm{B}}=0.044$, $h=0.7$ \citep{2013ApJS..208...19H}, consistent with the $N$-body simulations used. These parameter values differ from the cosmological parameter values used for the simulations used to generate the PS1 and PS2 datasets (see Section~\ref{sec:cosmo_evol}). The underlying DM density and velocity fields were taken from the {\tt PRACE4LOFAR} DM-only simulation suite \citep{2020MNRAS.493.4728G}, performed with the high-performance code {\tt CUBEP3M} \citep{Harnois12} . 
The simulation volume is a cube of 500\,$h^{-1}$\,cMpc on a side with a particle mass of $4.05\times10^{7}\,\rm M_\odot$, providing sufficient dynamic range to capture both large-scale cosmological modes ($k\simeq0.05\,h\,\mathrm{Mpc}^{-1}$) and the formation of haloes relevant for reionization. 
DM haloes were identified on-the-fly via a spherical-overdensity halo finder \citep{Watson2013TheAges}, retaining haloes with more than 25 particles, corresponding to a minimum resolved mass of $\sim10^{9}\,\rm M_\odot$. 
The density and velocity fields were interpolated onto a 3D uniform grid, with $300^3$ voxels (see e.g., \citealt{giri2019neutral, 2019JCAP...02..058G}), forming the basis for the subsequent ionization field generation step. 
These outputs provide a realistic cosmic density structure and halo clustering pattern from which ionizing sources were assigned.

\subsubsection{Source model and parameters}\label{sec:sourcemodel}
The four parameters in Table \ref{tab:cotss_params} define the model space controlling both the ionizing emissivity and the redshift evolution of the neutral fraction. 
Each DM halo with mass $M_{\mathrm{halo}}>M_{\mathrm{min}}$ was assumed to host an ionizing source whose photon production rate scales as
\begin{equation}
    \dot{N}_\gamma \propto M_{\mathrm{halo}}^{\,\alpha},
    \label{eq:n}
\end{equation}
where $\alpha$ sets the mass dependence of the ionizing emissivity. 
This simple power-law prescription captures the expected enhancement of star-formation efficiency in massive haloes. 

The global reionization history was described analytically as a smooth function of redshift following (see \citealt{2017PhRvD..95b3513H} for more details):
\begin{equation}
x_{\mathrm{HII}}(z) = \tfrac{1}{2}\left[1 + \tanh\!\left(\frac{z_{\mathrm{mid}} - z}{\Delta z}\right)\right],
\label{eq:xhii}
\end{equation}
where $z_{\mathrm{mid}}$ is the redshift at which $x_{\mathrm{HII}}=0.5$, and $\Delta z$ quantifies the duration of reionization.

\subsubsection{21-cm lightcone}
For an input $\theta$, coeval cubes of 21-cm signal at 8 different redshifts between 6.6 and 7.9 were first generated. 

First, $x_{\mathrm{HII}}(z_{i=1,8})$ is determined using Equation \ref{eq:xhii} for the input $\alpha,\,z_{\mathrm{mid}},\,\Delta z$ values. Given the input $M_{\rm min}$, the proportionality constant in Equation \ref{eq:n} is determined so that the coeval ionization fraction at $z_i$ matches with $x_{\mathrm{HII}}(z_{i})$ as estimated using Equation \ref{eq:xhii}. Next, using Equation \ref{eq:xhii}, the number of ionizing photons for each DM halo are estimated. The photon-conservation scheme of {\tt grizzly} was used to generate the coeval ionization cubes at those redshifts.  This scheme puts spherical ionized regions around the DM haloes, estimates the unused ionizing photons for each overlapping {\rm H{\sc ii }} region and distribute them around the overlapping regions such that each photon ionizes one hydrogen atom.

Next, the coeval $\delta T_{\mathrm{b}}$ cubes are generated using those coeval ionization fractions and density fields. Additionally, redshift-space-distortion on the coeval $\delta T_{\mathrm{b}}$ cubes are incorporated, a physical effect which was actually {\it not} used in the construction of the SDC3b datasets.  Finally, the $\delta T_{\mathrm{b}}$ brightness temperature lightcone is generated within the desired redshift range by using the coeval $\delta T_{\mathrm{b}}$ cubes and an interpolation scheme. From each cube, the two-dimensional power spectrum $P(k_\perp,k_\parallel)$ is computed using identical binning to the provided datasets.

\subsubsection{Neutral hydrogen fraction inference}
As generating a lightcone for an input $\theta$ is computationally costly, the lightcone generating framework wasn't used directly in the Bayesian analysis for constraining the neutral fraction at the given redshifts. Instead, $10^4$ 2D power spectra were first generated for different combinations of the model parameters. 
Ten bins were used for each parameter within the ranges in Table \ref{tab:cotss_params}, yielding a total of $10^4$ parameter sets. 

For the Bayesian analysis on the given 2D power spectra, the publicly available MCMC sampler code {\tt cobaya\footnote{\url{https://cobaya.readthedocs.io/en/latest/index.html}}}\citep{2002PhRvD..66j3511L} was used. For a chosen parameter $\theta$ in the parameter estimation step, the 2D power spectra, $P_\mathrm{model}^{(i,j)}$, were estimated by linearly interpolating from the previously generated power spectra set. 

The astrophysical parameters were inferred by comparing the modelled power spectra with the provided observations via a maximum-likelihood estimator:
\begin{equation}
    \mathcal{L} \propto 
    \exp\!\left[-\tfrac{1}{2}\sum_{i,j}
    \frac{\big(P_\mathrm{obs}^{(i,j)} - P_\mathrm{model}^{(i,j)}\big)^2}
    {\sigma_{ij}^2}\right],
\end{equation} where $\sigma_{ij}$ denotes the uncertainty in each $(k_\perp,k_\parallel)$ bin. By scanning the four-dimensional parameter space, the region maximizing $\mathcal{L}$ was identified, and marginalized confidence intervals for each parameter were extracted. 

For the MCMC step, the ranges of $\alpha,\,z_{\mathrm{mid}},\,\Delta z, \, \log(M_{\rm min}/\rm M_{\odot})$ were kept the same as above. From the results of the MCMC, the values of neutral fractions were derived at the three different redshifts using Equation \ref{eq:xhii}. 
Since the
MCMC analysis is done on the $\alpha,\,z_{\mathrm{mid}},\,\Delta z, \,
\log(M_{\rm min}/\rm M_{\odot})$ parameters, the chosen ranges on the model parameters are expected to have a significant impact on the constraints on the neutral fractions at those three redshifts.

\subsection{EoR-PIE}\label{sec:45}

\begin{table}
\renewcommand{\arraystretch}{1.25}
\centering
\caption{{\tt 21cmFAST} \citet{park2019inferring} astrophysical parameters (described in Table \ref{tab:21cmfast_parameters}) used by EoR-PIE and their adopted priors.\label{tab:eorpie_params}}
\begin{tabularx}{\linewidth}{@{} >{\raggedright\arraybackslash}m{4cm}
 >{\raggedright\arraybackslash}m{4cm}
@{}}
\toprule
\textbf{Parameter} &\textbf{Prior} \\
\midrule
$f_{\star,10}$ &Log$(-3.0,\;0.0)$ \\
$\alpha_{\star}$ &Flat$(-0.5,\;1.0$) \\
$f_{\rm esc,10}$ &Log$(-3.0,\;0.0)$ \\
$\alpha_{\rm esc}$ &Flat$(-1.0,\;0.5)$ \\
$t_{\star}$ &Flat$(0.1,\;1.0)$ \\
$M_{\rm turn}$ &Log$(10^{8},10^{10})$ \\
$L_X$ &Log($10^{38}, 10^{42}$) \\
$E_0$ &Flat$(0.1,\;1.5)$ \\
\bottomrule
\end{tabularx}
\end{table}

The EoR-PIE team employed an SBI approach for the PS1 and PS2 datasets. Using {\tt 21cmFAST\_v4}, a database of 11,628 lightcones was generated with a box length of 700\,cMpc and a cell size of 2\,cMpc, covering redshifts 6–12. To ensure a wide range of astrophysical models, the model parameters introduced in \cite{park2019inferring} were varied (see Table \ref{tab:eorpie_params}) and the random seed was changed to account for cosmic variance. These physical lightcones were then transformed into observational ones using {\tt tools21cm}. The {\tt ps\_eor} package was subsequently used to compute the 2D power spectra (2D PS) for the relevant frequency chunks, which were then rebinned to match the challenge data format. The challenge datasets PS1 and PS2 were preprocessed by subtracting the averaged noise power spectra provided by the SDC3b team, under the assumption that the systematic contribution was constant. During the course of the challenge, {\tt EoRFlow}~\citep{Pietschke2025} was developed. This is a conditional invertible neural network (cINN) based on normalizing flows, designed for fast and robust posterior estimation from 21-cm summary statistics. {\tt EoRFlow} was trained on the generated 2D PS with the objective of learning the neutral hydrogen fraction for each frequency chunk. To account for varying thermal noise realizations, Gaussian noise was added to the 2D PS during training, with the amplitude calibrated using the noise realizations provided by the challenge. Once trained, {\tt EoRFlow} was evaluated on unseen 2D PS to infer the joint posterior $P(x_\mathrm{HI}; z_1, z_2, z_3)$. The domain-shift error, occurring when applying the model to a different simulator, was tested on a set of {\tt Astraeus} simulations \citep{Hutter2021, Hutter2025}. Additionally, EoR-PIE submitted complementary classical MCMC-based results for comparison. For EoR-PIE MC simulation samples were drawn from uniform priors over the ranges in Table \ref{tab:eorpiemc_params}, then 2D PS and corresponding $x_{\rm HI}$ were derived and the likelihood evaluated based on the 2D PS for the
respective MC step. Note that instead of a minimum halo mass, this model uses a minimum virial temperature for the haloes.

\begin{table}
\renewcommand{\arraystretch}{1.25}
\centering
\caption{{\tt 21cmFAST} astrophysical parameters following the \citet{mesinger201121cmfast} parametrization used by EoR-PIE-MC and their adopted priors.\label{tab:eorpiemc_params}}
\begin{tabularx}{\linewidth}{@{} >{\raggedright\arraybackslash}m{1.0cm} >{\raggedright\arraybackslash}X >{\centering\arraybackslash}m{2.7cm} @{}}
\toprule
\textbf{Parameter} &
\multicolumn{1}{c}{\textbf{Description}} &
\textbf{Prior} \\
\midrule
$\zeta$ & Global ionizing efficiency&$\mathrm{Flat}(5,100)$\\
$T_{\rm vir}$& Minimum halo virial temperature [K]& $\mathrm{Log}(10^{3.8},10^{5.2})$\\
$R_{\rm max}$&Maximum radius of ionised bubbles [cMpc]&$\mathrm{Flat}(2,100)$\\
\bottomrule
\end{tabularx}
\end{table}

\subsection{HIMALAYA}\label{sec:46}
Team HIMALAYA used a FLOW-based machine learning approach, the Conditional Invertible Neural Network (cINN)~\citep{ardi2019,Zhang2023}, to infer the joint posterior of the reionization fraction at three redshifts. Latent variables drawn from standard normal distributions were transformed into the ionization fraction distributions via spline-based transformations~\citep{durkan2019neural}, conditioned on compressed statistics of the corresponding power spectra.

The compression was performed using a 1D Convolutional Neural Network (CNN) and a Multilayer Perceptron (MLP) head, the only trainable components. During training, fiducial ionization fractions were mapped to latent variables constrained to follow a standard normal distribution. The spline transformations ensure invertibility, while the CNN+MLP extracts features from the flattened power spectrum that capture the key statistical relationships needed for the transformation.

\begin{table}
\renewcommand{\arraystretch}{1.25}
\centering
\caption{{\tt 21cmFAST} \citet{mesinger201121cmfast} astrophysical parameters (described in Table \ref{tab:eorpiemc_params}) used by HIMALAYA and their adopted priors.\label{tab:himalaya_params}}

\begin{tabularx}{\linewidth}{@{} >{\raggedright\arraybackslash}m{4cm}
 >{\raggedright\arraybackslash}m{4cm}
@{}}

\toprule
\textbf{Parameter} &\textbf{Prior} \\

\midrule
$\zeta$ &$\mathrm{Flat}(10,60)$ \\

$T_\mathrm{vir}$ &$\mathrm{Flat}(10^4,10^{5.5})$ \\

$R_\mathrm{max}$  & $\mathrm{Flat}(5,30)$ \\ 
\bottomrule
\end{tabularx}
\end{table}

The training dataset initially comprised of 2,000 pairs of ionization fraction $x_{\rm HI}$ and corresponding power spectra generated from high-resolution lightcones produced by {\tt 21cmFAST v3.4}. The lightcones were created by uniformly sampling three key parameters of the model, as reported in Table \ref{tab:himalaya_params}. To augment the dataset, noise realizations accounting for cosmic variance were added to the power spectra, expanding the dataset to 20,000 samples.

IM1 and PS3 exhibit small foreground residuals; to mitigate these, principal component analysis (PCA) was applied on the image using the frequency–frequency covariance, and the first eigenmode map was subtracted. The suppression factor obtained from this procedure was then used to rescale the PS3 power spectrum accordingly. Subsequently, the maps IM, PS1, PS2, and PS3 were corrected by subtracting their respective average noise levels before being passed to the neural network for compression. The posterior distribution was evaluated by sampling latent variables from a standard normal distribution and transforming them into the corresponding ionization fraction fields. Tests on the validation dataset indicated that our model can accurately predict the ionization fractions when conditioned on the compressed representations of the corresponding power spectra.

\subsection{HIstorians}\label{sec:47}
The HIstorians team employed a hybrid framework that combines
machine-learning--based emulation with Bayesian inference to recover the
underlying astrophysical parameters governing the 
EoR. Three classes of emulators were considered: Artificial Neural
Networks (ANNs), Bayesian Neural Networks (BNNs), and Gaussian Process
Regression (GPR). The overall workflow consists of two main stages:
(i) generation of the training and validation datasets and construction
of the machine-learning emulators, and (ii) Bayesian parameter inference
using the trained emulators as computationally efficient surrogates for
the full 21-cm simulation pipeline. The first stage involves simulating
the 21-cm signal, constructing physical and observational lightcones,
modelling the instrumental response, incorporating relevant observational
effects, and estimating the 21-cm power spectrum.

To construct the training dataset, 600 reionization models were generated using {\tt 21cmFAST} V3.4.0 by varying eight astrophysical parameters, listed in Table~\ref{tab:historians_params} together with their ranges. The parameter space was sampled using Hammersley Sequence Sampling (HSS). These simulations establish the mapping between the astrophysical parameters and the corresponding 21-cm observables required to train and validate the three emulators.

\begin{table}
\renewcommand{\arraystretch}{1.25}
%\centering
\caption{{\tt 21cmFAST} \citet{park2019inferring} astrophysical parameters (described in Table \ref{tab:21cmfast_parameters}) used by HIstorians and their adopted priors.}
\label{tab:historians_params}

\begin{tabularx}{\linewidth}{@{} >{\raggedright\arraybackslash}m{4cm}
 >{\raggedright\arraybackslash}m{4cm}
@{}}
\toprule
\textbf{Parameter} &\textbf{Prior} \\
\midrule

$f_{\star,10}$ &Flat$(-3.0,\;0.0)$ \\

$\alpha_{\star}$ &Flat$(-0.5,\;1.0]$) \\

$f_{\rm esc,10}$ &Flat$(-3.0,\;0.0)$ \\

$\alpha_{\rm esc}$ &Flat$(-1.0,\;0.5)$ \\

$t_{\star}$ &Flat$(0.1,\;1.0)$ \\

$M_{\rm turn}$ &Log$(10^{8},10^{10})$ \\

$L_X$ &Log($10^{38}, 10^{42}$) \\

$E_0$ &Flat$(0.1,\;1.5)$ \\

%\hline
\bottomrule
\end{tabularx}
\end{table}

The initial conditions were generated in a $700\,\mathrm{Mpc}$ simulation
box. The simulations employed a spatial resolution of $2\,\mathrm{Mpc}$
for the ionization field and $0.67\,\mathrm{Mpc}$ for the underlying matter
density field. The cosmological parameters were kept fixed for all
simulation sets and correspond to the Planck 2018 cosmology, the same as used for the challenge data (see Section~\ref{sec:cosmo_evol}).
For each astrophysical parameter
combination, coeval differential brightness-temperature cubes were
generated at a sequence of redshifts. These coeval cubes were subsequently
combined using the {\tt make\_lightcone} routine, which linearly
interpolates between adjacent redshifts along the line of sight to
construct the evolving physical 21-cm lightcones.

For each resulting differential brightness-temperature cube, an
observational lightcone was generated using
{\tt tools21cm}.
Mock radio interferometric observations of the 21-cm signal were produced
using the {\tt 21cmE2E} pipeline
\footnote{\url{https://gitlab.com/samit-pal/21cme2e.git}}, which is based
on {\tt OSKAR}.
This synthetic radio-observation framework accounts for the effects of
the telescope layout, instrumental response, and relevant observational
systematics, thereby enabling a comprehensive, data-driven approach to
reliable inference of the 21-cm signal. These observational effects
include residual antenna-based gain calibration errors, residual
direction-dependent calibration errors associated with ionospheric
effects, and instrumental thermal noise corresponding to the anticipated
sensitivity of the SKA-Low AA4 configuration.

Visibilities were generated across the $151$--$196$ MHz frequency range
using $100$ kHz frequency channels and an integration time of $120$
seconds. The simulated visibilities were subsequently imaged using
{\tt wsclean}, following
the SDC3b guidelines. The reconstructed images were stacked along the
line-of-sight direction to form image cubes with a bandwidth of $15$ MHz
at each redshift. Finally, cylindrical two-dimensional 21-cm power
spectra were estimated from these image cubes using the
{\tt ps\_eor} package.
The resulting power spectra constitute the primary observables used for
training and testing the machine-learning emulators.

The trained emulators were subsequently incorporated into a Bayesian
inference framework. Three independent inference pipelines were
considered, namely ANN+MCMC, BNN+MCMC, and GPR+MCMC. In each analysis,
only one emulator was used at a time as a computationally efficient
surrogate for the full 21-cm simulation pipeline. For a given
astrophysical parameter vector $\boldsymbol{\theta}$, the emulator
provides the corresponding prediction of the 21-cm power spectrum,
$P_{\rm PS}^{\rm emu}(\boldsymbol{\theta})$.

For the Bayesian inference, the power spectrum from the SDC3b test
dataset was treated as the observed data, $P_{\rm PS}^{\rm obs}$. The
uncertainties associated with the observed power spectrum, provided as
part of the test dataset, were used as the observational uncertainties
in the likelihood. The averaged noise power spectrum provided with the
test dataset was used to characterize the instrumental noise and was not
directly treated as an additional uncertainty in the likelihood.

Assuming Gaussian observational uncertainties, the likelihood is given
by
\begin{equation}
\ln \mathcal{L}(\boldsymbol{\theta})
=
-\frac{1}{2}
\sum_i
\left[
\frac{
P_{{\rm PS},i}^{\rm obs}
-
P_{{\rm PS},i}^{\rm emu}(\boldsymbol{\theta})
}{
\sigma_{{\rm PS},i}
}
\right]^2
-
\frac{1}{2}
\sum_i
\ln\left(2\pi\sigma_{{\rm PS},i}^{2}\right),
\end{equation}
where $P_{{\rm PS},i}^{\rm obs}$ is the observed power spectrum,
$P_{{\rm PS},i}^{\rm emu}(\boldsymbol{\theta})$ is the emulator
prediction, and $\sigma_{{\rm PS},i}$ is the uncertainty associated with
the corresponding observed power-spectrum bin.

For emulators that provide predictive uncertainties, such as BNN and
GPR, the emulator uncertainty can additionally be propagated into the
likelihood. Assuming that the observational and emulator uncertainties
are independent, the total covariance can be written as
\begin{equation}
\boldsymbol{\Sigma}_{\rm tot}
=
\boldsymbol{\Sigma}_{\rm obs}
+
\boldsymbol{\Sigma}_{\rm emu}.
\end{equation}
For the ANN case, where an intrinsic predictive uncertainty is not
provided, the observational covariance from the SDC3b test dataset is
used.

The posterior distribution is given by
\begin{equation}
\ln \mathcal{P}(\boldsymbol{\theta}\mid P_{\rm PS}^{\rm obs})
=
\ln \mathcal{L}(\boldsymbol{\theta})
+
\ln \Pi(\boldsymbol{\theta}),
\end{equation}
where $\Pi(\boldsymbol{\theta})$ denotes the prior probability. Uniform
priors are adopted within the specified parameter ranges, while
parameter values outside these ranges are assigned zero prior
probability.

The posterior distributions are sampled using the affine-invariant
ensemble sampler {\tt emcee}. The resulting MCMC chains are used to
obtain marginalized posterior distributions, parameter uncertainties,
and parameter-to-parameter correlations, which are visualized using
corner plots. The ANN+MCMC, BNN+MCMC, and GPR+MCMC results are analysed
independently.

Finally, the inferred parameter sets obtained from the SDC3b test datasets, PS1 and PS2, using their corresponding power-spectrum uncertainties in the likelihood, are passed back to {\tt 21cmFAST} using the same initial conditions and simulation setup adopted for the original training dataset. This allows the corresponding reionization history to be reconstructed from the inferred astrophysical parameters and provides a physical interpretation of the Bayesian inference results. In particular, the analysis of both the PS1 and PS2 test datasets provides an estimate of the ionization fraction at each redshift, $x_{\rm HII}(z)$, together with its posterior probability distribution, $P(x_{\rm HII}\mid z)$, which quantifies the uncertainty in the inferred ionization fraction. This provides an estimate of the reionization history from the SDC3b test datasets.

\subsection{Imperial-Nottingham}\label{sec:48}
\begin{table}
\renewcommand{\arraystretch}{1.25}
\centering
\caption{{\tt 21cmFAST} \citet{mesinger201121cmfast} astrophysical parameters (described in Table \ref{tab:eorpiemc_params}) used by Imperial-Nottingham and their adopted priors. \label{tab:imperial_params}}

\begin{tabularx}{\linewidth}{@{} >{\raggedright\arraybackslash}m{4cm}
 >{\raggedright\arraybackslash}m{4cm}
@{}}

\toprule
\textbf{Parameter} &\textbf{Prior} \\

\midrule
$\zeta$ &$\mathrm{Flat}(10,50)$ \\

$T_\mathrm{vir}$ &$\mathrm{log}(10^{3.5},10^{5})$ \\

$R_\mathrm{max}$  & $\mathrm{Flat}(3,80)$ \\ 
\bottomrule
\end{tabularx}
\end{table}

Imperial-Nottingham followed an SBI pipeline based on the one outlined in \citet{2025arXiv250816329C}. The semi-numerical code {\tt 21cmFast} v3.3 \citep{mesinger201121cmfast, murray202021cmfast, qin2020tale, munoz2022impact, davies25} was used to build a database of simulated lightcones. This database was constructed by first drawing 20,000 samples via a Latin Hypercube with uniform priors of the parameters as in Table \ref{tab:imperial_params}. {\tt 21cmFast} was run for each sample with unique initial conditions, over the redshift range prescribed by the data challenge. The resulting lightcones were then mean subtracted before calculating the 2D PS. Finally, any runs where the 2D PS was zero in two or more redshift bins were filtered out, resulting in a final database of $\sim$ 9K simulations. This was due to scenarios allowed by the wide prior which had completed reionization before the redshift bins specified by the challenge. 

Noise was generated by sampling from the provided mean noise power spectra according to the equation \citep{mcquinn2006cosmological}
\begin{equation}\label{noise_dist_equation}
\delta P_{\mathrm{Noise}}(\textbf{k},z) =\mathcal{N}\left(\mu=0, \sigma^2 = \frac{2 \hat{P}_{\mathrm{Noise}}^{2}(\textbf{k},z)}{N_{\mathrm{modes}}(\textbf{k},z)}\right).
\end{equation}
Here, $\hat{P}_{\mathrm{Noise}}(\textbf{k},z)$ is the mean power spectra of the noise, and $N_{\mathrm{modes}}$ is the number of modes per pixel. 50 noise realizations were generated for each parameter combination to allow the SBI method to learn the uncertainty due to the spread of the noise. 

To improve training, the data was pre-processed as described in the following. Each noisy observation was divided by the standard deviation of the noise per pixel (given by Equation \ref{noise_dist_equation}). The data was clipped to a minimum value of 0.1, the logarithm was computed and the power spectrum scaled linearly to have values between 0 and 1. Parameter values were rescaled to be between 0 and 1. The dataset was then split into train, test and validation sets, ensuring noisy observations generated from the same simulation are kept within one set. These sets are kept consistent throughout the pipeline.

The core of the pipeline is training an SBI network to learn the astrophysical parameters from the simulated training set. Inference was performed using the Learning the Universe - Implicit Likelihood Inference package, \citep{ltu-ili}, with the SBI backend \citep{tejero-cantero2020sbi, BoeltsDeistler_sbi_2025}. The SBI framework consists of an embedding network \citep{BoeltsDeistler_sbi_2025}, trained jointly with a neural posterior estimator \citep{lueckmann2017flexible, papamakarios2019neural, cranmer2020frontier}.  Since they are trained jointly, the SBI framework still has access to the full information contained within the summary. The embedding network acts to reduce the dimensionality which the neural density estimator has to fit. The embedding network is a simple fully-connected dense network starting at 500 nodes and ending at 4. 

To optimise the setup, an architecture search was performed over ten neural density estimators and simulation based calibration \citep{talts2018validating} was used to assess the performance of each architecture. The final posterior density estimator consisted of a neural spline flow \citep{durkan2019neural} with 50 hidden features and 5 transformations. The neural density estimator was trained with a learning rate of $10^{-4}$, a batch size of 1000, and a stopping patience of 10.

To map the inferred astrophysical parameters to the ionization history, an emulator of the ionization history was developed. The emulator consisted of a fully connected dense network with 8 layers each with 64-512 nodes, which predicted the neutral fraction at 15 redshift values. This emulator was trained on our database of simulations with a variable learning rate, starting at $10^{-3}$ and dropping to $10^{-6}$.

To calculate the final posterior on the ionization history, 10,000 posterior samples of the astrophysical parameters were drawn using the neural density estimator and the ionization history was emulated at each of these parameter combinations. The neutral fraction was interpolated across the prediction at 15 redshifts, and then averaged within each frequency bin used in the challenge. Finally, the distribution of these samples was normalized to estimate the posterior in the three frequency bins.

\subsection{KUSANAGI}\label{sec:49}
Team KUSANAGI adopted a simple neural-network-based approach.
The public
semi-numerical code {\tt 21cmFAST\_v3} was first used to generate 1,000 lightcone realisations by varying the astrophysical
parameters over the ranges summarized in
Table~\ref{tab:21cmfast_parameters_kusanagi}.
For each simulated data cube, the cylindrical power
spectrum was computed by Fourier transforming the 21-cm brightness-temperature
cube and taking the squared Fourier amplitudes.
Noise-power realizations generated based on the provided
noise-power estimates were added. 

As input to the model, the cylindrical power spectra from
three frequency bands were used as three feature maps. To reduce the impact of
possible model misspecification at small scales, modes
with $k_\perp, k_{||} \gtrsim 0.3~{\rm Mpc}^{-1}$ were discarded, resulting in input
data with shape $(7,7,3)$. After subtracting the expected mean
cylindrical noise power from the power spectra, the resulting spectra were log-transformed and
standardised using the mean and variance of the
training set.

The neural network consisted of three convolutional layers with hidden
dimensions of 32, 64, and 128, respectively. The resulting feature
maps were flattened and passed through a linear layer to produce
logits with shape $(3,10)$, corresponding to 10 discretised bins of
$\bar{x}_{\rm HI}$ for each of the three frequency bands. The model
was trained with a negative log-likelihood loss, so that the output
can be interpreted as a binned posterior probability mass for
$\bar{x}_{\rm HI}$ under the training distribution. The
model was trained for 100 epochs with a batch size of 32, and applied to the
PS1 and PS2 datasets to obtain the binned posterior estimates. For the
final submission, the 10-bin posterior estimates were smoothed and
interpolated onto the required 100-bin format, while preserving the
total probability normalisation.

\begin{table}
%table OK
\renewcommand{\arraystretch}{1.25}
\centering
\caption{{\tt 21cmFAST} \citet{park2019inferring} astrophysical parameters (described in Table \ref{tab:21cmfast_parameters}) used by KUSANAGI and their adopted priors.}
\label{tab:21cmfast_parameters_kusanagi}
\begin{tabularx}{\linewidth}{@{} >{\raggedright\arraybackslash}m{4cm}
 >{\raggedright\arraybackslash}m{4cm}
@{}}
\toprule
\textbf{Parameter} &\textbf{Prior} \\
\midrule

$f_{\star,10}$ & 
$\mathrm{Log}~(0.004,0.16)$ \\
$\alpha_{\star}$ &
$\mathrm{Flat}~(0,1)$ \\
$f_{\mathrm{esc}, 10}$  & $\mathrm{Log}~(0.01,1)$ \\
$\alpha_\mathrm{esc}$ & 
$\mathrm{Flat}~(-1,0.5)$ \\
$M_\mathrm{turn}$ & $\mathrm{Log}~(10^8,10^{10})$ \\
$t_{\star}$ &
$\mathrm{Flat}~(0.1,1)$ \\
$L_\mathrm{X}/\mathrm{SFR}$  &
$\mathrm{Log}~(10^{37},10^{42})$ \\
$E_{0}$  &
$\mathrm{Flat}~(200,1500)$ \\
$\alpha_\mathrm{X}$ & $\mathrm{Flat}(-0.8,3.0)$ \\
\bottomrule
\end{tabularx}
\end{table}

\subsection{Loreli}\label{sec:410}

The Loreli team used two different approaches, labelled as \textsc{Loreli-A} and \textsc{Loreli-B}. Both approaches use the \textsc{Loreli II} database of radiation-hydrodynamic Licorice simulations presented in \cite{2025A&A...698A..80M} and \cite{2024A&A...683A..24M}. First, a learning sample of cylindrical power spectra in each of the frequency band were computed from the 21-cm signal lightcones for $\sim 8900$ simulations in the database. A Hann filter was applied to the lightcones in the frequency bands. The average neutral fraction of the lightcone slice corresponding to each power spectrum was computed. Note that the distribution of possible average neutral fractions for a given frequency band (and thus the effective prior for our inference) is determined by the shape and extent of the parameter-space region explored by the \textsc{Loreli II} database. This, of course, has a direct impact on the posterior.

The inference results of \textsc{Loreli-A} make use of an emulator of the power spectrum in a classical MCMC pipeline. 
An explicit diagonal Gaussian likelihood of the cylindrical power spectrum was used. This approach was applied to the 1D power spectrum in \cite{2025A&A...698A..80M} and led to well-calibrated posteriors (with errors of order $\lesssim 0.2 \sigma $) for a noise level corresponding to $\sim 100$ hours of integration. With a similar architecture (3 dense layers of 1024 neurons, with ReLu activation functions), an emulator of the 2D PS (sampled over $10\times 10\,k$ bins linearly spaced within $0.05-0.5$ h$^{-1}$ cMpc) was trained and reached comparably accurate posteriors over the astrophysical parameters of Licorice (using Licorice signals as targets). An emulator of the neutral IGM fraction was then trained for the \textsc{Loreli II} simulations at the redshifts of the SDC3b data, using a lighter network (3 layers of 128 neurons). The neutral fractions of \textsc{Loreli} were recovered by the emulator with a relative error of approximately $1\%$. To obtain posteriors on $x_{\rm HI}$, samples were drawn from the posteriors on the astrophysical parameters and used as inputs of the ionization fraction emulator. It was confirmed that the error on $x_{\rm HI}$ at a given redshift was typically very small compared to the width of the corresponding 1D marginalized posterior. 

For the \textsc{Loreli-B} approach, two SBI methods were explored: Neural Density Estimator (NDE) with Gaussian Mixture Models to discover the likelihood function, and Neural Posterior Estimator (NPE) to directly predict the posterior. Using the full information of the 100 individual bins in the target power spectra is difficult with NDE because it requires predicting a $100 \times 100$ covariance matrix for each component of the mixture model. When assessing the robustness of the result with simulation based calibration, less consistent results were obtained than with NPE. The results submitted to the challenge are those obtained with the NPE (with 2 Gaussian mixture components, and 4 layers of 32 neurons). Since the \textsc{Loreli II} database gives a flat prior only in the initial parameter space of the database, and not in the 1-dimensional parameter space of this challenge (the parameter being $x_{\mathrm{HI}}$), an effective prior was computed by binning the $x_{\mathrm{HI}}$ of the simulations in the database.

An important point is that the inference of \textsc{Loreli-B} was performed considering the ionization fraction in each band as independent variables, while \textsc{Loreli-A} performs the inference on all bands at once. Indeed, it was observed that, while one can find in the database a good match to the target signal in each band, no single simulation provides a good match in all three bands. The team's interpretation is that the source formation histories generated by the LICORICE code do not match those in the simulations used to create the target signals. It is therefore expected to lose some constraining power with this approach.

\subsection{ReionYuga}\label{sec:411}

The ReionYuga team implemented two distinct pipelines to generating the training data and performing the inference. The first pipeline was applied to PS1 and the second to PS2, which allowed testing both approaches. In both pipelines, the training datasets consisted of 1,000 models and were restricted to the mean neutral fraction interval $\bar{x}_{\rm HI}\in[0.1,0.9]$. Power-spectrum modes with $|k|<0.25\,h\,{\rm Mpc}^{-1}$ were retained, corresponding to 13 $(k_\perp,k_\parallel)$ bins, while the remaining error-dominated modes were excluded from the analysis. Both pipelines used XGBoost-based emulators and adopted an 80/20 split for training and testing, respectively. Emulator performance was assessed on the independent test sets using the mean squared error (MSE) and the coefficient of determination ($R^2$), with $R^2$ values of approximately 0.99.

The first pipeline (called ReionYuga PS1 in Table \ref{tab:methods}) utilized the ReionYuga semi-numerical code \citep{Mondal2017}. In this framework, a particle-mesh $N$-body code\footnote{\url{https://github.com/rajeshmondal18/N-body}} \citep{Mondal2015} was used to generate the underlying DM distribution. The simulations were carried out within a comoving volume of $286.72\,{\rm Mpc}^3$. The cosmological parameters were fixed at $h=0.6704$, $\Omega_m=0.3183$, $\Omega_\Lambda=0.6817$, $\Omega_b=0.04902$, $\sigma_8=0.8347$, and $n_s=0.9619$, slightly different from the cosmology used in the challenge data (see Section~\ref{sec:cosmo_evol}). DM haloes were identified using a Friends-of-Friends (FoF) halo finder\footnote{\url{https://github.com/rajeshmondal18/FoF-Halo-finder}} \citep{Mondal2015} using a linking length equal to 0.2 times the mean interparticle separation. The minimum halo mass in the simulations was $M_{\mathrm{min}}=1.09\times10^{9}\,M_\odot$. The ionisation field was then calculated from the density and halo fields using the excursion-set formalism.

For this pipeline, the training models were generated by varying the ionizing photon efficiency parameter $N_{\rm ion}$ over the range 10 to 40. To accelerate the inference, an XGBoost emulator was constructed to predict the corresponding power spectrum. Posterior sampling was subsequently performed using an affine-invariant MCMC ensemble with 40 walkers and 20,000 steps.

The second pipeline (called ReionYuga PS2 in Table \ref{tab:methods}) used {\tt 21cmFAST} v3.3.1 \citep{mesinger201121cmfast} to generate the brightness-temperature fields. Of the available parameters (see e.g. Table \ref{tab:21cmfast_parameters_kusanagi}) only the global ionizing-efficiency parameter $\zeta$ was varied, while the remaining astrophysical parameters $T_{\rm vir}\simeq5\times10^4\,{\rm K}$, and $R_{\rm max}=15\,{\rm Mpc}$ were kept fixed at their default value. $\zeta$ was varied logarithmically, with prior ranges separate for the redshift intervals ([8,90] for $z_1$, [5,57] for $z_2$ and [3,39] for $z_3$). The cosmological parameters were kept identical to those adopted in the first pipeline throughout these runs.

For the second pipeline, an Isolation Forest algorithm \citep{liu2008isolation} with a contamination fraction of 0.05 was applied only to the training subset, removing approximately 5\% of the samples identified as statistical outliers. The independent test set was left unchanged. The XGBoost emulator used $\bar{x}_{\rm HI}$ as input and predicted the corresponding power spectrum at the retained 13 $(k_\perp,k_\parallel)$ bins using a multi-output regression framework. For inference, an MCMC with 24 walkers was run for 20,000 iterations per walker, with the first 2,000 steps discarded as burn-in, to constrain $\bar{x}_{\rm HI}$ at specific redshifts.

\subsection{RISE- \textit{R}eionization epoch \textit{I}nference with efficient 
\textit{S}CRIPT \textit{E}valuations}\label{sec:412}

\begin{table}
\renewcommand{\arraystretch}{1.25}
\centering
\caption{Parameters used by RISE and their  priors (see Sec. \ref{sec:412} for more details).}
\label{tab:rise_params}
\begin{tabularx}{\linewidth}{@{} >{\raggedright\arraybackslash}m{1.0cm} >{\raggedright\arraybackslash}X >{\centering\arraybackslash}m{2.7cm} @{}}
\toprule
\textbf{Parameter} &
\multicolumn{1}{c}{\textbf{Description}} &
\textbf{Prior} \\
\midrule
$M_\mathrm{min,0}$ &Normalization for minimum mass for collapsed haloes [M$_\odot$]&$\mathrm{Log}(10^{7},10^{11})$ \\
$\alpha_{\rm M}$ &Spectral index for minimum mass $M_{\rm min}$ &$\mathrm{Flat}(-5,5)$ \\
$\zeta_0$ &Normalization for ionization efficiency $\zeta$& $\mathrm{Log}(1,10^{4})$ \\
$\alpha_\zeta$ & Spectral index for ionization efficiency $\zeta$& $\mathrm{Flat}(-4,4)$ \\
\bottomrule
\end{tabularx}
\end{table}

The RISE team consists of researchers working with the semi-numerical reionisation model {\tt SCRIPT}\footnote{\url{https://bitbucket.org/rctirthankar/script/src/master/}} \citep{choudhury2018,maity2023,choudhury2024}. The inference results were submitted for the datasets PS1 and PS2, applying a Bayesian analysis technique for the 2D cylindrical power spectrum.

{\tt SCRIPT} models produce the ionization field based on an explicit photon-conserving algorithm, given the input density ($\Delta$) and collapsed fraction ($f_{\mathrm{coll}}$) field. This also requires information about the astrophysical free parameters, namely, ionization efficiency ($\zeta$) and minimum threshold mass for collapsed haloes ($M_{\mathrm{min}}$), which are varied during parameter exploration. A power-law dependence on redshift was assumed for both the parameters, i.e., $\zeta(z)=\zeta_0\left(\frac{1+z}{8}\right)^{\alpha_{\mathrm{\zeta}}}$ and $M_{\mathrm{min}}(z)=M_{\mathrm{min,0}}\left(\frac{1+z}{8}\right)^{\alpha_M}$, which together give a total of four free parameters, quoted in Table \ref{tab:rise_params} together with the adopted ranges.

Simulation boxes of size $256~h^{-1}\mathrm{cMpc}$ were utilised in the redshift range $z=6.2-8.5$ with an interval $\Delta z=0.2$ (the first two bins with $\Delta z=0.1$ to match the interval range), which covers the redshifts corresponding to the challenge dataset. Specifically, second-order Lagrangian perturbation theory (2LPT) was employed to compute the density field \citep{music_2011}, while the collapsed mass fraction was calculated assuming a conditional ellipsoidal collapse mass function \citep{sheth_tormen2002}. The spatial resolution is $2~h^{-1}\mathrm{cMpc}$ throughout the boxes. The {\tt SCRIPT} model provides the ionisation fields at each of these redshifts, which are subsequently used to construct the light-cone volume using our own software based on the methodology in \citet{datta2012light}. These lightcones are then used to compute the 2D cylindrical power spectra corresponding to the three different frequency/redshift intervals provided by the data. Similarly, the required global neutral fractions ($x_{\mathrm{HI}}$) are calculated using the corresponding volume segment of the lightcones.

For inference, the publicly available MCMC sampler \texttt{COBAYA} \citep{cobaya2021} was deployed, with the standard assumption of a Gaussian likelihood. The provided averaged thermal noise power spectra were added to the simulated power spectra to make them comparable with the noisy dataset, while the total noise was estimated by adding the given average instrumental (thermal) noise and sample variance noise in quadrature. The log-likelihood function is defined as 
\footnotesize\begin{equation}
   -2\ln \mathcal{L}
=
\sum_{\alpha,\beta, \gamma}
\left[
\frac{
\Delta^{2}_{21,\mathcal{D}}\!\left(k_{\alpha \beta}, z_{\gamma}\right)
- \Delta^{2}_{21,\mathcal{M}}\!\left(k_{\alpha \beta}, z_{\gamma}\right)
- \Delta^{2}_{21,\mathrm{N}}\!\left(k_{\alpha \beta}, z_{\gamma}\right)
}{
\sigma^{\mathrm{tot}}_{\mathcal{D}}\!\left(k_{\alpha \beta}, z_{\gamma}\right)
}
\right]^{2}
\end{equation}\normalsize
with $\sigma^{\mathrm{tot}}_{\mathcal{D}} = \sqrt{(\Delta^{2}_{21,\mathrm{N}})^2 + (\Delta^{2}_{21,\mathrm{CV}})^2}$; $k_{\alpha \beta}$ is the $k$-mode element of 2D spectra, and $z_{\gamma}$ denotes the corresponding redshift.  Here, $\Delta^{2}_{21,\mathcal{D}}$ is the provided noisy power spectra and $\Delta^{2}_{21,\mathcal{M}}$ is the model prediction. Furthermore, $\Delta^{2}_{21,\mathrm{N}}$ denotes the average noise while $\Delta^{2}_{21,\mathrm{CV}}$ is the cosmic variance uncertainties (provided with the challenge dataset). The effect of varying the initial seeds of the simulation on the inference was additionally checked with a low resolution, i.e., $8 h^{-1}\mathrm{cMpc}$ box.  Consequently, a few lower $k$-modes ($k=0.05$ and 0.1, corresponding to 20 data points) were removed from the final analysis for robust inference. The convergence of MCMC chains was checked using the Gelman-Rubin $R - 1$ statistic \citep{Gelman_Rubin1992}, where a value $\le0.01$ is assumed as the convergence criterion. The first $\approx$ 30\% of samples were removed as "burn-in" and the inference was performed with the rest.

\subsection{SEarCH}\label{sec:413}

\begin{table}
\renewcommand{\arraystretch}{1.25}
\centering
\caption{
{\tt 21cmFAST} \citet{park2019inferring} astrophysical parameters (described in Table \ref{tab:21cmfast_parameters}) used by SEarCH and their adopted priors for constructing the training dataset.
}
\label{tab:search_params}
\begin{tabularx}{\linewidth}{@{} >{\raggedright\arraybackslash}m{4cm}
 >{\raggedright\arraybackslash}m{4cm}
@{}}
\toprule
\textbf{Parameter} &\textbf{Prior} \\
\midrule

$f_{\star,10}$ & 
$\mathrm{Log}~(0.004,0.16)$ \\
$\alpha_{\star}$ &
$\mathrm{Flat}~(0,1)$ \\
$f_{\mathrm{esc}, 10}$  & $\mathrm{Log}~(0.01,1)$ \\
$\alpha_\mathrm{esc}$ & 
$\mathrm{Flat}~(-1,0.5)$ \\
$M_\mathrm{turn}$ & $\mathrm{Log}~(10^8,10^{10})$ \\
$t_{\star}$ &
$\mathrm{Flat}~(0.1,1)$ \\
% $L_\mathrm{X}/\mathrm{SFR}$  &
% $\mathrm{Log}~(10^{37},10^{42})$ \\
% $E_{0}$  &
% $\mathrm{Flat}~(200,1500)$ \\
% $\alpha_\mathrm{X}$ & $\mathrm{Flat}(-0.8,3.0)$ \\
\bottomrule
\end{tabularx}
\end{table}

SEarCH (\textbf{S}vea's \textbf{E}ndeavour in \textbf{A}I and \textbf{R}eionization \textbf{C}osmology with \textbf{H}elvetia) is a Swedish-Swiss team of cosmologists and data scientists that submitted results for all four datasets. The inference framework employed by SEarCH relies on a simulation dataset of $\sim$15,900 models of 21-cm signal lightcones produced with the {\tt 21cmFAST\_v3} code \citep{murray202021cmfast}. These simulations were run with a simulation volume of 500\,Mpc on a side, with 128 cells along each coordinate direction. 
% {\bf Introduce Table \ref{tab:search_params} and complete it?}

While these simulations are dependent on the astrophysical parameters introduced in \citet{park2019inferring}, the inference problem posed in this work is defined on derived parameters, namely the neutral fraction values at three redshifts. To obtain a balanced dataset in the latter parameter space, the SEarCH team followed the strategy described in sec. 2.2 of \citet{cerardi2026implicit}. This strategy begins assuming a flat prior on the astrophysical parameters presented in Table \ref{tab:search_params}, and then adjusts the distribution of samples such that they provide approximately uniform distribution of neutral fraction values at the three redshifts explored here. These simulated lightcones were forward-modelled with instrumental noise calculated using the \texttt{tools21cm} package \citep{giri2020tools21cm}. These forward models include stochastic effects of $\sim$25 independent noise realizations. 

This team employed two different inference pipelines, as described in Sec.~\ref{sec:traditional_search} and \ref{sec:modern_search}, for the power spectra datasets PS1, PS2 and PS3. See \citet{de2025exploring} for a detailed comparison of several inference pipelines. 
The impact of foreground residuals and antenna beam chromaticity was ignored in these models, which is consistent with the setup of PS1 and PS2. To infer the results for PS3 and IM1, the team mitigated the foreground contaminants using \texttt{SERENet}, which subtracts foreground signal with PCA followed by applying a neural network trained to reduce the remaining residuals \citep[see][for details]{bianco2025deep}. 
Sec.~\ref{sec:segunet_search} presents the third method applied by the SEarCH team to the IM1 data.

\subsubsection{Traditional SEarCH}\label{sec:traditional_search}
First, the SEarCH team developed a pipeline based on SBI that trains a neural network on the joint distribution of parameter $\pmb{\theta}$ and summary statistics $\pmb{t}$ \citep[e.g.,][]{cranmer2020frontier}. $\pmb{t}$ is the power spectrum data provided in the data challenge and contains the impact of instrumental effects through the several independent forward models. In this approach, a neural density estimator was trained on the posterior distribution $p(\pmb{\theta},\pmb{t})$. To optimally extract information from $\pmb{t}$, the training process employed variational mutual information maximisation \citep{jeffrey2021likelihood}. For a detailed description of this framework, see sec. 3 in \citet{cerardi2026implicit}. Once trained, this network can swiftly estimate the posterior distribution of any given data.

\subsubsection{Modern SEarCH}\label{sec:modern_search}
The second pipeline developed by the SEarCH team employed a multilayer perceptron (MLP) architecture called MLP-Mixer \citep{tolstikhin2021mlp}. This method assumes the cylindrical power spectrum data as a 2D image and trains a mapping from this image to the neutral fraction values. MLP-Mixer contains two types of layers: 1) MLPs applied independently to image patches (i.e. ``mixing" the per-location features), and 2) MLPs applied across patches (i.e. ``mixing" spatial information). The detailed architecture and the training process are presented in \citet{de2025exploring}. 

\subsubsection{SegU-Net SEarCH}\label{sec:segunet_search}
For the image data IM1, the team formulated the estimation of the neutral fraction as an image segmentation task, leveraging the fact that these images contain distinct, identifiable features corresponding to neutral regions \citep{giri2018optimal,giri2019neutral}. The model employs a U-Net architecture to classify pixels as either ionized or neutral; the input image is progressively compressed through convolutional layers with varying kernel sizes into a low-dimensional latent space before being decoded back to the original image dimensions. This framework, detailed in \citet{Bianco2021,bianco2025deep}, allows the mean neutral fraction to be derived directly from the resulting segmentation map. To quantify uncertainty arising from image noise, the team estimated error bars by passing multiple augmented versions of the input data through the trained model \citep[see e.g.,][]{wang2021regularizing}. These augmentations involved several rotational transformations of the data. % using a suite of rotational augmentations.

\subsection{Shuimu-Tianlai}\label{sec:414}

\begin{table}
%table OK
\renewcommand{\arraystretch}{1.25}
\centering
\caption{{\tt 21cmFAST} \citet{mesinger201121cmfast} astrophysical parameters (described in Table \ref{tab:eorpiemc_params}) used by Shuimu-Tianlai and their adopted priors.\label{tab:tianlai_params}}

\begin{tabularx}{\linewidth}{@{} >{\raggedright\arraybackslash}m{4cm}
 >{\raggedright\arraybackslash}m{4cm}
@{}}

\toprule
\textbf{Parameter} &\textbf{Prior} \\

\midrule
$\zeta$ &$\mathrm{Flat}(10,250)$ \\

$T_\mathrm{vir}$ &$\mathrm{log}(10^{4},10^{6})$ \\

\bottomrule
\end{tabularx}
\end{table}

Shuimu-Tianlai is a team from Tsinghua University, the National Astronomical Observatories of the Chinese Academy of Sciences, and a number of other institutions in China. The team developed a lightweight forward-modelling pipeline to generate mock SKA-Low observations and combined it with SBI to constrain the ionization history using the PS1 and PS2 datasets.

For the forward pipeline, 1,000 pairs of astrophysical parameters, $\zeta$ and $\mathrm{log}_{10} T_{\mathrm{vir}}$, were uniformly sampled (with ranges in Table \ref{tab:tianlai_params}) while the $R_{\rm max}$ parameter was fixed at the default value of 15\,cMpc. The corresponding lightcones were generated using {\tt 21cmFAST} v3.4 over the redshift range $z=7$--$10$. These physical lightcones were converted into observational lightcones using {\tt tools21cm}. The same package was also used to generate thermal-noise lightcones, assuming the SKA-Low telescope configuration adopted for the data challenge. The resulting lightcones were post-processed through padding and interpolation to match exactly the field of view and spatial resolution of the target datasets. Each lightcone was then divided into three frequency (redshift) intervals, and cylindrical power spectra were computed for each interval using {\tt ps\_eor}.

For the inference, the \texttt{ltu-ili}\footnote{\textcolor{blue}{https://github.com/maho3/ltu-ili}}\citep{ltu-ili} package and neural posterior estimation were employed. Two preprocessing steps were applied to the input data. First, the 2D power spectrum in each redshift bin was normalized by the power at the smallest $k$-mode. Second, to account for potential model mismatch between our signal model ({\tt 21cmFAST}) and the target simulations ({\tt C$^2$Ray)}, as well as for observational effects not included in the forward model, an additional modelling uncertainty was introduced. This uncertainty was implemented by perturbing the 2D power spectra with a Gaussian noise model with standard deviation $r\,\Delta^2_{21}$, where $\Delta^2_{21}$ denotes the 2D power spectrum and $r$ controls the magnitude of the uncertainty. Three values, $r \in \{0, 0.25, 0.50\}$, were considered. These preprocessing steps define the final data products used for inference.

Two inference strategies were explored. The first is a \emph{direct inference} approach, in which the neutral fraction $x_{\mathrm{HI}}$ at each redshift is inferred directly from the input 2D power spectra. The second is a \emph{two-step inference} approach, in which the astrophysical parameters are inferred first, and posterior samples are then propagated through new simulations to estimate the neutral fraction at each redshift. For each combination of uncertainty level $r$ and inference strategy, an ensemble of 10 masked autoregressive flow (MAF) models was trained to approximate the posterior. This resulted in six inference scenarios in total, corresponding to the two inference strategies and three model-uncertainty levels.

Since it was found that the level of uncertainty did not significantly affect the results, the three most conservative results were submitted. These correspond to three of the six inference scenarios:
(1) \texttt{Shuimu-Tianlai A}: two-step inference with $r = 0.5$;
(2) \texttt{Shuimu-Tianlai B}: direct inference with $r = 0.25$;
(3) \texttt{Shuimu-Tianlai C}: direct inference with $r = 0.5$.

The fully documented code for this analysis is available at \url{https://github.com/suicee/21cmPipeline/tree/main/sdc}.

\subsection{ToSKA teams}\label{sec:415}
\begin{table}
%table OK
\renewcommand{\arraystretch}{1.25}
\centering
\caption{{\tt 21cmFAST} \citet{park2019inferring} astrophysical parameters (described in Table \ref{tab:21cmfast_parameters}) used by ToSKA and their adopted priors.}
\label{tab:toska_params}
\begin{tabularx}{\linewidth}{@{} >{\raggedright\arraybackslash}m{4cm}
 >{\raggedright\arraybackslash}m{4cm}
@{}}
\toprule
\textbf{Parameter} &\textbf{Prior} \\
\midrule

$f_{*,10}$ & $\mathrm{Log}~(10^{-3},1)$ \\
$f_{\mathrm{esc}, 10}$  & $\mathrm{Log}~(10^{-3},1)$ \\
$\alpha_\mathrm{esc}$ & $\mathrm{Flat}~(-1,0.5)$ \\
$M_\mathrm{turn}$ & $\mathrm{Log}~(10^8, 10^{10})$ \\

\bottomrule
\end{tabularx}
\end{table}

The ToSKA group comprised three teams:
ToSKA - Explicit Likelihood, ToSKA - Model Selection, ToSKA - SBI General.
Each team tackled the challenge with different methodologies, performing inferences on the PS1 and PS2 datasets using the spherical power spectrum compressed from the original cylindrical power spectrum.

{\tt 21cmFAST} (v4) \citep{mesinger201121cmfast,davies25} and {\tt C2Ray} \citep{Mellema:2005ht,Hirling:2023cjx} were used to create two separate lightcone and $x_{\rm HI}$ datasets assuming the same source model and density field.
The {\tt 21cmFAST} and {\tt C2Ray} datasets contain $5 \times 10^4$ and $5000$ samples respectively.
In both datasets four astrophysical parameters were varied, listed in Table \ref{tab:toska_params}.
%$f_{\star, 10}$, 
%$f_{\rm{esc}, 10}$,
%$\alpha_{\rm esc}$ and $M_{\rm turn}$.
The rest of source parameters are either largely degenerate with these ones or irrelevant during EoR and were therefore fixed.
Each simulation used a box length of 350\,cMpc and resolution of $175^3$.
The power spectrum from the lightcones were then computed using {\tt tools21cm} \citep{giri2020tools21cm}, adopting the same frequency binning as the PS1 and PS2 datasets.

These paired power spectrum and $x_{\rm HI}$ samples were then shared among all ToSKA teams.
Specifically,
two machine learning emulators which predict the power spectrum and $x_{\rm HI}$ for {\tt 21cmFAST} and {\tt C2Ray} were trained.
These emulators were used by the ToSKA - Explicit Likelihood and ToSKA - Model Selection teams,
which performed inferences using an explicit likelihood.
On the other hand,
the ToSKA - SBI General team used these dataset samples to perform simulation-based inference.
The strategies adopted by each ToSKA team are reviewed below.

\subsubsection{ToSKA - Model Selection}
The ToSKA - Model Selection and ToSKA - Explicit Likelihood teams both used a Gaussian likelihood $\mathcal{L}$ of the form,
\begin{equation}
\ln \mathcal{L}
=
\frac{-1}{2}
\sum_{k, z}
\left[
\ln (2 \pi \sigma^2)
+
\frac{(\Delta^2_{21, {\rm signal}} - \Delta^2_{21, {\rm model}})^2}
{\sigma^2}
\right],
\label{Eq_ToSKA_Gaussian_Likelihood}
\end{equation}
where $\Delta^2_{21}$ denotes the power spectrum,
$\Delta^2_{21, {\rm signal}} \equiv \Delta^2_{21, {\rm data}} - \Delta^2_{21, {\rm noise}}$ is the noise subtracted signal,
$\Delta^2_{21, {\rm data}}$ and $\Delta^2_{21, {\rm noise}}$ represent the simulated data and noise,
respectively.
The sum is taken over all $k$ modes and redshifts,
and only modes with signal to noise ratio above unity were used 
(i.e.,
$\Delta^2_{21, {\rm signal}}/\Delta^2_{21, {\rm noise}} \geq 1$).
$\sigma^2$ is the likelihood variance and was modelled differently between the Explicit Likelihood and Model Selection teams.

The ToSKA - Model Selection team modelled the likelihood variance $\sigma^2$ in equation (\ref{Eq_ToSKA_Gaussian_Likelihood}) as,
\begin{equation}
\sigma^2 = 
\sigma^2_{\rm noise}
+
\sigma^2_{\rm CV}
+
\sigma^2_{\rm SV},
\end{equation}
where $\sigma_{\rm noise} = \Delta^2_{21, {\rm noise}}$ and $\sigma_{\rm CV} = \Delta^2_{21, {\rm error}}$ are the noise and cosmic variance (CV) of the challenge datasets,
respectively.
$\sigma_{\rm SV} = \Delta^2_{21, {\rm model}}/\sqrt{n_{\rm modes}}$ is the sample variance (SV),
$n_{\rm modes}$ is the number of modes in each $k$ bin.
The likelihoods were sampled using the {\tt MultiNest} \citep{Feroz:2008xx} nested sampler,
which produced both Bayesian evidences and parameter posteriors. 
Posterior samples for the model parameters in Table \ref{tab:toska_params} were used to derive joint posterior distributions of $x_{\rm HI}$ at each redshift.
Each of the PS1 and PS2 likelihoods were sampled using both the {\tt 21cmFAST} and the {\tt C2Ray} emulators,
and results were selected by Bayesian evidence ratio.

Specifically, for the PS1 dataset a Bayes factor of
$\mathcal{Z}_{\tt C2Ray}/\mathcal{Z}_{\tt 21cmFAST} = 1.3 \times 10^6$ was obtained,
where \(\mathcal{Z}_{\tt C2Ray}\) and \(\mathcal{Z}_{\tt 21cmFAST}\) are the Bayesian evidences computed using the {\tt C2Ray} and {\tt 21cmFAST} emulators, respectively. For the PS2 dataset, the corresponding evidence ratio was $\mathcal{Z}_{\tt C2Ray}/\mathcal{Z}_{\tt 21cmFAST} = 0.73.$
These results indicate that the PS1 and PS2 datasets are better described by the {\tt C2Ray} and {\tt 21cmFAST} simulators, 
respectively. 
Therefore for PS1 and PS2 
the $x_{\rm HI}$ posteriors obtained with {\tt C2Ray} and {\tt 21cmFAST} emulators, respectively, are reported. 

\subsubsection{ToSKA - Explicit Likelihood}
The ToSKA -- Explicit Likelihood team sampled both the PS1 and PS2 likelihoods using only the {\tt 21cmFAST} emulator.
Similar to the Model Selection team,
{\tt MultiNest} \citep{Feroz:2008xx} was used to compute posteriors for the model parameters, 
from which posterior distributions for $x_{\rm HI}$ were derived.
The likelihood variance $\sigma^2$ was modelled as,
\begin{equation}
\sigma^2 = 
\sigma^2_{\rm noise}
+
\sigma^2_{\rm CV}
+
\sigma^2_{\rm SV}
+
x \sigma^2_{\rm SimErr}.
\end{equation}
Compared to the Model Selection team,
to account for uncertainties in reionization simulators (i.e., {\tt 21cmFAST} and {\tt C2Ray}),
a simulation error term $\sigma_{\rm SimErr}$ was included,
which was set to $\Delta^2_{21, {\rm model}}$ computed by the {\tt 21cmFAST} emulator for different model parameters.
$x$ is a free parameter which was varied in the inferences.
For the PS1 dataset, the posterior was $\log_{10} x = -1.02^{+0.24}_{-0.36}$ (mean and 68\% C.I.),
whereas for PS2 it peaked at zero with 68\% C.I. upper limit of $\log_{10} x < -2.45$,
indicating that PS1 requires the simulation error term whereas PS2 disfavors it.
For PS2 the $x_{\rm HI}$ posteriors are similar to that of the Model Selection team,
and for PS1  the $x_{\rm HI}$ posteriors peak at higher values but exhibit local peaks around the input truths.

\subsubsection{ToSKA - SBI General}
The ToSKA - SBI General team modelled the likelihood implicitly using a neural ratio estimation (NRE) approach \citep{Hermans+2019, Cranmer+2015, Durkan+2020, Miller+2022, Thomas+2016}. Given the data ($\Delta^2_{21, {\rm data}}$) and parameter ($x_{\rm HI }$) pairs \footnote{$f_{\star, 10}$, 
$f_{\rm{esc}, 10}$,
$\alpha_{\rm esc}$ and $M_{\rm turn}$ are ignored, and so, effectively, marginalized}, it holds that:

\begin{equation}
\frac{p(x_{\rm HI },\Delta^2_{21, {\rm data}})}{p(
x_{\rm HI })p(\Delta^2_{21, {\rm data}})} = \frac{p(\Delta^2_{21, {\rm data}}|x_{\rm HI })}{p(\Delta^2_{21, {\rm data}})}.
\label{eq: NRE}
\end{equation}
Knowing the likelihood-to-evidence ratio (RHS of equation \ref{eq: NRE}) allows for a posterior estimation after a multiplication with the prior. The NRE method allows for an approximation of the LHS.

The dataset was evenly split into two subsets and 10 realizations of the noise ($\sigma^2_{\rm noise}$) were added to the data, effectively increasing the size of the dataset by the same factor. 
 The parameter and data pairs were shuffled for one of the subsets while keeping the other one as is, acting like independently-drawn and jointly-drawn samples respectively. 
Then, a neural network (NN) classifier was trained to distinguish between the two, allowing for an estimation of the LHS's ratio.

The team used a fully connected architecture for the NN
with nodes [128, 64, 8, 1] with a ReLU activation function after each hidden layer and a sigmoid for the last layer. A batch normalization was performed after the first two layers and a dropout layer of 10\% was added between all layers. Commonly,  the dataset was split into training (70\%), validation (15\%) and test sets (15\%) and used an Adam optimizer with a learning rate of $10^{-3}$. A binary cross-entropy loss function was trained for 100 epochs with a batch size of 256 until the training and validation loss curves had converged. 
Load balancing techniques between {\tt 21cmFAST} and {\tt C2Ray} datasets, i.e., using only a small percentage of the first one or drawing more noise realizations for the other, did not show an improvement in the validation or test losses.
The final NN was trained with the full {\tt 21cmFAST} and {\tt C2Ray} dataset.

For the prior, a Gaussian kernel density estimator (KDE) was fitted to the parameter samples using the Silverman's rule of thumb \citep{silverman+1986} to set the bandwidth.
The posteriors were sampled with the nested sampling Monte Carlo algorithm
{\tt MLFriends} \citep{Buchner+2016, Buchner+2019} using the
{\tt UltraNest}\footnote{\url{https://johannesbuchner.github.io/UltraNest/}} package \citep{Buchner+2021}.

%%%%%%%%%%%%%%%%%%%%%%%%%%%%%%%%%%%%%%%%%%%

\subsection{YEYE}\label{sec:416}
YEYE is a team from the Shanghai Astronomical Observatory, Chinese Academy of Sciences.  YEYE can be pronounced as "Y-Eye", but in Chinese pronunciation as "Yeah-Yeah", it means grandfather. This team submitted inference results to PS1, PS2, PS3, and IM1.

For PS1, PS2 and PS3, YEYE used the same method, described step-by-step below. First, the provided 2D power spectrum is compressed into the spherically-averaged 1D power spectrum, with the data pixel in each 2D power spectrum weighted by its associated noise. Second, the 1D power spectrum error is calculated by combining the standard deviation of the 2D power spectrum pixels and the related noise together. Third, a JAX\footnote{\url{https://docs.jax.dev}} compatible model is calculated using GPR, based on {\tt 21cmEMU} \citep{21cmemu}, which creates a fast map between astrophysical parameters and the 1D power spectrum. Fourth, The No-U-Turn Sampler (NUTS) provided in {\tt numpyro}\footnote{\url{https://num.pyro.ai}} is used to perform an MCMC for parameter inference. Fifth, the inferred $x_{\rm HI}$ parameter is calculated from the previous inferred astrophysical parameters with {\tt 21cmEMU}.\footnote{\url{https://21cmemu.readthedocs.io/en/latest/}} 

Priors on the nine {\tt 21cmEMU} astrophysical parameters are presented in Table~\ref{tab:yeye_params}. The description of the parameters is identical to the {\tt 21cmFAST} convention presented in Table~\ref{tab:21cmfast_parameters}. Note that the analysis has been performed on normalized parameters rather than the physical parameters. $x_{\rm HI}$ is then inferred from the outputs. 

For the IM1 dataset, YEYE performed a fully model-independent algorithm. First, the provided data cube is divided into 30 frequency bins. For each bin, the number of pixels falling in each pixel-value interval is counted inside a circular aperture (radius of 400 pixels centred on the field) to build a histogram. Second, each histogram is fitted with a double-Gaussian model, in which one Gaussian has its mean fixed at zero (the noise component) and the other has a free mean and width (the signal component). Third, the integrated area of the zero-centred Gaussian divided by the total integrated area is identified with $1-x_{\rm HI}$, so that the free Gaussian's area fraction gives $x_{\rm HI}$ in each of the 30 bins, with the per-bin uncertainty propagated from the covariance matrix of the fit. Finally, a Gaussian Process (with a Gaussian kernel and the per-bin fitting uncertainties adopted as the noise level) is used to smooth the resulting $x_{\rm HI}$ measurements along the frequency axis and to infer the central value and $1\sigma$ uncertainty of $x_{\rm HI}$ in the three frequency bins defined by the challenge. This procedure is conservative, since the Gaussian Process prediction uncertainty is dominated by the scatter among the 30 individual bins, which forms the results we have submitted. Better overall performance is obtained by tuning the number of frequency bins as a trade-off between the number of pixels in each bin and the evolution of the signal within the bin. A more detailed description of the methods and more test results can be found in \cite{2025arXiv251113370Z}.

Note that no additional foreground mitigation was applied for either PS3 or IM1 beyond the default noise-bias subtraction provided with the power spectrum dataset.

\begin{table}
\renewcommand{\arraystretch}{1.25}
\centering
\caption{Priors on the nine {\tt 21cmEMU} astrophysical parameters used by the ACG ({\tt acg}/v1) emulator by YEYE (see Table \ref{tab:21cmfast_parameters} for their definition). The emulator actually implemented flat priors between 0 and 1 for normalized parameters.}
\label{tab:yeye_params}
\begin{tabularx}{\linewidth}{@{} >{\raggedright\arraybackslash}m{4cm}
 >{\raggedright\arraybackslash}m{4cm}
@{}}
\toprule
\textbf{Parameter} &\textbf{Prior} \\
\midrule

$f_{*,10}$ & $\mathrm{Log}~(10^{-3},10^{0})$ \\
$\alpha_{*}$  & $\mathrm{Flat}~(-0.5,1)$ \\
$f_{\mathrm{esc},10}$  & $\mathrm{Log}~(10^{-3},10^{0})$ \\
$\alpha_{\mathrm{esc}}$ &$\mathrm{Flat}~(-1,0.5)$ \\
$M_{\mathrm{turn}}$ & $\mathrm{Log}~(10^{8},10^{10})$ \\
$t_{*}$ &  $\mathrm{Flat}~(0.01,1)$ \\
$L_{\mathrm{X}}/\mathrm{SFR}$ &  $\mathrm{Log}~(10^{38},10^{42})$ \\
$E_{0}$  & $\mathrm{Flat}~(100,1500)$ \\
$\alpha_{\mathrm{X}}$&  $\mathrm{Flat}~(-1,3)$ \\
\bottomrule
\end{tabularx}
\end{table}

%%%%%%%%%%%%%%%%%%%%%%%%%%%%%%%%%%%%%%%%%%%%%%%%%%%
%%%%%%%%%%%%%%%%%%%%%%%%%%%%%%%%%%%%%%%%%%%%%%%%%%%%

\section{Assessment of results} \label{sec:results}

In this section we describe the results from the different teams for the data challenge. However, before we do so we introduce the metrics which will be used to evaluate the results, as well as the types of figures used to present the results.

\begin{figure}
    \centering
    \includegraphics[
        width=\columnwidth,
        trim=2cm 3cm 2cm 3cm,
        clip
    ]{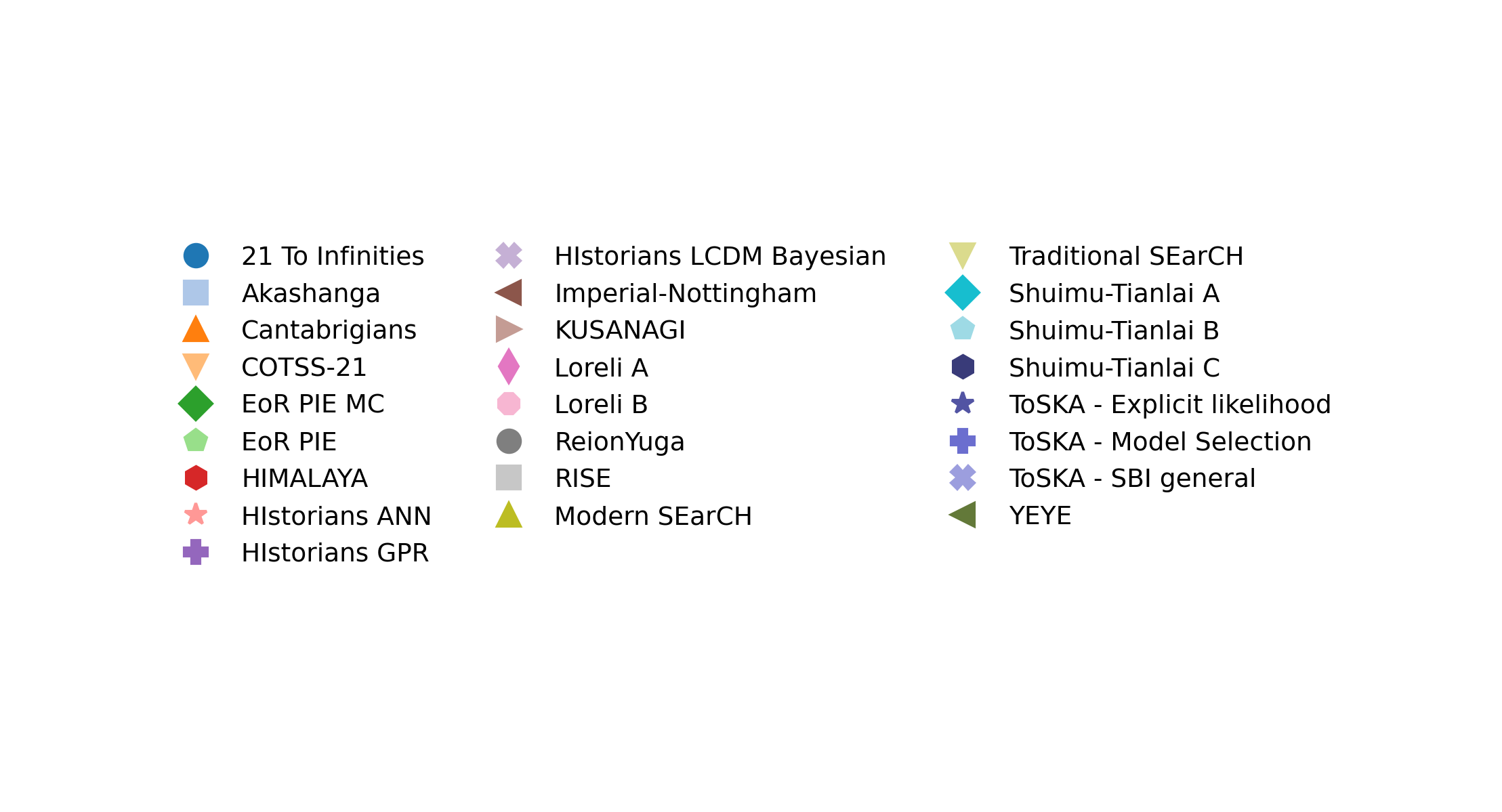}
    \caption{Legend showing the markers for all SDC3b teams used in the figures in this section. \label{fig:legend}}
\end{figure}

\begin{figure*}
\includegraphics[width=15cm]{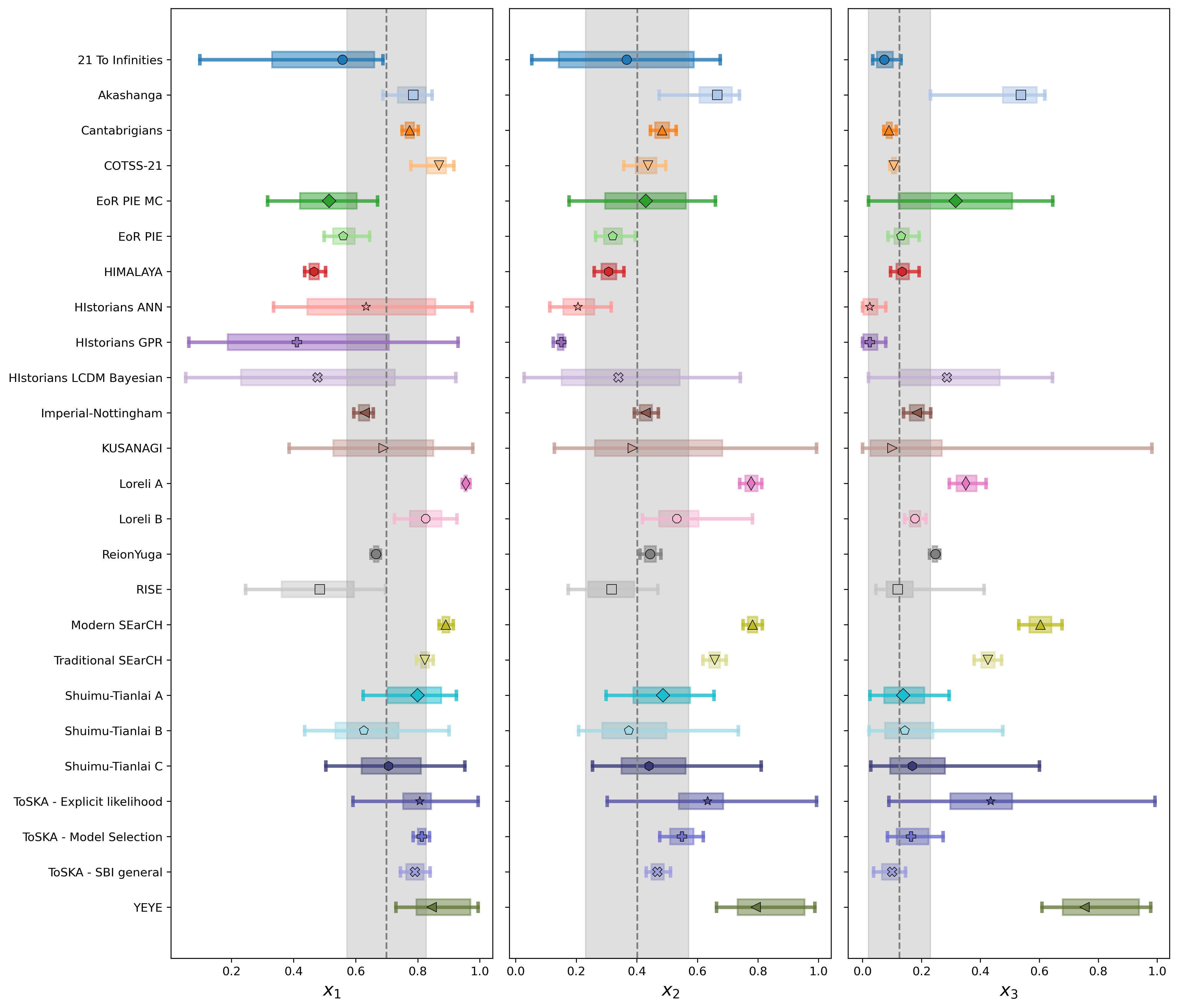}
\caption{Boxplot (see Sec. \ref{sec:forest}) for the PS1 results. The boxes and bars represent $\Delta_{68}$ and $\Delta_{95}$, respectively. The symbols are the marginal median $\bar{x}$, the vertical dashed line is the central input value $x'$ and the shaded area its evolution range $[a,b]$ across the light-cone.
\label{fig:forest1}}
\end{figure*}

\begin{figure}
\includegraphics[width=8cm]{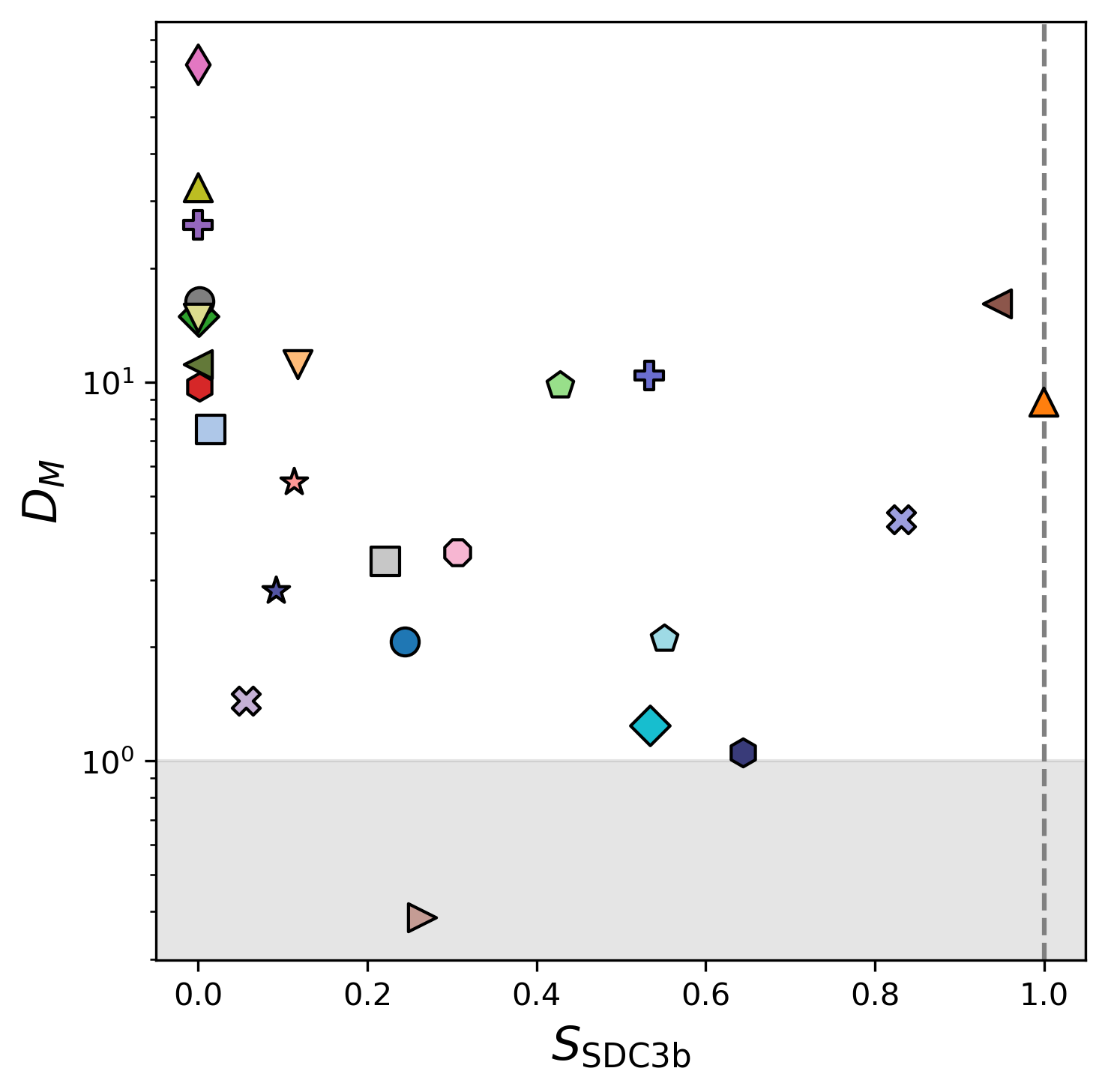}
\caption{Comparison of the SDC3b score $S_\mathrm{SDC3b}$ and the Mahalanobis distance $D_\mathrm{M}$ for the PS1 results. The markers are defined in Fig. \ref{fig:legend}. The gray horizontal band represents the $D_\mathrm{M}<1$ area, outside of which solutions are overconfident. The dashed vertical line is the maximum $S_\mathrm{SDC3b}$ score. \label{fig:scores1}}
\end{figure}

\begin{figure*}
\includegraphics[width=15cm]{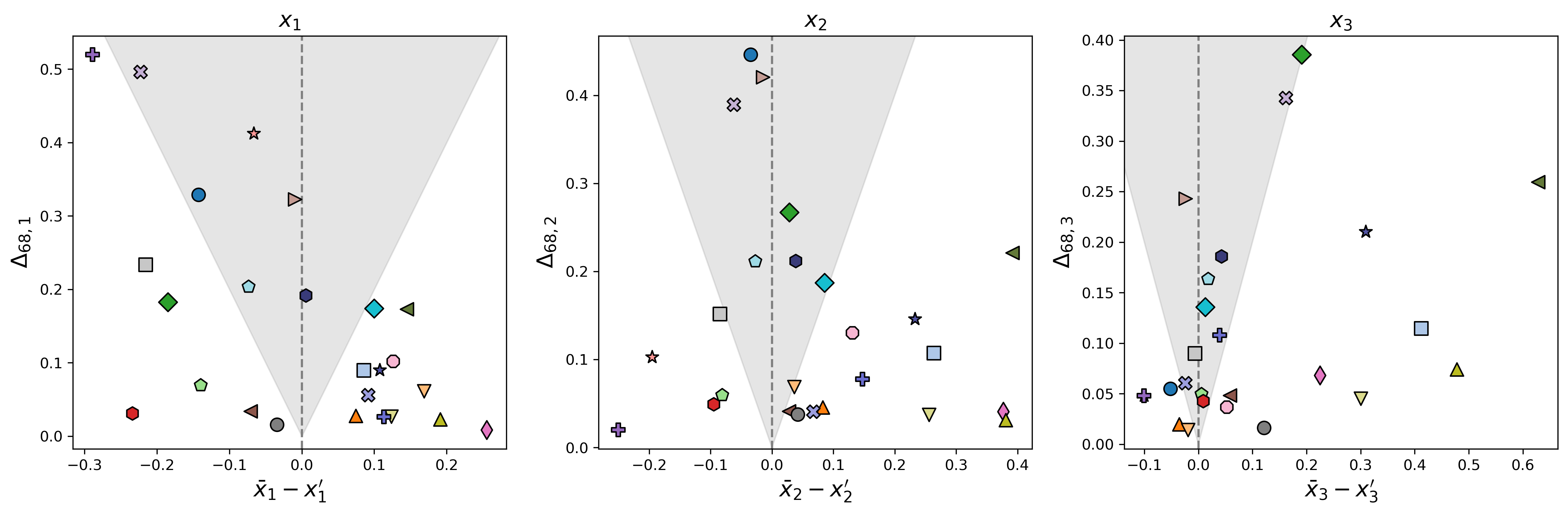}

\caption{Comparison of the bias $\bar{x}-x'$ and 68\% confidence interval $\Delta_{68}$ of the PS1 posteriors for the three redshift intervals (from left to right $z=z_1,z_2,z_3$). The grey area indicates where the bias is smaller than 68\% CI. The markers are as in the legend in Fig. \ref{fig:legend}.
\label{fig:pa1}}
\end{figure*}

\begin{figure}
\includegraphics[width=\columnwidth]{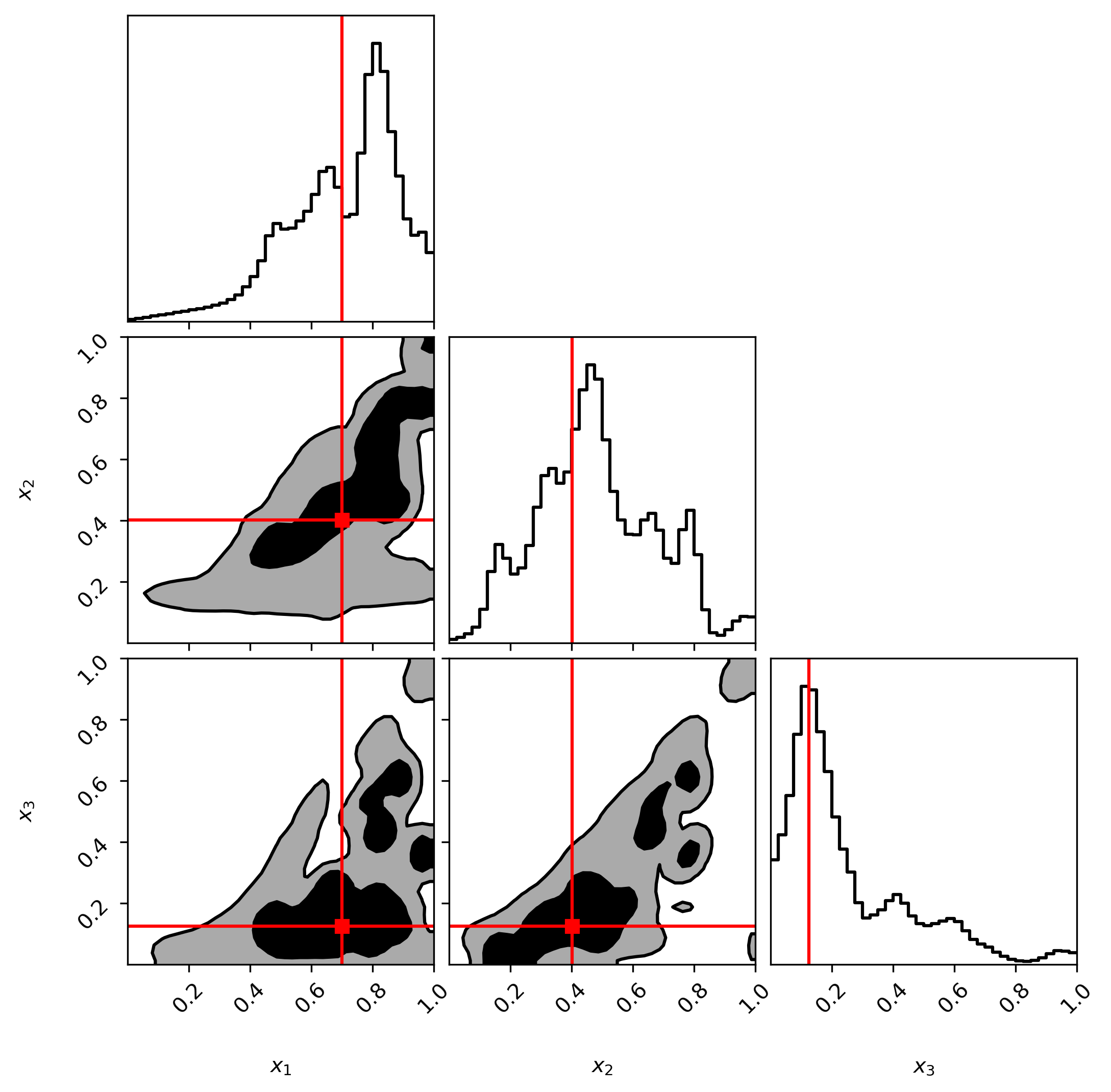}
\caption{Corner plot showing the average posteriors of all teams for PS1 (see \ref{sec:corner}). The gray and black contours give the 1 and 2\,$\sigma$ of this average posterior and the red line shows the input values. \label{fig:corner1}}
\end{figure}

\begin{figure*}
\includegraphics[width=15cm]{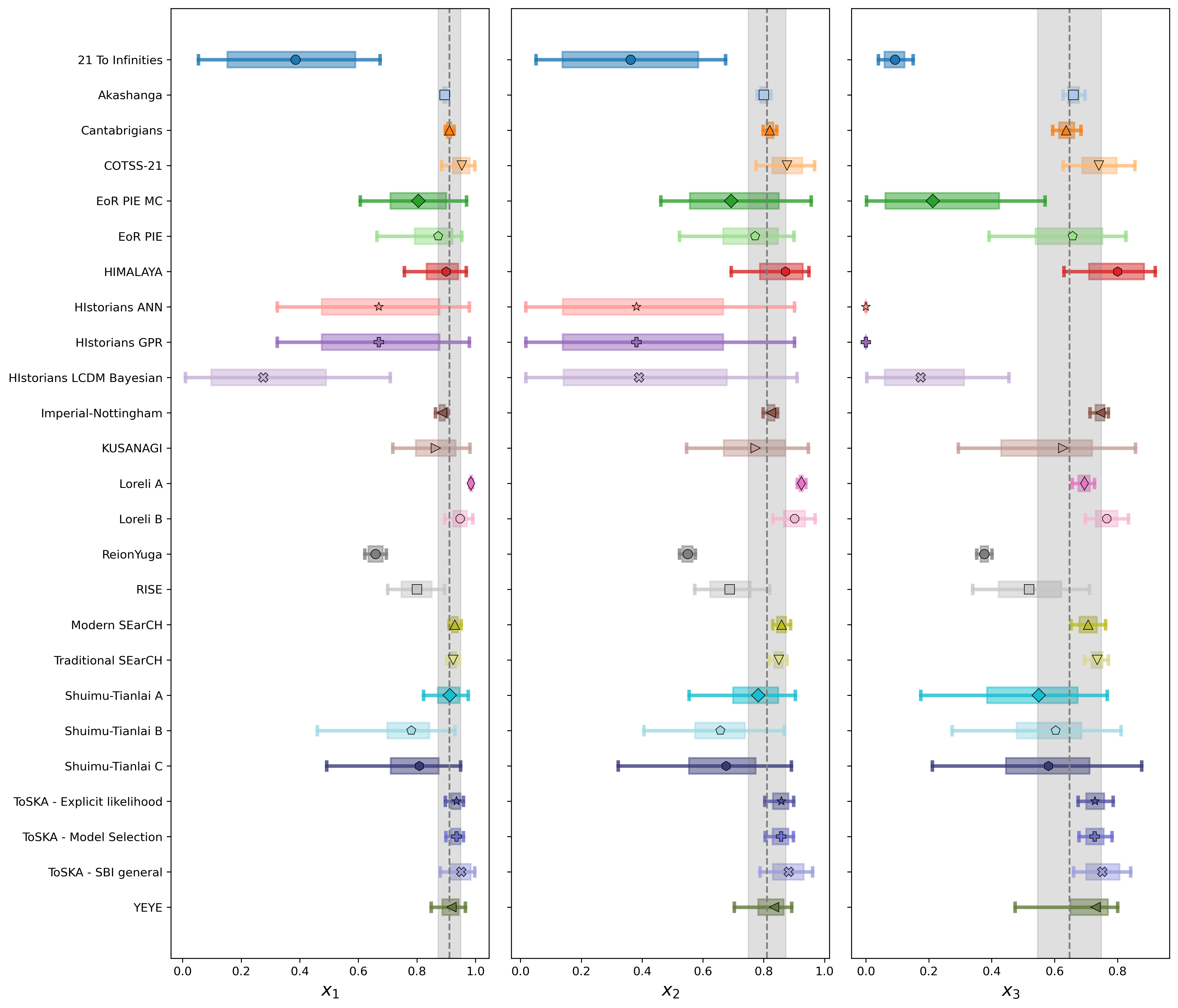}
\caption{Boxplot for the PS2 results, otherwise similar to Fig~\ref{fig:forest1}.
\label{fig:forest2}}
\end{figure*}

\begin{figure}
\includegraphics[width=\columnwidth]{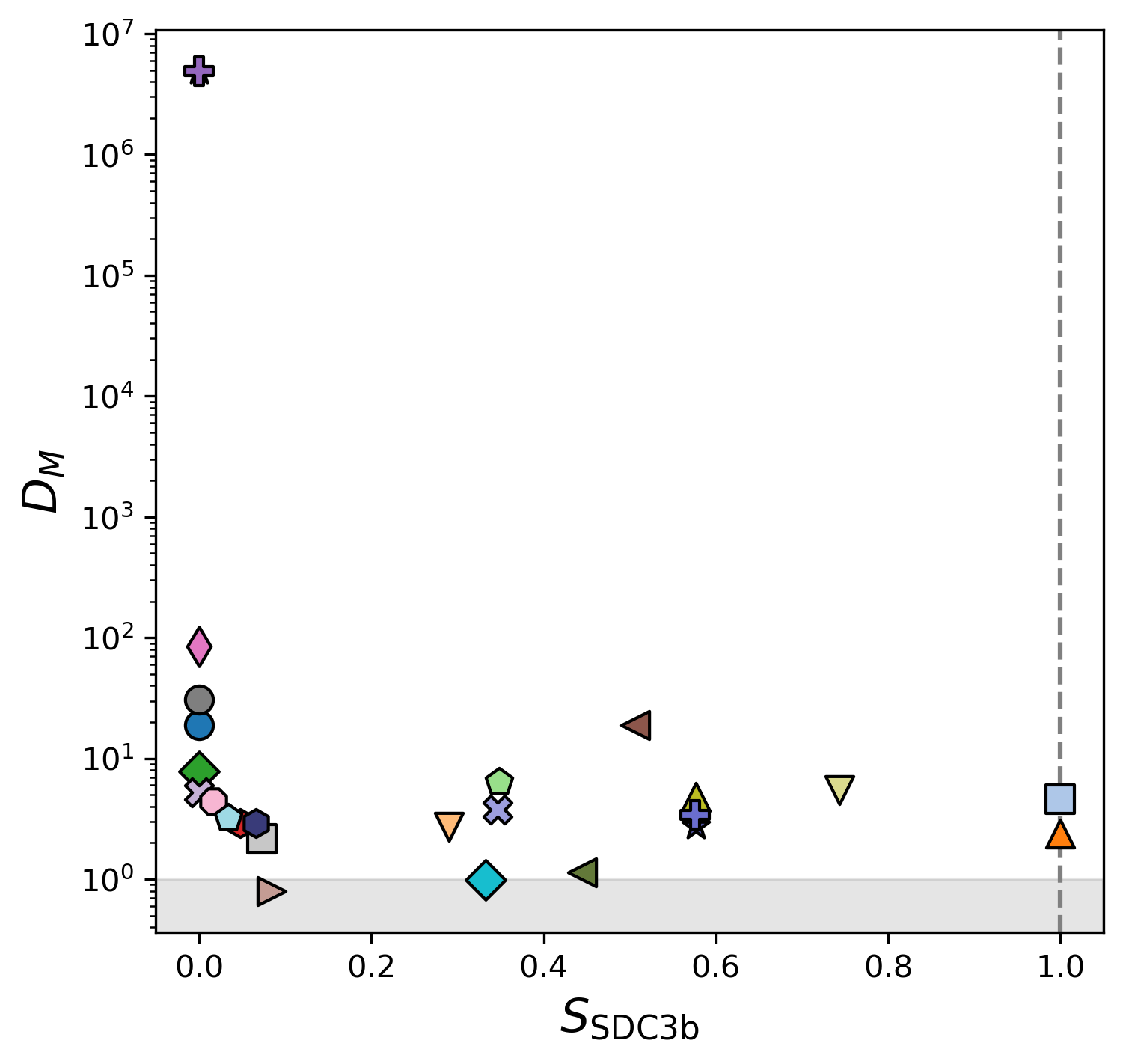}
\caption{As Fig.~\ref{fig:scores1} but for dataset PS2. \label{fig:scores2}}
\end{figure}

\begin{figure*}
\includegraphics[width=15cm]{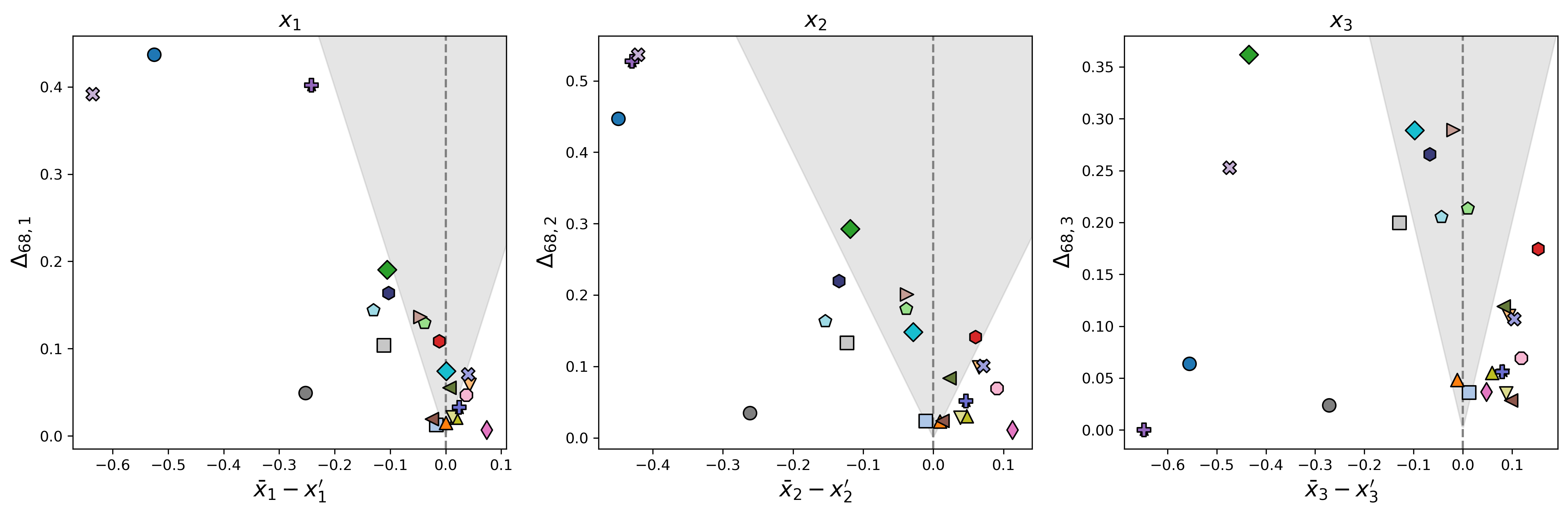}

\caption{As Fig.~\ref{fig:pa1} but for dataset PS2. \label{fig:pa2}}
\end{figure*}

\begin{figure}
\includegraphics[width=\columnwidth]{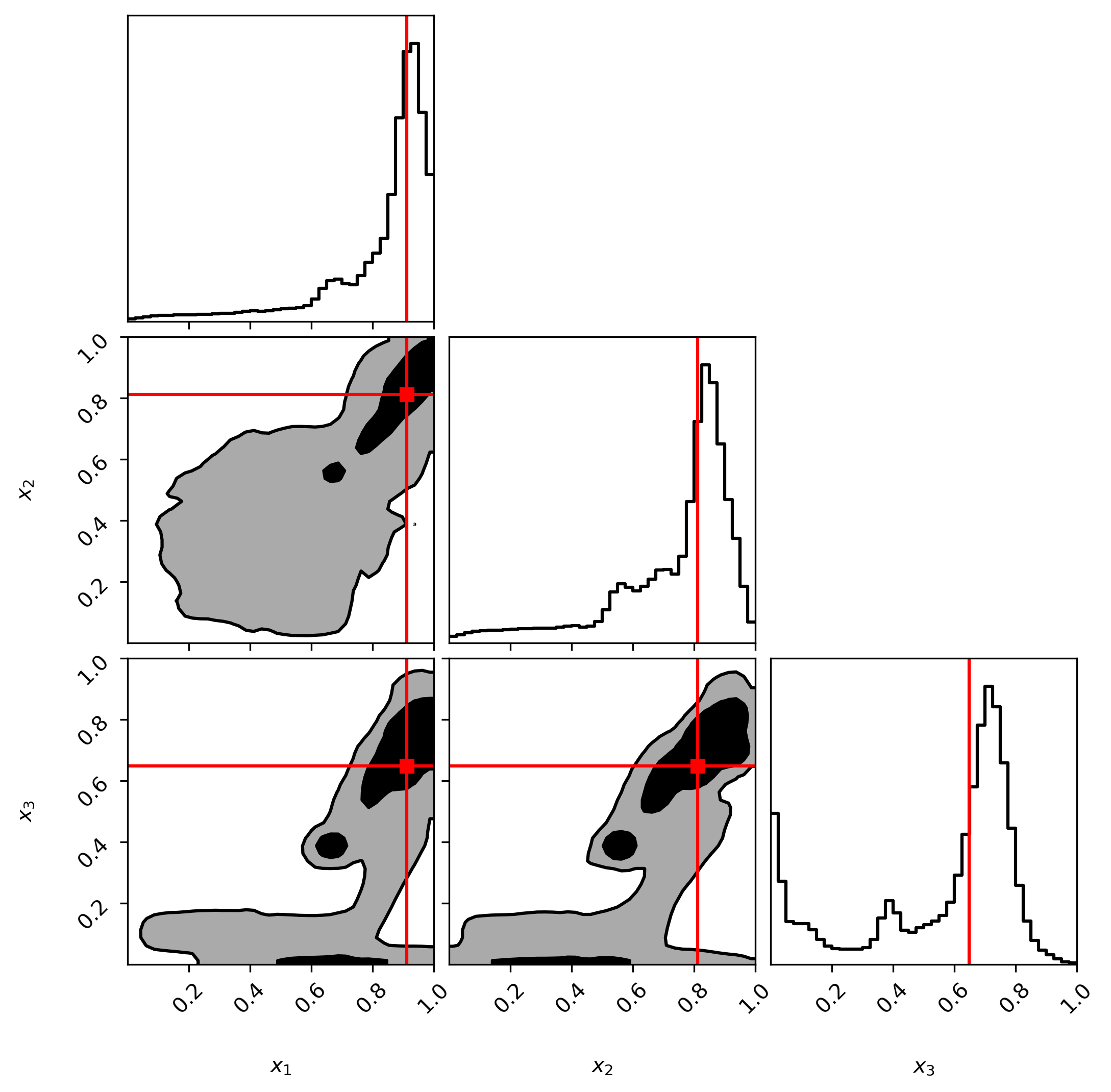}
\caption{As Fig.~\ref{fig:corner1} but for dataset PS2. \label{fig:corner2}}
\end{figure}

\subsection{Metrics and plots}\label{sec:metrics}
As already mentioned in Sec. \ref{sec:defined}, for each dataset, the results submitted by the participating teams are posterior distributions for the neutral fraction over three redshift ranges. We would aim to convert these distributions into performance scores, but due to the nature of the data there is no unique way to do this. In this section we describe the different metrics we use to evaluate the performance, as well as graphical representations that we use to present the results.

\subsubsection{Quality metrics}\label{sec:qmetrics}

We will denote the posterior distributions which we want to evaluate as $P(x_{\rm HI}; z_1,z_2,z_3)$. To simplify the notation, we adopt: 
\begin{itemize}
\item $x_1=x_{\rm HI}(z_1)$ where $z_1=8.41$--7.56
\item $x_2=x_{\rm HI}(z_2)$ where $z_2=7.56$--6.85
\item $x_3=x_{\rm HI}(z_3)$ where $z_3=6.85$--6.25
\end{itemize}
and, in vector form, $\mathbf{x}=[x_1,x_2,x_3]$. 

The corresponding true inputs are the neutral fraction values of the simulations for the three scenarios (PS1, PS2, PS3/IM1). As shown in Fig. \ref{fig:modelcomp}, the neutral fraction evolves smoothly within the given redshift ranges. With $\mathbf{x'}=[x'_1, x'_2,x'_3]$ we refer to the value at the centre of the redshift intervals. The ranges including evolutions are called $[a'_1,b'_1]$,  $[a'_2,b'_2]$ and $[a'_3,b'_3]$; in 3D they define a cuboid $R=[a'_1,b'_1]\times [a'_2,b'_2] \times [a'_3,b'_3]$

When considering the $x_1, x_2, x_3$ parameters separately, we define the following metrics
\begin{itemize}
\item the \emph{bias} $\bar{x}-x'$, where $\bar{x}$ is the marginal median of $x$;
\item the \emph{confidence intervals} (CI) $\Delta_{68}$ and $\Delta_{95}$, corresponding to the 68\% and 95\,\% confidence levels computed from the posterior (with no assumption of Gaussianity);
\item the \emph{standard deviation} $\sigma$ is estimated assuming a Gaussian approximation and is approximately half of $\Delta_{68}$;
\end{itemize}
From these quantities we construct the following metric to characterize the performance for a single redshift
\begin{equation}
    Z=\frac{|\bar{x}-x'|}{\sigma}
    \label{eq:zscore}
\end{equation}
We refer to this as the \emph{Z-score}. A $Z <1$ indicates a good recovery while $Z>2$ indicates a significant tension between input and recovered value; 

When evaluating all three neutral fraction parameters together, we consider two different metrics, namely the SDC3b score and the Mahalanobis distance. The former is defined as
\begin{equation}
    S_{\rm SDC3b}=\iiint _R P(x_1,x_2,x_3) dV\, , 
    \label{eq:scb3bscore}
\end{equation}
which is the integrated 3D posterior probability within the cuboid $R$ defined previously. This score was used at the end of the challenge to rank the team's performance. Its formulation was communicated to teams in advance of the end of the challenge. The optimal result for this metric is a score of 1, which indicates that the full posterior is within the cuboid $R$. The minimum score is zero, which means the posterior is entirely outside $R$. This is indicative of a large bias and a small CI (i.e. a large z score) for at least one parameter. 

The \emph{Mahalanobis distance} is instead defined as 
\begin{equation}
    D_{M}=\sqrt{(\mathbf{x}-\mathbf{x}')^T\mathbf{C}^{-1}(\mathbf{x}-\mathbf{x}')}\,,
    \label{eq:mdistance}
\end{equation} 
which is a multivariate Z-score accounting for parameter correlations ($\mathbf{C}$ is the covariance matrix). $D_M \lesssim 1$ indicates that the truth is within the joint 1\,$\sigma$ region. The optimal distance is zero, indicating that there is no bias in the recovery of the parameters. A broad posterior distribution will also reduce this metric. 

Both the SDC3b score and the Mahalanobis distance depend on the shape of the full posterior which encodes the estimated parameters and the relative confidence level. However, the formulation is different, which typically results in a different assessment of the team's performance.  The SDC3 score accommodates a bias as long as the solution is within the bounds of $R$. Solutions partially or completely outside these bounds are heavily penalised.  However, the ionization state of the central slice of the volume taken from a light-cone ($\mathbf{x}'$) typically has the strongest correlation with the power spectrum \citep{datta2012light}. Therefore the posterior of the neutral fraction inferred from the power spectrum really should include $\mathbf{x}'$, which is not ensured by a high SDC3b score. Also, there is no good physical motivation for a strong penalty on the parts of the posterior outside of the bounds of $R$. %{\bf sentence on Garrelth's paper on the fact that the bounds [a,b] are not important in inference, so this metric maybe overstates their importance?}

The Mahalanobis distance considers the central true values $\mathbf{x}'$ and makes no explicit accommodation for the true evolution within the redshift ranges. % It accommodates any bias as long as it is  reflected in the CI, so it particularly penalises underestimated error bars. 
If this distance is less than a certain value $q$, it means that the truth lies with q-$\sigma$ confidence interval of the inferred posterior. However, a broad posterior can easily capture the true value, even when the data is informative enough for a better constraint. Therefore this metric alone cannot provide a good assessment of the results from a team. %{\bf however the larger the error bar the better the metric looks?}

Hence both metrics have their advantages and drawbacks and we choose to present results in terms of both of these to provide a more balanced evaluation of the results. Other metrics could be chosen and result in a different scoring, so the performance of the different teams in terms of these two metrics should only be considered indicative of their actual performance.

\subsubsection{Boxplots and metric combination plots}\label{sec:forest}
The basic results are presented below using so-called box plots. In these plots, each team is represented by the $\Delta_{68}$ (rectangular box) and $\Delta_{95}$ (solid horizontal line) confidence interval error bars and a marker for the marginal median $\bar{x}$.  This compact representation allows us to display all teams on the same diagram and comparing their 1D marginalised posterior constraints with the truth $x'$ and the true range $[a,b]$ for the three redshift ranges.

The subsequent figures compare different performance metrics: the  SDC3b score $S_{\rm SDC3b}$ vs the Mahalanobis distance $D_{\rm M}$ and, for each redshift, the bias $\bar{x}-x'$ vs $\Delta_{68}$. For the latter the ideal outcome lies close to (0, 0). Results should preferably fall in the shaded area, which is where the bias is smaller than the error bar. Markers outside this region are indicative of an overconfident solution.

\subsubsection{Corner plots}\label{sec:corner}
The 3D posterior $P(x_1,x_2,x_3)$ can be visualised using a so-called "corner plot", which displays all three 2D posteriors and 1D posteriors that are obtained after marginalising over the other parameters.  Multiple entries can usually be overlaid on a single corner plot, however the large number of submissions for PS1 and PS2 do not allow a clear enough figure, so this is included only for PS3/IM1 (see Fig.\ref{fig:corner3} below). For PS1 and PS2 we instead present the corner plot of the \emph{average} posterior of all teams (Figs. \ref{fig:corner1} and \ref{fig:corner2}). 
Although this average posterior does not have a direct interpretation in Bayesian statistics, where independent likelihoods or posterior distributions would ordinarily be combined through multiplication rather than addition, the corner plot is useful for visualising how the ${x_{1},x_{2},x_{3}}$ parameter space has been populated by different teams. In addition, \url{https://abonaldi.github.io/SDC3b} allows interactively visualising corner plots for all teams, either individually or in comparison with any other number of teams, for all SDC3b datasets.

\subsection{Results for dataset PS1}
The boxplot in Fig. \ref{fig:forest1} provides an overview of the results for PS1. It shows the spread among the solutions obtained by individual teams, and the overall lack of overlap. The different median values $x^\prime$ show considerable differences, as do the CIs. Some teams obtained very narrow posterior distributions, whereas others obtained distributions spanning nearly the entire range. Some teams do very well on some of the redshift ranges, but not on others. 
Intriguingly, some cases show very little evolution of $x$ with redshift.
There are, however, no apparent trends in the sense that it is  easier to get close to the true value for one redshift interval than for another.

On the basis of the posteriors, the two different overall scores $S_{\rm SDC3b}$ and $D_{\rm M}$ were calculated, as well as the Z-scores for the three different redshift ranges. The numerical values are listed in Table \ref{tab:leader1}. Teams with their marker and error bars overlapping significantly with the vertical grey band (representing the range $R$) in Fig. \ref{fig:forest1} achieve the best $S_{\rm SDC3b}$; the 68\,\% and 95\,\% CIs touching the vertical dashed line representing the input $x'$ means a Z score less than 1 or 2 respectively; if this occurs in all three panels, the team obtains a good Mahalanobis score $D_{\rm M}$. Five teams obtained an $S_{\rm SDC3b}$ score greater than 0.68, which means that most of their posterior is within the cuboid $R$; the top three teams for this metric are Cantabrigians, Imperial-Nottingham and ToSKA-SBI general. One team (KUSANAGI) has $D_{\rm M}<1$;  Shuimu-Tianlai C and Shuimu-Tianlai A are just above 1 and two more teams (HIstorians LCDM Bayesian and 21 To Infinities) are below 2. The results of all remaining teams are in significant tension with the input values.

Fig. \ref{fig:scores1} compares $S_{\rm SDC3b}$ and $D_{\rm M}$ for all teams. First of all we note that we have used the logarithmic scale for $D_{\rm M}$ due to the high dynamic range for this metric. 
There is no clear anti-correlation between the two metrics, which confirms that they test different things. 
The best results in terms of $D_{\rm M}$ correspond to large error bars. However, this means that a significant portion of the posterior would be outside the bounds $[a,b]$ which leads to a lower $S_{\rm SDC3b}$ score. The optimal performance corresponds to  $S_{\rm SDC3b}=1$ and $D_{\rm M}=0$. There is a cluster of markers for $D_{\rm M}>>1$ and $S_{\rm SDC3b}=0$ , corresponding to poor performance according to both metrics. Those results are offset from the input and have a narrow posterior distribution. 

Fig. \ref{fig:pa1} presents $\bar{x}-x'$ vs $\Delta_{68}$ for each of the redshift ranges. The ideal outcome is close to $(0,0)$ and any location within the shaded area means that the bias is smaller than the error bar. Markers outside this region are indicative of an overconfident solution. Most results fall outside this shaded area, which implies that, even when the inferred values are close to the true values, the error bars are too small. 

To conclude we show in Fig. \ref{fig:corner1} the corner plot for the average posterior between all teams (see Sec.~\ref{sec:corner}). Interestingly, the average posterior is in agreement with the true inputs, albeit with a large spread. The presence of multiple peaks in both the 2D and 1D average posteriors is, however, indicative of individual solutions not being in agreement with each other. It is worth noting that there exists no overlap between all posteriors, so combining the posteriors in the Bayesian sense would return a null posterior.

\subsection{Results for dataset PS2}

Looking at the boxplot (Fig. \ref{fig:forest2}) of the results for dataset PS2 and comparing it with the PS1 results, 
we find better agreement across teams with the input values, $x'$, especially for the highest redshift, $z_1$. Looking at Fig. \ref{fig:modelcomp}, PS2 has a very high neutral fraction $x_{\rm HI}\sim 0.9$ for the highest redshift bin, with little evolution. In this scenario, the signal should be dominated by the matter power spectrum rather than by the reionisation parameters, which should make the neutral fraction easier to recover. Still we see considerable spread in the width of the posteriors, and some results quite far away from the truth.

Table \ref{tab:leader2} lists the metrics $Z_1$, $Z_2$, $Z_3$, $D_{\rm M}$ and $S_{\rm SDC3b}$ for all teams. In terms of $S_{\rm SDC3b}$, the best results are achieved by Cantabrigians and Akashanga; Traditional SEarCH, Modern SEarCH and Shuimu-Tianlai C also perform above 0.68. Two teams (KUSANAGI and Shuimu-Tianlai A) have $D_{\rm M}<1$ and a further one (YEYE) $D_{\rm M}<2$. The results of all other teams are in significant tension with the input values.

Comparing the SDC3b score and the Mahalanobis distance in Fig. \ref{fig:scores2} confirms the lack of a clear anti-correlation between the two metrics, meaning that teams performing well according to one metric generally do not perform well according to the other. Strikingly, although Fig.~\ref{fig:forest2} gives the general impression that the average performance is better than for PS1, only two teams achieve $S_\mathrm{SDC3b}>0.8$, whereas this was three for PS1. Actually comparing Figs. \ref{fig:scores1} and  Figs. \ref{fig:scores2} does not support the conclusion that the overall performance is better for the PS2 dataset than for the PS1 one.

However, Fig. \ref{fig:pa2} does show a larger number of markers falling within the grey areas than for PS1, which points to a more faithful recovery, even though significant outliers remain present. These different results point to the challenge of comparing the performance of different teams using simple metrics.

As for PS1, the corner plot for the average PS2 posterior for all teams, Fig. \ref{fig:corner2},  shows that the truth is consistent with the average posterior, which however exhibits multiple peaks and generally a large spread. Also for this dataset there exists no overlap between all the posteriors returned by the teams.

\subsection{PS3/IM1}
Submissions for the datasets PS3 and IM1 (containing foreground residuals) were optional for the challenge and were only performed by three teams: HIMALAYA, SEarCH, and YEYE. Since PS3 is the power spectrum derived from IM1, the two datasets have the same input values and are analysed together in this section.

\begin{figure*}
\includegraphics[width=15cm]{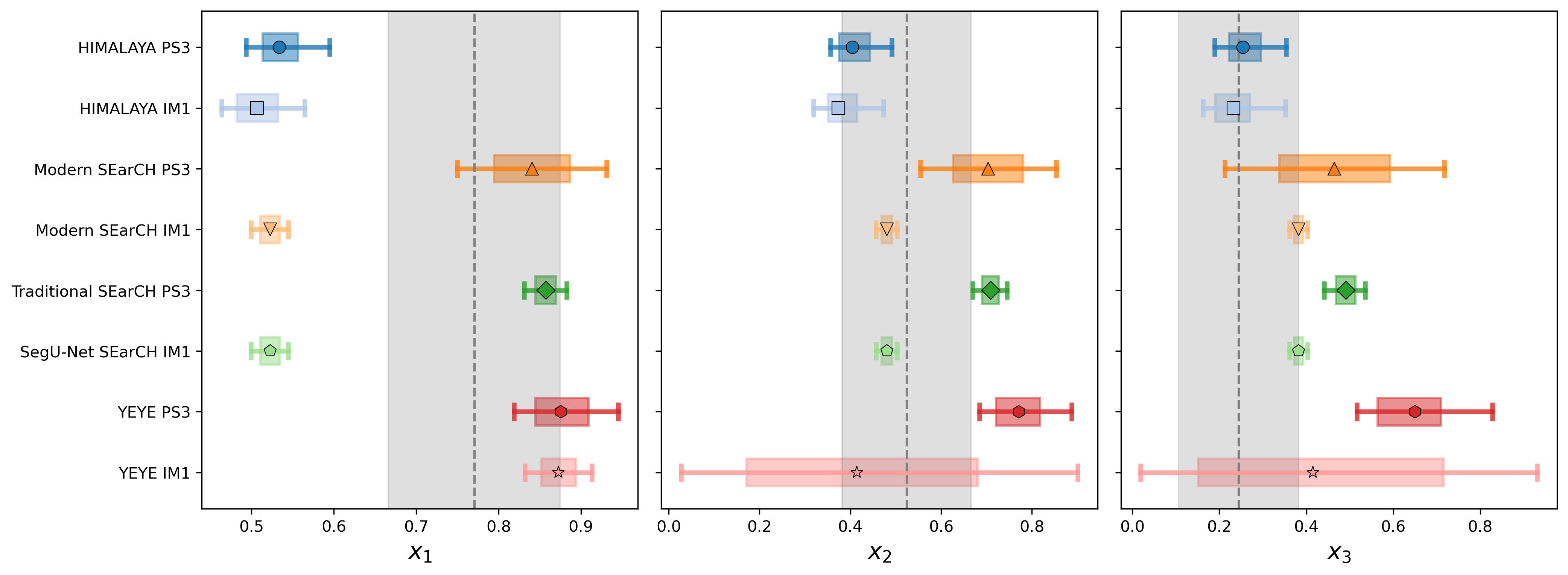}
\caption{Boxplot (see Sec. \ref{sec:forest}) for the PS3/IM1 results,  otherwise similar to Fig~\ref{fig:forest1}. 
\label{fig:forest3}}
\end{figure*}

The boxplot in Fig. \ref{fig:forest3} summarises the results. It is interesting to note that while HIMALAYA and YEYE present overlapping solutions for PS3 and IM1, SEarCH reach distinct conclusions based on the power spectrum and the image information. This is because an independent method was applied on the imaging dataset, as detailed in Sec. \ref{sec:413}. For PS3/IM1 we do not include the metric comparison figures because the small number of data points does not allow to draw any further conclusions. 

\begin{figure*}
\includegraphics[width=\columnwidth]{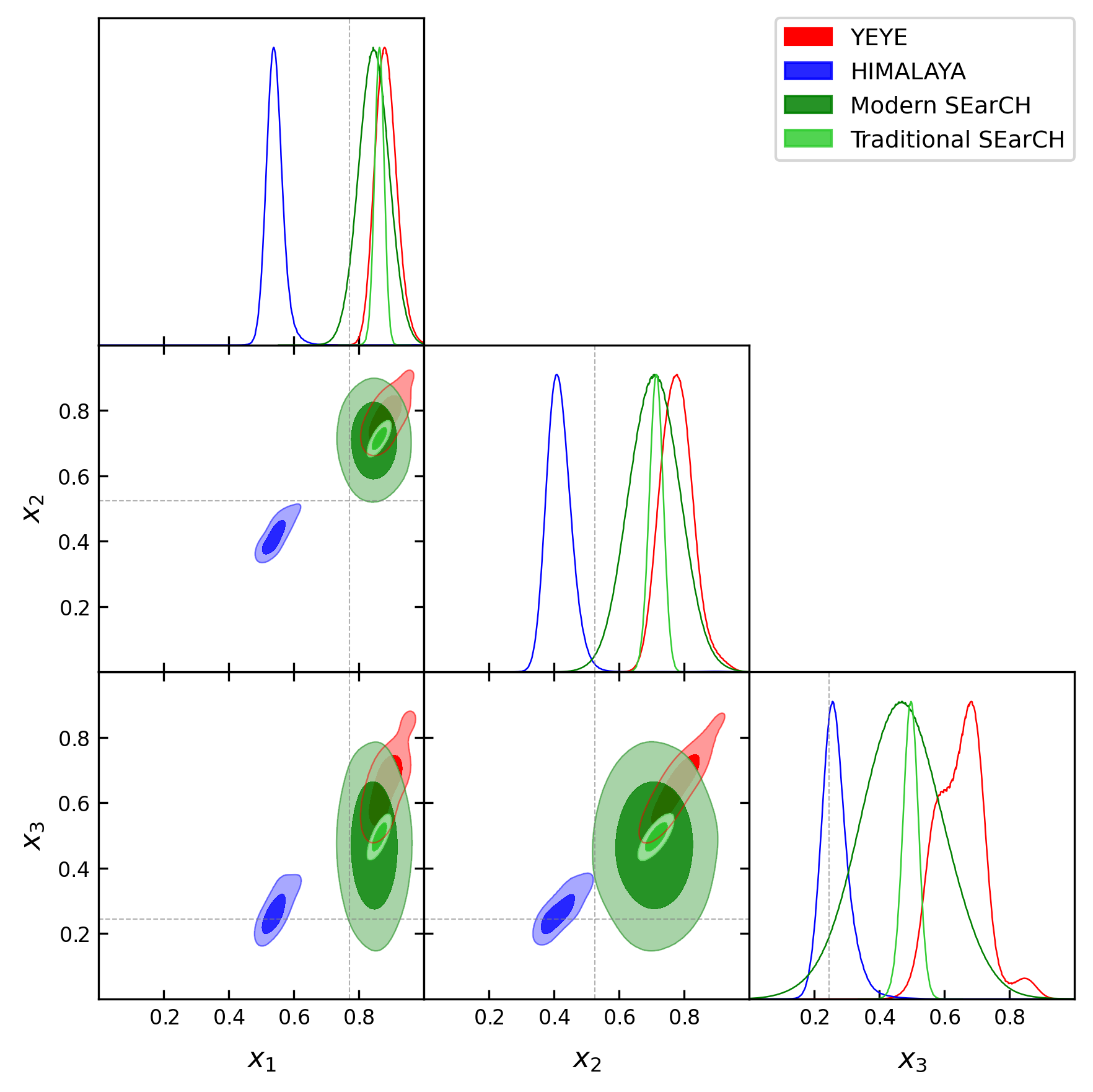}
\includegraphics[width=\columnwidth]{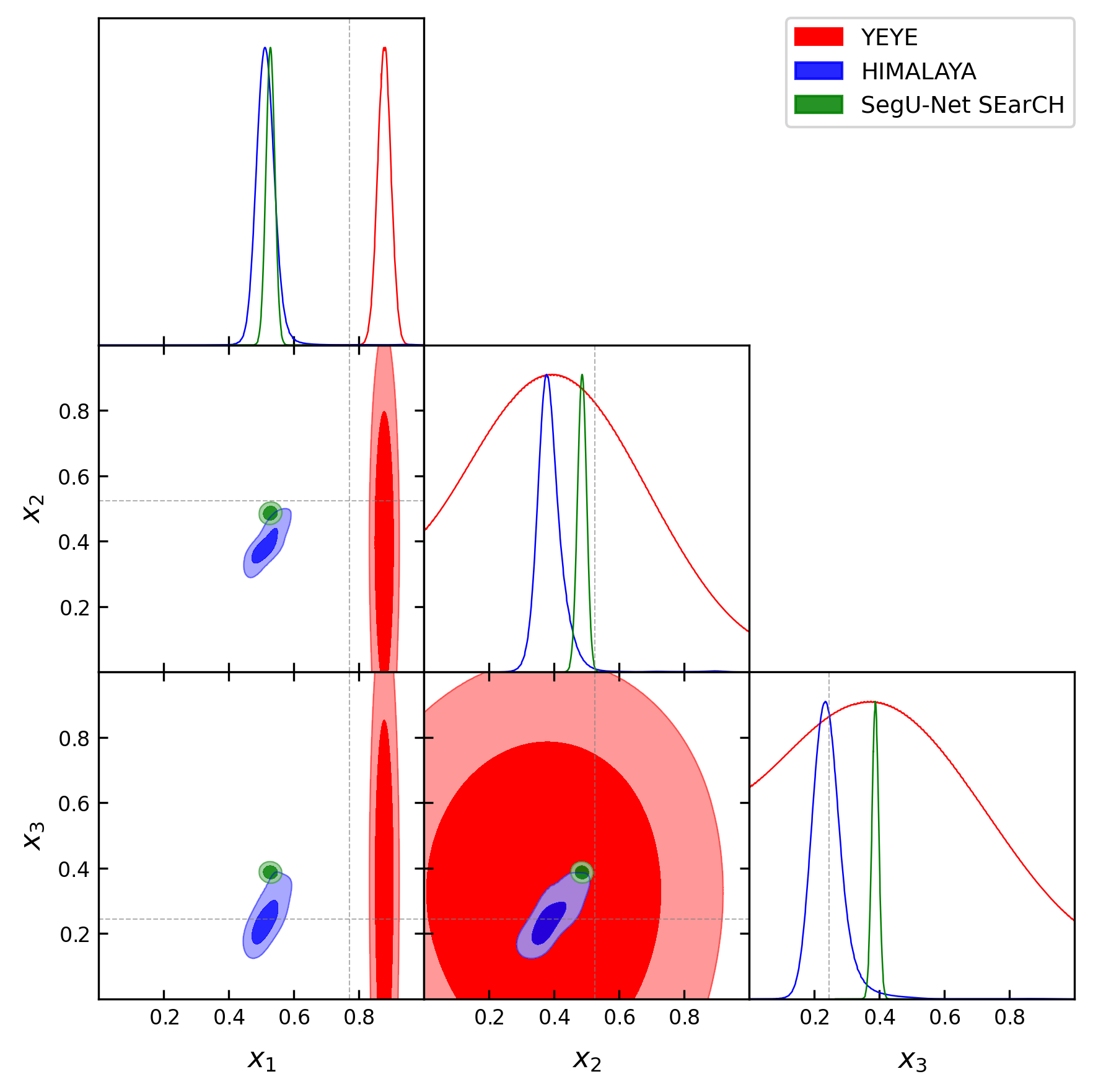}
\caption{Corner plots showing the posteriors for PS3 (left) and IM1 (right). \label{fig:corner3}}
\end{figure*}

As only three teams produced results for this dataset, it is possible to show the corner plots for all the teams superimposed. Fig. \ref{fig:corner3} shows the result for PS3 in the left panel and for IM1 on the right panel. For PS3, three results have some overlapping regions in posteriors but one result (HIMALAYA) reaches a completely different solution. For IM1, the posterior from HIMALAYA and SEarCH show some overlap but the YEYE team's posterior is biased in the opposite direction for $x_1$ and very broad for the other frequency bins. 

In both datasets, there is no good agreement with the true inputs. As a result, the scores (in Table \ref{tab:leader3}) do not indicate a good performance. The poor performance could be due to the higher complexity in having to deal with foreground contamination \citep[][]{giri2026impact}. However the number of participating teams is not enough to draw any firm conclusions, especially since the teams' performance for PS1 and PS2 were not particularly remarkable.

\section{Discussion and conclusions} \label{sec:conclu}

One of the main findings of this challenge is the substantial variation in parameter recovery among participating teams, even in the simpler cases in which only instrumental noise and minimal calibration errors were added to the 21-cm signal (PS1 and PS2). The recovered astrophysical constraints exhibit significant inter-team variability. This finding highlights that differences in inference methodology, statistical treatment, emulator construction, and implementation details can introduce systematic effects that are comparable to, or even larger than, the statistical uncertainties associated with the observations themselves.

An important implication of this result is that the uncertainties reported by individual pipelines may not fully capture the total uncertainty. The quoted confidence intervals typically account for statistical uncertainties within a given inference framework, but often do not include contributions arising from modelling choices, emulator construction, likelihood assumptions, or implementation-specific details. Consequently, the reported error bars may underestimate the total uncertainty. This is reflected in the typically large Z-scores, which are often greater than one (the median for $(z_1,z_2,z_3)$ between all teams is $(2.1,2.2,1.6)$ for PS1 and $(1.3,1.5,2.1)$ for PS2). Future analyses would benefit from explicitly quantifying these methodological uncertainties and incorporating them alongside conventional statistical errors in order to obtain more realistic confidence intervals.

One possible contributor to the inter-team variation is the treatment of the EoR signal in the inference pipelines.
EoR inference relies on a simulator code to produce the expected signal for different astrophysical parameter sets. Multiple simulators were used, as reported in Table \ref{tab:methods}, which may differ substantially in their physical assumptions and numerical approximations. However, within each inference framework, uncertainties in signal simulations or modelling are not typically accounted for. Shuimu-Tianlai and ToSKA are the only approaches making specific allowances for this error component, either by explicitly parametrising this uncertainty or by combining two independent simulators for the inference. 

However, if differences in EoR signal modelling were the primary driver of the observed inter-team scatter, one might expect teams employing the same simulator to reach similar conclusions. This does not appear to be the case. Overall, the most widely used simulator is {\sc 21cmFAST}, employed by 11 teams (Akashganga, EoR-PIE, EoR-PIE-MC, HIMALAYA, HIstorians, Imperial-Nottingham, KUSANAGI, ReionYuga, SEarCH, Shuimu-Tianlai, ToSKA). However, not all teams used the same version of the code, or the same source model, which should be taken into account. The Akashganga, EoR-PIE, HIstorians, KUSANAGI, SearCH, ToSKA teams used the \citet{park2019inferring}, whereas the EoR-PIE-MC, HIMALAYA, Imperial-Nottingham, ReionYuga and Shuimu-Tianlai used the \citet{mesinger201121cmfast} source model (see also Table \ref{tab:methods}). Nevertheless, even after accounting for these differences, no obvious trends emerge in the results, neither from  PS1 inference, that used {\tt C2Ray} for the mock data generation, nor from PS2, that used {\tt 21cmFAST}. This can be seen by looking e.g. Figs. \ref{fig:forest1} and  \ref{fig:forest2}. This suggests that, although simulator-dependent effects are likely present, they are not the sole contributor to the observed spread of results.

For those teams that performed a joint analysis of the three redshifts ranges, one possible source of modelling problems could be the flexibility of the model. Experiments by the {\sc Loreli} teams showed that 
they achieved better results when considering each redshift range separately, which they attributed to the fact that their simulator could not reproduce the reionization histories in PS1 and PS2. This is an important observation, particularly for the analysis of the real data, as we do not a priori know the reionization history of the Universe. If the reionization simulator used cannot reproduce the reionization history of the data, the inference would produce poor constraints, for one or for all three of the redshift ranges.

Almost all teams used machine-learning techniques that require training on large numbers of simulations, e.g. for an emulator or for SBI. The quality of this training could potentially be another source of discrepancies. Both under- and overtraining could be issues, as well as the sampling strategy and the dimensionality and extent of the parameter space. 
Different teams made substantially different choices in this respect, and this could also contribute to the different outcomes. 

We note that some of the teams used a parameter space of larger dimensionality than was strictly necessary as it included parameters which control the evolution of the spin temperature (EoR-PIE, Kusanagi, YEYE). The data challenge chose the spin temperature to be in the saturated regime, when it does not impact the 21-cm signal.

Others chose parameter spaces which were probably too small, by fixing key parameters such as the minimum or turn-over mass to values different from what was used to generate the challenge data (ReionYUGA), or even fixing the shape of the 21-cm power spectrum to be identical to the matter power spectrum (21 To Infinities). Since the minimum mass of haloes is known to have a large impact on the shapes of ionized regions and the 21-cm power spectrum \citep{2007MNRAS.377.1043M}, and the 21-cm power spectrum typically is not a scaled version of the matter power spectrum \citep{2022MNRAS.513.5109G}, neither of these two approaches should yield reasonable answers. Yet, their results are not appreciably worse than many of the others.

Apart from the dimension of the parameter space, the choice of prior ranges can also influence the outcome. For PS1 the minimum mass used was $10^{10}$~M$_\odot$. For nearly half of the teams, this was at the edge of their prior range (Akashganga, EoR-PIE, HIstorians, KUSANAGI, SEarCH, ToSKA, YEYE) and for some outside it (EoR-PIE MC, Imperial-Nottingham, ReionYUGA). As already stated, the choice of minimum mass has a large impact on the 21-cm power spectrum and having it outside or at the edge of the prior range can be expected to negatively impact the inference attempts. In addition it should be noted that $M_\mathrm{turn}$ and $M_\mathrm{min}$ do not impact the simulations in the same way, as the former implements an exponential, but still continuous reduction in the halo population, whereas the latter constitutes a hard limit.

%The results suggest that these modelling choices play a significant role in determining the final parameter constraints. Future comparisons should therefore focus not only on the inference machinery itself but also on quantifying the impact of different signal models when applied to identical datasets. Controlled experiments in which multiple inference frameworks employ a common EoR signal model would be particularly valuable for disentangling modelling uncertainties from inference-related effects.

Another possibly important factor to consider is the different approach to modelling instrumental effects. 
Some teams adopted simpler approaches, such as modelling or subtracting the noise bias, while others attempted to reproduce the mock observations in greater detail (Cantabrigians, HIstorians, Shiumu-Tianlai). We note that 
the results from the latter teams show an overall overlap with the ground truth, which may indicate the value of end-to-end forward modelling approaches that incorporate instrumental systematics as realistically as possible. However, this trend is not sufficiently strong. Several teams with more detailed instrumental modelling do not necessarily outperform those adopting simpler approaches. One complication is that teams modelling instrumental effects relied on different assumptions and approximations, some of which may have been more successful than others. 

The inclusion of foreground contamination in PS3 and IM1 undoubtedly further complicates the inference problem. Consistent with this expectation, the teams that analysed the foreground-contaminated dataset (HIMALAYA, SEarCH, YEYE) generally achieved less accurate parameter recovery than they did for the corresponding signal+noise case. However, the number of teams that submitted results for the foreground-contaminated dataset is substantially smaller than for the foreground-free case, limiting the statistical power of any comparison. It should also be noted that one of the teams (YEYE) did not attempt any actual residual foreground removal/mitigation, and therefore its results should not really be compared to the others.

Overall, the challenge highlights the need for further work to disentangle the relative importance of instrumental modelling, signal modelling, and inference methodology in determining the robustness of EoR parameter recovery in the signal+noise case. Once these contributions are better understood, a similar analysis can be extended to the foreground-contaminated regime.

Another notable outcome is the potential benefit of employing multiple independent inference pipelines. While individual teams occasionally produce biased parameter estimates, the ensemble average of all submitted solutions is found to be close to the true input values. This behaviour suggests that different methods may exhibit complementary strengths and weaknesses, and that combining independent analyses could help mitigate method-specific biases. Future observational analyses may therefore benefit from coordinated multi-pipeline inference efforts, rather than reliance on a single framework, in order to obtain more robust and representative parameter constraints.

What do these results imply for the inference on real 21-cm data once a detection has been achieved? Given the large spread in results it would seem safe to say that any inference on the mean neutral fraction should be considered indicative at best. As already suggested above, repeating the inference with a range of different methods may give a better estimate of the intrinsic uncertainties.

It is also important to remember that all current simulators rely on a relatively small number of parametrized models for the leakage of ionizing photons from galaxies. A challenge such as the one presented here could be an interesting test for how dependent any inference would be on the choice of galaxy model. Unfortunately the current challenge was not controlled enough to perform this test. Our poor knowledge of the actual properties of the galaxies driving reionization, many of them too faint to even be observable with current telescopes, also implies that we do not know how well the challenge data represents the actual 21-cm data. In view of this, even a good performance on challenge data does not necessarily imply a good performance on real data. Still a good performance on a controlled experiment, such as the one studied here, should be the starting point for any reliable inference tool. It will therefore be useful to perform future, more tightly restricted challenges in order to establish the origins of the wide range of outcomes that the current challenge has yielded.

\section*{Author contributions}
O. Bait, A. Bonaldi, P. Hartley, S. Purser, and R. Braun organised the challenge and prepared the SDC3b dataset. M. Bianco contributed to the production of the dataset by testing and developing a software interface to the simulators used and assisted with preparing the paper. S.\,K. Giri led the comparative analysis of the results from teams, produced the interactive web portal for exploring the team submissions and assisted with preparing the paper. While Bianco and Giri were also members of the SEarCH team, the dataset was blind to them, and the comparison tool was developed on fake data. G. Mellema provided guidance on the challenge design from its inception, led the overall interpretation of the results and co-led the writing of the paper together with A. Bonaldi. All remaining authors are SDC3b team participants, listed alphabetically in two tiers according to their level of contribution to the team effort, as determined by each team.

% here are team's acknowledgments
\section*{Acknowledgements}

Omkar Bait was supported by the National Science Foundation under Cooperative Agreement 2421782 and the Simons Foundation grant MPS-AI-00010515 awarded to the NSF-Simons AI Institute for Cosmic Origins — CosmicAI, https://www.cosmicai.org/.
\paragraph*{21 To Infinites}
21 To Infinites acknowledges the contribution of:
G B  Raghavkrishna, Deeptangshu Banik

\paragraph*{Akashganga}
Akashganga acknowledges the contribution of:
Abinash Kumar Shaw, Debanjan Sarkar, Suman Chatterjee, Arnab Chakraborty

AKS acknowledge the support by the U. S. National Science Foundation grant \#2206602. AKS also acknowledge the use of Bridges-2 at the Pittsburgh Computing Center through allocation PHY240002P from the Advanced Cyberinfrastructure Coordination Ecosystem: Services \& Support (ACCESS) program, which is supported by U. S. National Science Foundation grants. DS acknowledges the support of the Canada $150$ Chairs program, the Fonds de recherche du Qu\'{e}bec Nature et Technologies (FRQNT) and Natural Sciences and Engineering Research Council of Canada (NSERC) joint NOVA grant, and the Trottier Space Institute Postdoctoral Fellowship program. SC acknowledges computational resources that were provided by the Centre for High Performance Computing (NICIS-CHPC) under programme ASTR0945 and financial support from the South African National Research Foundation (Grant No. 84156) and the Inter-University Institute for Data Intensive Astronomy (IDIA)
\paragraph*{Cantabrigians}
Cantabrigians acknowledges the contribution of:
Yuchen Liu, Simon Pochinda, Charles Walker, Melis Irfan, Jiten Dhandha, Rennan Barkana, Harry T. J. Bevins, Oscar S.D. O’Hara, Fred Dulwich, Anastasia Fialkov, Dominic Anstey, Jeremy Coles, Eloy de Lera Acedo

\paragraph*{COTSS-21}
COTSS-21 acknowledges the contribution of:
Anshuman Acharya, Raghunath Ghara, Léon V.E. Koopmans, Arghyadeep Basu, Florent Mertens, Benedetta Ciardi, Satyapan Munshi

\paragraph*{EoR-PIE}
EoR-PIE acknowledges the contribution of:
Yannic Pietschke, Tom  Schlenker, Caroline Heneka, Anne  Hutter

YP's work is funded by the Volkswagen Foundation. CH's work is funded by the Volkswagen Foundation, and supported through Germany's Excellence Strategy through EXC~2181/1 -- 390900948 (the Heidelberg STRUCTURES Excellence Cluster) and with support by the Federal Ministry of Education and Research (BMBF) and the Ministry of Science, Research and the Arts of Baden-Württemberg. CH also acknowledges funding via the SciFM consortium (05D25VH3) funded by the German Federal Ministry of Research, Technology, and Space (BMFTR) in the ErUM-Data action plan.
\paragraph*{HIMALAYA}
HIMALAYA acknowledges the contribution of:
Junzhou Zhang, Le  Zhang

LZ is supported by the National Science and Technology Major Project of China (No.2024ZD1100601), National SKA Program of China (2025SKA0160100), National Science Foundation of China (12473097),  Guangdong Basic and Applied Basic Research Foundation (2024A1515012309).
\paragraph*{HIstorians}
HIstorians acknowledges the contribution of:
Anshuman Tripathi, Samit Kumar Pal, Leon Noble, Chandra Shekhar  Murmu, Yashrajsinh  Mahida, Manas Mohit Dosibhatla, Suman Majumdar, Abhirup Datta, Narendranath Patra Patra, Eeshan Beohar, Rashmi Sagar

AT acknowledges financial support from the Indian Institute of Technology Indore through a Teaching Assistantship and from the Indian Institute of Technology Kharagpur through the Institute Postdoctoral Fellowship. SM, MMD, LN, and AD acknowledge financial support through the Core Research Grant titled “Observing the Cosmic Dawn in Multicolour using Next Generation Telescopes” from the Science and Engineering Research Board (SERB) and the Department of Science and Technology (DST), Government of India. EB and MMD acknowledge financial support from the Indian Institute of Technology Indore. SKP, LN, and YM acknowledge financial support from the Department of Science and Technology, Government of India, through the INSPIRE Fellowship. CSM acknowledges financial support from the ARCO Prize Fellowship.
\paragraph*{Imperial-Nottingham}
Imperial-Nottingham acknowledges the contribution of:
Jonathan R. Pritchard, Nadia Cooper, Carina Norregaard, Emma Chapman, Jennifer Feron, Luke Conaboy

\paragraph*{KUSANAGI}
KUSANAGI acknowledges the contribution of:
Teppei Minoda, Kana Moriwaki, Shintaro Yoshiura, Hayato Shimabukuro, Tsutomu T.  Takeuchi, Takumi Ito

TM was supported by The National Natural Science Foundation of China (NSFC) (Grant No. W2433009). KM was supported by JSPS KAKENHI (Grant Number, 23K03446, 23K20035, 24H00004). SY was supported by JSPS KAKENHI (Grant Number, 24K17098, 24K00625). HS was supported by the Yunnan Provincial Key Laboratory of Survey Science (Project No. 202449CE340002), the National SKA Program of China (Grant No. 2020SKA0110401), the National Natural Science Foundation of China (Grant No. 12103044), and the “High-End Foreign Expert of the Yunnan Revitalization Talents Support Plan of Yunnan Province”. TTT was supported by JSPS Grant-in-Aid for Scientific Research (24H00247), JST CREST (JPMJCR24Q1), and the Joint Research Program of the Institute of Statistical Mathematics, ``Machine Learning Cosmology: From Structure Formation to Galaxy Evolution''.
. TI was supported by JST SPRING, Grant Number JPMJSP2127 (Kumamoto University Cross disciplinary Doctoral Human Resource Development Program to Lead the Well-being Society), JSPS Overseas Challenge Program for Young Researchers.
\paragraph*{Loreli}
Loreli acknowledges the contribution of:
Romain Meriot, Benoit Semelin, Aristide Doussot, David Cornu, Florent Mertens

\paragraph*{ReionYuga}
ReionYuga acknowledges the contribution of:
Aadarsh  Ojha, Anoop Krishna, Deepthi Moorkanat, Hiten Hiten, Rajesh Mondal, Samir Choudhuri, Khandakar Md Asif Elahi, Srijita Pal

\paragraph*{RISE}
RISE acknowledges the contribution of:
Anirban  Chakraborty, Tirthankar Roy Choudhury, Barun Maity, Divesh  Jain, Prakash  Gaikwad, Aseem  Paranjape, Suman  Pramanick

\paragraph*{SEarCH}
SEarCH acknowledges the contribution of:
Michele Bianco, Sambit K. Giri, Nicolas  Cerardi, Davide  Piras, Emmanuel de Salis, Massimo De Santis, Merve  Selcuk-Simsek, Philipp  Denzel, Kelley M. Hess, M. Carmen  Toribio, Franz  Kirsten, Hatem  Ghorbel

The SEarCH team acknowledges support from the Sweden SKA Regional Centre (sweSRC) node operated by Onsala Space Observatory in collaboration with Chalmers e-Commons and the Swiss National Supercomputing Centre (CSCS) computing allocation under the project \texttt{sk014}. The Onsala Space Observatory national research infrastructure is funded through Swedish Research Council grant No 2019-00208. SKG acknowledges support from the Olle Engkvist Stiftelse (grant no. 232-0238).
\paragraph*{Shuimu-Tianlai
}
Shuimu-Tianlai
 acknowledges the contribution of:
Yi Mao, Ce Sui, Xiaosheng Zhao, Shifan Zuo, Xuelei Chen, Xin Zhang, Richard Grumitt, Kangning Diao, Jun-Qing Xia, Xianchuan Yu, Yepeng Yan, Zongyao Yin, Ruiqing Yan, Guo-Jian Wang, Hayato Shimabukuro, Qingbo Ma, Li Zhang, Yan Gong, Yidong Xu, Bin Yue, Qiao Wang, Kaifeng Yu, Xingchen Zhou, Yuer Jiang, Furen Deng, Chengliang Xu, Xin Wang, Jiacheng Ding, Yue Wang, Yichao Li, Zewei Zhao, Liyang Gao, Pengju Wu, Yue Shao, Yuxin Wang, Qi Niu

X.Z. has been supported by a grant from Schmidt Sciences. We acknowledge support from the National Natural Science Foundation of China (No. 12303004), the Guizhou Provincial Basic Research Program(Natural Science)(No. QNA[2026]005), the National Natural Science Foundation of China (No.12273007),  the Guizhou Provincial Leading Talent Development Program (No. KJLYRC-[2026]007)
\paragraph*{ToSKA}
ToSKA acknowledges the contribution of:
Junsong Cang, Andrei Mesinger, Adélie Gorce, Steven Murray, Daniela Breitman, Nikolaos Triantafyllou, Roberto Trotta, Yuxiang Qin

\paragraph*{YEYE}
YEYE acknowledges the contribution of:
Jiajun Zhang

\section*{Data availability statement}

The challenge inputs are available on \url{https://www.skao.int/en/science-users/170/simulated-data-products/ska-low-eor}.
For the main data simulation procedure and underlying software repositories, we refer the reader to \cite{sdc3a1}. All challenge outputs, the code to reproduce the 21-cm signal simulation, and the analysis scripts are available at \url{https://github.com/abonaldi/SDC3b}. This repository also includes code that powers an interactive web portal to explore team submissions within and beyond the figures presented here (\url{https://abonaldi.github.io/SDC3b/}).

% The challenge outputs are available on \url{https://github.com/abonaldi/SDC3b}, together with code to interactively display team's results according to and beyond the plots presented in this paper (\url{https://abonaldi.github.io/SDC3b/}). 

% For the main data simulation procedure, including software repositories, we refer the reader to \cite{sdc3a1}. The code to reproduce the 21-cm signal in this simulation and the analysis of the results are available in \url{https://github.com/abonaldi/SDC3b}. 

\bibliography{sdc3b}

@ARTICLE{sdc3a1,
       author = {{Bonaldi}, Anna and {Hartley}, Philippa and {Purser}, Simon and {Bait}, Omkar and {Lee}, Eunseong and {Braun}, Robert and {Mertens}, Florent and {Bracco}, Andrea and {Williams}, Wendy and {Trott}, Cath},
        title = "{SKA-Low simulations for a cosmic dawn/epoch of reionisation deep field}",
      journal = {The Open Journal of Astrophysics},
         year = 2026,
        month = feb,
       volume = {9},
        pages = {57763},
          doi = {10.33232/001c.157763},
       adsurl = {https://ui.adsabs.harvard.edu/abs/2026OJAp....957763B}
}

@ARTICLE{giri2018optimal,
       author = {{Giri}, Sambit K. and {Mellema}, Garrelt and {Ghara}, Raghunath},
        title = "{Optimal identification of H II regions during reionization in 21-cm observations}",
      journal = {\mnras},
         year = 2018,
        month = oct,
       volume = {479},
       number = {4},
        pages = {5596-5611},
          doi = {10.1093/mnras/sty1786},
archivePrefix = {arXiv},
       eprint = {1801.06550},
 primaryClass = {astro-ph.CO},
       adsurl = {https://ui.adsabs.harvard.edu/abs/2018MNRAS.479.5596G}
}

@ARTICLE{cerardi2026implicit,
       author = {{Cerardi}, Nicolas and {Giri}, Sambit K. and {Bianco}, Michele and {Piras}, Davide and {de Salis}, Emmanuel and {De Santis}, Massimo and {Selcuk-Simsek}, Merve and {Denzel}, Philipp and {Hess}, Kelley M. and {Toribio}, M. Carmen and {Kirsten}, Franz and {Ghorbel}, Hatem},
        title = "{Implicit inference of the reionization history with higher-order statistics of the 21-cm signal}",
      journal = {\mnras},
         year = 2026,
        month = jun,
       volume = {549},
       number = {1},
          eid = {stag779},
        pages = {stag779},
          doi = {10.1093/mnras/stag779},
       adsurl = {https://ui.adsabs.harvard.edu/abs/2026MNRAS.549ag779C}
}

@INPROCEEDINGS{de2025exploring,
       author = {{de Salis}, Emmanuel and {De Santis}, Massimo and {Piras}, Davide and {Giri}, Sambit K. and {Bianco}, Michele and {Cerardi}, Nicolas and {Denzel}, Philipp and {Selcuk-Simsek}, Merve and {Hess}, Kelley M. and {Toribio}, M. Carmen and others},
        title = "{Exploring the Early Universe with Deep Learning}",
    booktitle = {Progress in Artificial Intelligence (EPIA 2025)},
         year = 2025,
        pages = {426-438},
    publisher = {Springer}
}

@INPROCEEDINGS{tolstikhin2021mlp,
       author = {{Tolstikhin}, Ilya and {Houlsby}, Neil and {Kolesnikov}, Alexander and {Beyer}, Lucas and {Zhai}, Xiaohua and {Unterthiner}, Thomas and {Yung}, Jessica and {Steiner}, Andreas and {Keysers}, Daniel and {Uszkoreit}, Jakob and {Lucic}, Mario and {Dosovitskiy}, Alexey},
        title = "{MLP-Mixer: An all-MLP Architecture for Vision}",
    booktitle = {Advances in Neural Information Processing Systems},
         year = 2021,
       volume = {34},
        pages = {24261-24272},
archivePrefix = {arXiv},
       eprint = {2105.01601},
 primaryClass = {cs.CV},
       adsurl = {https://ui.adsabs.harvard.edu/abs/2021arXiv210501601T}
}

@ARTICLE{datta2012light,
       author = {{Datta}, Kanan K. and {Mellema}, Garrelt and {Mao}, Yi and {Iliev}, Ilian T. and {Shapiro}, Paul R. and {Ahn}, Kyungjin},
        title = "{Light-cone effect on the reionization 21-cm power spectrum}",
      journal = {\mnras},
         year = 2012,
        month = aug,
       volume = {424},
       number = {3},
        pages = {1877-1891},
          doi = {10.1111/j.1365-2966.2012.21293.x},
archivePrefix = {arXiv},
       eprint = {1109.1284},
 primaryClass = {astro-ph.CO},
       adsurl = {https://ui.adsabs.harvard.edu/abs/2012MNRAS.424.1877D}
}

@ARTICLE{jeffrey2021likelihood,
       author = {{Jeffrey}, Niall and {Alsing}, Justin and {Lanusse}, Fran{\c{c}}ois},
        title = "{Likelihood-free inference with neural compression of DES SV weak lensing map statistics}",
      journal = {\mnras},
         year = 2021,
        month = jan,
       volume = {501},
       number = {1},
        pages = {954-969},
          doi = {10.1093/mnras/staa3594},
archivePrefix = {arXiv},
       eprint = {2009.08459},
 primaryClass = {astro-ph.CO},
       adsurl = {https://ui.adsabs.harvard.edu/abs/2021MNRAS.501..954J}
}

@ARTICLE{davies25,
       author = {{Davies}, James E. and {Mesinger}, Andrei and {Murray}, Steven G.},
        title = "{Efficient simulation of discrete galaxy populations and associated radiation fields over the first billion years}",
      journal = {\aap},
         year = 2025,
        month = sep,
       volume = {701},
          eid = {A236},
        pages = {A236},
          doi = {10.1051/0004-6361/202554951},
archivePrefix = {arXiv},
       eprint = {2504.17254},
 primaryClass = {astro-ph.CO},
       adsurl = {https://ui.adsabs.harvard.edu/abs/2025A&A...701A.236D}
}

@ARTICLE{1996Natur.380..603B,
       author = {{Bond}, J. Richard and {Kofman}, Lev and {Pogosyan}, Dmitry},
        title = "{How filaments of galaxies are woven into the cosmic web}",
      journal = {\nat},
         year = 1996,
        month = apr,
       volume = {380},
       number = {6575},
        pages = {603-606},
          doi = {10.1038/380603a0},
archivePrefix = {arXiv},
       eprint = {astro-ph/9512141},
 primaryClass = {astro-ph},
       adsurl = {https://ui.adsabs.harvard.edu/abs/1996Natur.380..603B}
}

@ARTICLE{2007ApJ...669..663M,
       author = {{Mesinger}, Andrei and {Furlanetto}, Steven},
        title = "{Efficient Simulations of Early Structure Formation and Reionization}",
      journal = {\apj},
         year = 2007,
        month = nov,
       volume = {669},
       number = {2},
        pages = {663-675},
          doi = {10.1086/521806},
archivePrefix = {arXiv},
       eprint = {0704.0946},
 primaryClass = {astro-ph},
       adsurl = {https://ui.adsabs.harvard.edu/abs/2007ApJ...669..663M}
}

@ARTICLE{sheth_tormen2002,
       author = {{Sheth}, Ravi K. and {Tormen}, Giuseppe},
        title = "{An excursion set model of hierarchical clustering: ellipsoidal collapse and the moving barrier}",
      journal = {\mnras},
         year = 2002,
        month = jan,
       volume = {329},
       number = {1},
        pages = {61-75},
          doi = {10.1046/j.1365-8711.2002.04950.x},
archivePrefix = {arXiv},
       eprint = {astro-ph/0105113},
 primaryClass = {astro-ph},
       adsurl = {https://ui.adsabs.harvard.edu/abs/2002MNRAS.329...61S}
}

@ARTICLE{Zhang2023,
       author = {{Zhang}, Haolin and {Zuo}, Shifan and {Zhang}, Le},
        title = "{Map Reconstruction of Radio Observations with Conditional Invertible Neural Networks}",
      journal = {Research in Astronomy and Astrophysics},
         year = 2023,
        month = jul,
       volume = {23},
       number = {7},
          eid = {075011},
        pages = {075011},
          doi = {10.1088/1674-4527/acd0ee},
archivePrefix = {arXiv},
       eprint = {2306.09217},
 primaryClass = {astro-ph.IM},
       adsurl = {https://ui.adsabs.harvard.edu/abs/2023RAA....23g5011Z}
}

@ARTICLE{ardi2019,
       author = {{Ardizzone}, Lynton and {L{\"u}th}, Carsten and {Kruse}, Jakob and {Rother}, Carsten and {K{\"o}the}, Ullrich},
        title = "{Guided Image Generation with Conditional Invertible Neural Networks}",
      journal = {arXiv e-prints},
         year = 2019,
        month = jul,
          eid = {arXiv:1907.02392},
        pages = {arXiv:1907.02392},
          doi = {10.48550/arXiv.1907.02392},
archivePrefix = {arXiv},
       eprint = {1907.02392},
 primaryClass = {cs.CV},
       adsurl = {https://ui.adsabs.harvard.edu/abs/2019arXiv190702392A}
}

@ARTICLE{21cmemu,
       author = {{Breitman}, Daniela and {Mesinger}, Andrei and {Murray}, Steven G. and {Prelogovi{\'c}}, David and {Qin}, Yuxiang and {Trotta}, Roberto},
        title = "{21CMEMU: an emulator of 21CMFAST summary observables}",
      journal = {\mnras},
         year = 2024,
        month = feb,
       volume = {527},
       number = {4},
        pages = {9833-9852},
          doi = {10.1093/mnras/stad3849},
archivePrefix = {arXiv},
       eprint = {2309.05697},
 primaryClass = {astro-ph.CO},
       adsurl = {https://ui.adsabs.harvard.edu/abs/2024MNRAS.527.9833B}
}

@ARTICLE{giri2020tools21cm,
       author = {{Giri}, Sambit and {Mellema}, Garrelt and {Jensen}, Hannes},
        title = "{Tools21cm: A python package to analyse the large-scale 21-cm signal from the Epoch of Reionization and Cosmic Dawn}",
      journal = {The Journal of Open Source Software},
         year = 2020,
        month = aug,
       volume = {5},
       number = {52},
          eid = {2363},
        pages = {2363},
          doi = {10.21105/joss.02363},
       adsurl = {https://ui.adsabs.harvard.edu/abs/2020JOSS....5.2363G}
}

@INPROCEEDINGS{oskar,
       author = {{Dulwich}, F. and {Mort}, B.~J. and {Salvini}, S. and {Zarb Adami}, K. and {Jones}, M.~E.},
        title = "{OSKAR: Simulating Digital Beamforming for the SKA Aperture Array}",
    booktitle = {Wide Field Astronomy \& Technology for the Square Kilometre Array},
         year = 2009,
        month = jan,
          eid = {31},
        pages = {31},
          doi = {10.22323/1.132.0031},
       adsurl = {https://ui.adsabs.harvard.edu/abs/2009wska.confE..31D}
}

@ARTICLE{vogel,
       author = {{Vogel}, Helmut},
        title = "{A better way to construct the sunflower head}",
      journal = {Mathematical Biosciences},
         year = 1979,
        month = jun,
       volume = {44},
       number = {3-4},
        pages = {179-189},
          doi = {10.1016/0025-5564(79)90080-4}
}

@ARTICLE{mesinger201121cmfast,
       author = {{Mesinger}, Andrei and {Furlanetto}, Steven and {Cen}, Renyue},
        title = "{21CMFAST: a fast, seminumerical simulation of the high-redshift 21-cm signal}",
      journal = {\mnras},
         year = 2011,
        month = feb,
       volume = {411},
       number = {2},
        pages = {955-972},
          doi = {10.1111/j.1365-2966.2010.17731.x},
archivePrefix = {arXiv},
       eprint = {1003.3878},
 primaryClass = {astro-ph.CO},
       adsurl = {https://ui.adsabs.harvard.edu/abs/2011MNRAS.411..955M}
}

@ARTICLE{2011MNRAS.414..727Z,
       author = {{Zahn}, Oliver and {Mesinger}, Andrei and {McQuinn}, Matthew and {Trac}, Hy and {Cen}, Renyue and {Hernquist}, Lars E.},
        title = "{Comparison of reionization models: radiative transfer simulations and approximate, seminumeric models}",
      journal = {\mnras},
         year = 2011,
        month = jun,
       volume = {414},
       number = {1},
        pages = {727-738},
          doi = {10.1111/j.1365-2966.2011.18439.x},
archivePrefix = {arXiv},
       eprint = {1003.3455},
 primaryClass = {astro-ph.CO},
       adsurl = {https://ui.adsabs.harvard.edu/abs/2011MNRAS.414..727Z}
}

@ARTICLE{2014MNRAS.444..606O,
       author = {{Offringa}, A.~R. and {McKinley}, B. and {Hurley-Walker}, N. and {Briggs}, F.~H. and {Wayth}, R.~B. and {Kaplan}, D.~L. and {Bell}, M.~E. and {Feng}, L. and {Neben}, A.~R. and {Hughes}, J.~D. and {Rhee}, J. and {Murphy}, T. and {Bhat}, N.~D.~R. and {Bernardi}, G. and {Bowman}, J.~D. and {Cappallo}, R.~J. and {Corey}, B.~E. and {Deshpande}, A.~A. and {Emrich}, D. and {Ewall-Wice}, A. and {Gaensler}, B.~M. and {Goeke}, R. and {Greenhill}, L.~J. and {Hazelton}, B.~J. and {Hindson}, L. and {Johnston-Hollitt}, M. and {Jacobs}, D.~C. and {Kasper}, J.~C. and {Kratzenberg}, E. and {Lenc}, E. and {Lonsdale}, C.~J. and {Lynch}, M.~J. and {McWhirter}, S.~R. and {Mitchell}, D.~A. and {Morales}, M.~F. and {Morgan}, E. and {Kudryavtseva}, N. and {Oberoi}, D. and {Ord}, S.~M. and {Pindor}, B. and {Procopio}, P. and {Prabu}, T. and {Riding}, J. and {Roshi}, D.~A. and {Shankar}, N. Udaya and {Srivani}, K.~S. and {Subrahmanyan}, R. and {Tingay}, S.~J. and {Waterson}, M. and {Webster}, R.~L. and {Whitney}, A.~R. and {Williams}, A. and {Williams}, C.~L.},
        title = "{WSCLEAN: an implementation of a fast, generic wide-field imager for radio astronomy}",
      journal = {\mnras},
         year = 2014,
        month = oct,
       volume = {444},
       number = {1},
        pages = {606-619},
          doi = {10.1093/mnras/stu1368},
archivePrefix = {arXiv},
       eprint = {1407.1943},
 primaryClass = {astro-ph.IM},
       adsurl = {https://ui.adsabs.harvard.edu/abs/2014MNRAS.444..606O}
}

@ARTICLE{park2019inferring,
       author = {{Park}, Jaehong and {Mesinger}, Andrei and {Greig}, Bradley and {Gillet}, Nicolas},
        title = "{Inferring the astrophysics of reionization and cosmic dawn from galaxy luminosity functions and the 21-cm signal}",
      journal = {\mnras},
         year = 2019,
        month = mar,
       volume = {484},
       number = {1},
        pages = {933-949},
          doi = {10.1093/mnras/stz032},
archivePrefix = {arXiv},
       eprint = {1809.08995},
 primaryClass = {astro-ph.GA},
       adsurl = {https://ui.adsabs.harvard.edu/abs/2019MNRAS.484..933P}
}

@ARTICLE{skao2026overview,
       author = {{CD Science Working Group} and {Bernardi}, Gianni and {Breitman}, Daniela and {Datta}, Abhirup and {Fialkov}, Anastasia and {Koopmans}, Leon V.~E. and {Liu}, Adrian and {Mao}, Yi and {Mellema}, Garrelt and {Mertens}, Florent and {Mesinger}, Andrei and {Pritchard}, Jonathan and {Schneider}, Aurel and {Trott}, Cathryn M. and {Xu}, Yidong and {Acharya}, Anshuman and {Bag}, Satadru and {Barkana}, Rennan and {Barry}, Nichole and {Basquette}, Oliver and {Bianco}, Michele and {Brackenhoff}, Stefanie A. and {Burigana}, Carlo and {Carucci}, Isabella P. and {Ceccotti}, Emilio and {Chakraborty}, Arnab and {Choudhuri}, Samir and {Choudhury}, Tirthankar Roy and {Ciardi}, Benedetta and {Cruz}, Hector Afonso G. and {Dasgupta}, Saswata and {Datta}, Kanan K. and {Dayal}, Pratika and {de Lera Acedo}, Eloy and {Elahi}, Khandakar Md Asif and {Georgiev}, Ivelin and {Singh Gill}, Sukhdeep and {Giri}, Sambit K. and {Gorce}, Adelie and {Guo}, Quan and {Heneka}, Caroline and {Hothi}, Ian and {Hutter}, Anne and {Kittiwisit}, Piyanat and {Li}, Bohua and {Mahida}, Yashrajsinh and {Maity}, Barun and {Majumdar}, Suman and {Meiksin}, Avery and {Meriot}, Romain and {Mishra}, Arnab and {Mittal}, Shikhar and {Mondal}, Rajesh and {Moscardini}, Lauro and {Munoz}, Julian and {Munshi}, Satyapan and {Natwariya}, Pravin Kumar and {Noble}, Leon and {O'Hara}, Oscar Sage David and {Ocvirk}, Pierre and {Pal}, Samit Kumar and {Pietschke}, Yannic and {Piyanat}, Steven G. and {Sagar}, Rashmi and {Samir}, Rasha M. and {Santos}, Mario G. and {Semelin}, Benoit and {Shah}, Rahul and {Shao}, Yali and {Shaw}, Abinash Kumar and {Shimabukuro}, Hayato and {Sims}, Peter H. and {Soltinsky}, Tomas and {Sun}, Guochao and {Tripathi}, Anshuman and {Trombetti}, Tiziana and {Ulusoy}, Ceren and {Umberto}, Maio and {Weller}, Jochen and {Yatawatta}, Sarod and {Yoshiura}, Shintaro and {Yue}, Bin and {Zaroubi}, Saleem and {Zheng}, Qian},
        title = "{Overview of 21cm Experiments at high redshift with SKAO}",
      journal = {arXiv e-prints},
         year = 2026,
        month = jun,
          eid = {arXiv:2606.26435},
        pages = {arXiv:2606.26435},
          doi = {10.48550/arXiv.2606.26435},
archivePrefix = {arXiv},
       eprint = {2606.26435},
 primaryClass = {astro-ph.CO},
       adsurl = {https://ui.adsabs.harvard.edu/abs/2026arXiv260626435C}
}

@ARTICLE{2012RPPh...75h6901P,
       author = {{Pritchard}, Jonathan R. and {Loeb}, Abraham},
        title = "{21 cm cosmology in the 21st century}",
      journal = {Reports on Progress in Physics},
         year = 2012,
        month = aug,
       volume = {75},
       number = {8},
          eid = {086901},
        pages = {086901},
          doi = {10.1088/0034-4885/75/8/086901},
archivePrefix = {arXiv},
       eprint = {1109.6012},
 primaryClass = {astro-ph.CO},
       adsurl = {https://ui.adsabs.harvard.edu/abs/2012RPPh...75h6901P}
}

@ARTICLE{2026ApJ...998...33A,
       author = {{Abdurashidova}, Zuhra and {Adams}, Tyrone and {Aguirre}, James E. and {Baartman}, Rushelle and {Barkana}, Rennan and {Berkhout}, Lindsay M. and {Bernardi}, Gianni and {Billings}, Tashalee S. and {Bizarria}, Bruno B. and {Bowman}, Judd D. and {Breitman}, Daniela and {Bull}, Philip and {Burba}, Jacob and {Byrne}, Ruby and {Carey}, Steven and {Chandra}, Rajorshi Sushovan and {Chen}, Kai-Feng and {Choudhuri}, Samir and {Cox}, Tyler and {DeBoer}, David R. and {de Lera Acedo}, Eloy and {Dexter}, Matt and {Dhandha}, Jiten and {Dillon}, Joshua S. and {Dynes}, Scott and {Eksteen}, Nico and {Ely}, John and {Ewall-Wice}, Aaron and {Fagnoni}, Nicolas and {Fialkov}, Anastasia and {Furlanetto}, Steven R. and {Gale-Sides}, Kingsley and {Garsden}, Hugh and {Gorce}, Adelie and {Gorthi}, Deepthi and {Halday}, Ziyaad and {Hazelton}, Bryna J. and {Hewitt}, Jacqueline N. and {Hickish}, Jack and {Huang}, Tian and {Jacobs}, Daniel C. and {Josaitis}, Alec and {Kern}, Nicholas S. and {Kerrigan}, Joshua and {Kittiwisit}, Piyanat and {Kolopanis}, Matthew and {Lanman}, Adam and {La Plante}, Paul and {Liu}, Adrian and {Ma}, Yin-Zhe and {MacMahon}, David H.~E. and {Malan}, Lourence and {Malgas}, Cresshim and {Malgas}, Keith and {Marero}, Bradley and {Martinot}, Zachary E. and {McBride}, Lisa and {Mesinger}, Andrei and {Mirocha}, Jordan and {Mohamed-Hinds}, Nicel and {Molewa}, Mathakane and {Morales}, Miguel F. and {Mu{\~n}oz}, Julian B. and {Murray}, Steven G. and {Nikolic}, Bojan and {Nuwegeld}, Hans and {Parsons}, Aaron R. and {Pascua}, Robert and {Patra}, Nipanjana and {Pochinda}, Simon and {Qin}, Yuxiang and {Rath}, Eleanor and {Razavi-Ghods}, Nima and {Riley}, Daniel and {Rosie}, Kathryn and {Santos}, Mario G. and {Singh}, Saurabh and {Storer}, Dara and {Swarts}, Hilton and {Tan}, Jianrong and {Th{\'e}lie}, Emilie and {van Wyngaarden}, Pieter and {Wilensky}, Michael J. and {Williams}, Peter K.~G. and {Zheng}, Haoxuan and {HERA Collaboration}},
        title = "{First Results from HERA Phase II}",
      journal = {\apj},
         year = 2026,
        month = feb,
       volume = {998},
       number = {1},
          eid = {33},
        pages = {33},
          doi = {10.3847/1538-4357/ae2d54},
archivePrefix = {arXiv},
       eprint = {2511.21289},
 primaryClass = {astro-ph.CO},
       adsurl = {https://ui.adsabs.harvard.edu/abs/2026ApJ...998...33A}
}

@ARTICLE{2025A&A...699A.109G,
       author = {{Ghara}, R. and {Zaroubi}, S. and {Ciardi}, B. and {Mellema}, G. and {Giri}, S.~K. and {Mertens}, F.~G. and {Mevius}, M. and {Koopmans}, L.~V.~E. and {Iliev}, I.~T. and {Acharya}, A. and {Brackenhoff}, S.~A. and {Ceccotti}, E. and {Chege}, K. and {Georgiev}, I. and {Ghosh}, S. and {Hothi}, I. and {H{\"o}fer}, C. and {Ma}, Q. and {Munshi}, S. and {Offringa}, A.~R. and {Shaw}, A.~K. and {Pandey}, V.~N. and {Yatawatta}, S. and {Choudhury}, M.},
        title = "{Constraints on the state of the intergalactic medium at z{\ensuremath{\sim}}8 {\ensuremath{-}} 10 using redshifted 21 cm observations with LOFAR}",
      journal = {\aap},
         year = 2025,
        month = jul,
       volume = {699},
          eid = {A109},
        pages = {A109},
          doi = {10.1051/0004-6361/202554163},
archivePrefix = {arXiv},
       eprint = {2505.00373},
 primaryClass = {astro-ph.CO},
       adsurl = {https://ui.adsabs.harvard.edu/abs/2025A&A...699A.109G}
}

@ARTICLE{skao2026inference,
       author = {{Acharya}, Anshuman and {Bag}, Satadru and {Berger}, Sabrina and {Bianco}, Michele and {Breitman}, Daniela and {Crascall-Kennedy}, Lilian and {Dasgupta}, Saswata and {Datta}, Abhirup and {Douspis}, Marian and {Georgiev}, Ivelin and {Giri}, Sambit K. and {Gorce}, Ad{\'e}lie and {Heneka}, Caroline and {Hiegel}, Julien and {Hothi}, Ian and {Kapahtia}, Akanksha and {Kern}, Nicholas and {Mahida}, Yashrajsinh and {Maity}, Barun and {Majumdar}, Suman and {M{\'e}riot}, Romain and {Mondal}, Rajesh and {Murray}, Steven G. and {Noble}, Leon and {Pal}, Samit Kumar and {Schneider}, Aurel and {Semelin}, Benoit and {Shaw}, Abinash Kumar and {Shimabukuro}, Hayato and {Th{\'e}lie}, Emilie and {Tripathi}, Anshuman and {Trott}, Cathryn M.},
        title = "{Inferring Cosmology and Astrophysics from the High-redshift 21cm Signal with SKA-Low}",
      journal = {arXiv e-prints},
         year = 2026,
        month = jun,
          eid = {arXiv:2606.27858},
        pages = {arXiv:2606.27858},
          doi = {10.48550/arXiv.2606.27858},
archivePrefix = {arXiv},
       eprint = {2606.27858},
 primaryClass = {astro-ph.CO},
       adsurl = {https://ui.adsabs.harvard.edu/abs/2026arXiv260627858A}
}

@ARTICLE{2021MNRAS.500.3821B,
       author = {{Bonaldi}, A. and {An}, T. and {Br{\"u}ggen}, M. and {Burkutean}, S. and {Coelho}, B. and {Goodarzi}, H. and {Hartley}, P. and {Sandhu}, P.~K. and {Wu}, C. and {Yu}, L. and {Zhoolideh Haghighi}, M.~H. and {Ant{\'o}n}, S. and {Bagheri}, Z. and {Barbosa}, D. and {Barraca}, J.~P. and {Bartashevich}, D. and {Bergano}, M. and {Bonato}, M. and {Brand}, J. and {de Gasperin}, F. and {Giannetti}, A. and {Dodson}, R. and {Jain}, P. and {Jaiswal}, S. and {Lao}, B. and {Liu}, B. and {Liuzzo}, E. and {Lu}, Y. and {Lukic}, V. and {Maia}, D. and {Marchili}, N. and {Massardi}, M. and {Mohan}, P. and {Morgado}, J.~B. and {Panwar}, M. and {Prabhakar}, P. and {Ribeiro}, V.~A.~R.~M. and {Rygl}, K.~L.~J. and {Sabz Ali}, V. and {Saremi}, E. and {Schisano}, E. and {Sheikhnezami}, S. and {Vafaei Sadr}, A. and {Wong}, A. and {Wong}, O.~I.},
        title = "{Square Kilometre Array Science Data Challenge 1: analysis and results}",
      journal = {\mnras},
         year = 2021,
        month = jan,
       volume = {500},
       number = {3},
        pages = {3821-3837},
          doi = {10.1093/mnras/staa3023},
archivePrefix = {arXiv},
       eprint = {2009.13346},
 primaryClass = {astro-ph.IM},
       adsurl = {https://ui.adsabs.harvard.edu/abs/2021MNRAS.500.3821B}
}

@ARTICLE{2023MNRAS.523.1967H,
       author = {{Hartley}, P. and {Bonaldi}, A. and {Braun}, R. and {Aditya}, J.~N.~H.~S. and {Aicardi}, S. and {Alegre}, L. and {Chakraborty}, A. and {Chen}, X. and {Choudhuri}, S. and {Clarke}, A.~O. and {Coles}, J. and {Collinson}, J.~S. and {Cornu}, D. and {Darriba}, L. and {Veneri}, M. Delli and {Forbrich}, J. and {Fraga}, B. and {Galan}, A. and {Garrido}, J. and {Gubanov}, F. and {H{\r{a}}kansson}, H. and {Hardcastle}, M.~J. and {Heneka}, C. and {Herranz}, D. and {Hess}, K.~M. and {Jagannath}, M. and {Jaiswal}, S. and {Jurek}, R.~J. and {Korber}, D. and {Kitaeff}, S. and {Kleiner}, D. and {Lao}, B. and {Lu}, X. and {Mazumder}, A. and {Mold{\'o}n}, J. and {Mondal}, R. and {Ni}, S. and {{\"O}nnheim}, M. and {Parra}, M. and {Patra}, N. and {Peel}, A. and {Salom{\'e}}, P. and {S{\'a}nchez-Exp{\'o}sito}, S. and {Sargent}, M. and {Semelin}, B. and {Serra}, P. and {Shaw}, A.~K. and {Shen}, A.~X. and {Sj{\"o}berg}, A. and {Smith}, L. and {Soroka}, A. and {Stolyarov}, V. and {Tolley}, E. and {Toribio}, M.~C. and {van der Hulst}, J.~M. and {Sadr}, A. Vafaei and {Verdes-Montenegro}, L. and {Westmeier}, T. and {Yu}, K. and {Yu}, L. and {Zhang}, L. and {Zhang}, X. and {Zhang}, Y. and {Alberdi}, A. and {Ashdown}, M. and {Bom}, C.~R. and {Br{\"u}ggen}, M. and {Cannon}, J. and {Chen}, R. and {Combes}, F. and {Conway}, J. and {Courbin}, F. and {Ding}, J. and {Fourestey}, G. and {Freundlich}, J. and {Gao}, L. and {Gheller}, C. and {Guo}, Q. and {Gustavsson}, E. and {Jirstrand}, M. and {Jones}, M.~G. and {J{\'o}zsa}, G. and {Kamphuis}, P. and {Kneib}, J.-P. and {Lindqvist}, M. and {Liu}, B. and {Liu}, Y. and {Mao}, Y. and {Marchal}, A. and {M{\'a}rquez}, I. and {Meshcheryakov}, A. and {Olberg}, M. and {Oozeer}, N. and {Pandey-Pommier}, M. and {Pei}, W. and {Peng}, B. and {Sabater}, J. and {Sorgho}, A. and {Starck}, J.~L. and {Tasse}, C. and {Wang}, A. and {Wang}, Y. and {Xi}, H. and {Yang}, X. and {Zhang}, H. and {Zhang}, J. and {Zhao}, M. and {Zuo}, S.},
        title = "{SKA Science Data Challenge 2: analysis and results}",
      journal = {\mnras},
         year = 2023,
        month = aug,
       volume = {523},
       number = {2},
        pages = {1967-1993},
          doi = {10.1093/mnras/stad1375},
archivePrefix = {arXiv},
       eprint = {2303.07943},
 primaryClass = {astro-ph.IM},
       adsurl = {https://ui.adsabs.harvard.edu/abs/2023MNRAS.523.1967H}
}

@ARTICLE{sdc3a,
       author = {{Bonaldi}, A. and {Hartley}, P. and {Braun}, R. and {Purser}, S. and {Acharya}, A. and {Ahn}, K. and {Aparicio Resco}, M. and {Bait}, O. and {Bianco}, M. and {Chakraborty}, A. and {Chapman}, E. and {Chatterjee}, S. and {Chege}, K. and {Chen}, H. and {Chen}, X. and {Chen}, Z. and {Conaboy}, L. and {Cruz}, M. and {Darriba}, L. and {De Santis}, M. and {Denzel}, P. and {Diao}, K. and {Feron}, J. and {Finlay}, C. and {Gehlot}, B. and {Ghosh}, S. and {Giri}, S.~K. and {Grumitt}, R. and {Hong}, S.~E. and {Ito}, T. and {Jiang}, M. and {Jordan}, C. and {Kim}, S. and {Kim}, M. and {Kim}, J. and {Krishna}, S.~P. and {Kulkarni}, A. and {L{\'o}pez-Caniego}, M. and {Labadie-Garc{\'\i}a}, I. and {Lee}, H. and {Lee}, D. and {Lee}, N. and {Line}, J. and {Liu}, Y. and {Mao}, Y. and {Mazumder}, A. and {Mertens}, F.~G. and {Munshi}, S. and {Nasirudin}, A. and {Ni}, S. and {Nistane}, V. and {Norregaard}, C. and {Null}, D. and {Offringa}, A. and {Oh}, M. and {Oh}, S.-H. and {Parkinson}, D. and {Pritchard}, J. and {Ruiz-Granda}, M. and {Salvador L{\'o}pez}, V. and {Shan}, H. and {Sharma}, R. and {Trott}, C. and {Yoshiura}, S. and {Zhang}, L. and {Zhang}, X. and {Zheng}, Q. and {Zhu}, Z. and {Zuo}, S. and {Akahori}, T. and {Alberto}, P. and {Allys}, E. and {An}, T. and {Anstey}, D. and {Baek}, J. and {Basavraj} and {Brackenhoff}, S. and {Browne}, P. and {Ceccotti}, E. and {Chen}, T. and {Choudhuri}, S. and {Choudhury}, M. and {Coles}, J. and {Cook}, J. and {Cornu}, D. and {Cunnington}, S. and {Das}, S. and {de Lera Acedo Acedo}, E. and {Delouis}, J.-M. and {Deng}, F. and {Ding}, J. and {Elahi}, K.~M.~A. and {Fernandez}, P. and {Fern{\'a}ndez}, C. and {Fern{\'a}ndez Alc{\'a}zar}, A. and {Galluzzi}, V. and {Gao}, L.-Y. and {Garain}, U. and {Garrido}, J. and {Gendron-Marsolais}, M.-L. and {Gessey-Jones}, T. and {Ghorbel}, H. and {Gong}, Y. and {Guo}, S. and {Hasegawa}, K. and {Hayashi}, T. and {Herranz}, D. and {Holanda}, V. and {Holloway}, A.~J. and {Hothi}, I. and {H{\"o}fer}, C. and {Jeli{\'c}}, V. and {Jiang}, Y. and {Jiang}, X. and {Kang}, H. and {Kim}, J.-Y. and {Koopmans}, L.~V. and {Lacroix}, R. and {Lee}, E. and {Leeney}, S. and {Levrier}, F. and {Li}, Y. and {Ma}, Q. and {Meriot}, R. and {Mesinger}, A. and {Mevius}, M. and {Minoda}, T. and {Miville-Desch{\^e}nes}, M.-A. and {Moldon}, J. and {Mondal}, R. and {Murmu}, C. and {Murray}, S. and {Nirmala}, Sr. and {Niu}, Q. and {Nunhokee}, C. and {O'Hara}, O. and {Pal}, S.~K. and {Pal}, S. and {Park}, J. and {Parra}, M. and {Patra}, N.~N. and {Pindor}, B. and {Remazeilles}, M. and {Rey}, P. and {Rubino-Martin}, J.~A. and {Saha}, S. and {Selvaraj}, A. and {Semelin}, B. and {Shah}, R. and {Shao}, Y. and {Shaw}, A.~K. and {Shi}, F. and {Shimabukuro}, H. and {Singh}, G. and {Sohn}, B.~W. and {Stagni}, M. and {Starck}, J.-L. and {Sui}, C. and {Swinbank}, J.~D. and {S{\'a}nchez}, J. and {S{\'a}nchez-Exp{\'o}sito}, S. and {Takahashi}, K. and {Takeuchi}, T. and {Tripathi}, A. and {Verdes-Montenegro}, L. and {Vielva}, P. and {Vitello}, F.~R. and {Wang}, G.-J. and {Wang}, Q. and {Wang}, X. and {Wang}, Y. and {Wang}, Y.-X. and {Wiegert}, T. and {Wild}, A. and {Williams}, W.~L. and {Wolz}, L. and {Wu}, X. and {Wu}, P. and {Xia}, J.-Q. and {Xu}, Y. and {Yan}, R. and {Yan}, Y.-P. and {Yin}, Z. and {You}, Z. and {Yu}, X. and {Yu}, K. and {Yue}, B. and {Zhao}, Z. and {Zhao}, X. and {Zhou}, X.},
        title = "{Square Kilometre Array Science Data Challenge 3a: foreground removal for an EoR experiment}",
      journal = {\mnras},
         year = 2025,
        month = oct,
       volume = {543},
       number = {2},
        pages = {1092-1119},
          doi = {10.1093/mnras/staf1466},
archivePrefix = {arXiv},
       eprint = {2503.11740},
 primaryClass = {astro-ph.IM},
       adsurl = {https://ui.adsabs.harvard.edu/abs/2025MNRAS.543.1092B}
}

@TECHREPORT{ska_low_conf,
       author = {{Dewdney}, P.~E. and {Braun}, R.},
        title = "{SKA-TEL-SKO-0000422}",
  institution = {SKA Organisation},
         year = 2016
}

@ARTICLE{2020A&A...641A...6P,
       author = {{Planck Collaboration} and {Aghanim}, N. and {Akrami}, Y. and {Ashdown}, M. and {Aumont}, J. and {Baccigalupi}, C. and {Ballardini}, M. and {Banday}, A.~J. and {Barreiro}, R.~B. and {Bartolo}, N. and {Basak}, S. and {Battye}, R. and {Benabed}, K. and {Bernard}, J. -P. and {Bersanelli}, M. and {Bielewicz}, P. and {Bock}, J.~J. and {Bond}, J.~R. and {Borrill}, J. and {Bouchet}, F.~R. and {Boulanger}, F. and {Bucher}, M. and {Burigana}, C. and {Butler}, R.~C. and {Calabrese}, E. and {Cardoso}, J. -F. and {Carron}, J. and {Challinor}, A. and {Chiang}, H.~C. and {Chluba}, J. and {Colombo}, L.~P.~L. and {Combet}, C. and {Contreras}, D. and {Crill}, B.~P. and {Cuttaia}, F. and {de Bernardis}, P. and {de Zotti}, G. and {Delabrouille}, J. and {Delouis}, J. -M. and {Di Valentino}, E. and {Diego}, J.~M. and {Dor{\'e}}, O. and {Douspis}, M. and {Ducout}, A. and {Dupac}, X. and {Dusini}, S. and {Efstathiou}, G. and {Elsner}, F. and {En{\ss}lin}, T.~A. and {Eriksen}, H.~K. and {Fantaye}, Y. and {Farhang}, M. and {Fergusson}, J. and {Fernandez-Cobos}, R. and {Finelli}, F. and {Forastieri}, F. and {Frailis}, M. and {Fraisse}, A.~A. and {Franceschi}, E. and {Frolov}, A. and {Galeotta}, S. and {Galli}, S. and {Ganga}, K. and {G{\'e}nova-Santos}, R.~T. and {Gerbino}, M. and {Ghosh}, T. and {Gonz{\'a}lez-Nuevo}, J. and {G{\'o}rski}, K.~M. and {Gratton}, S. and {Gruppuso}, A. and {Gudmundsson}, J.~E. and {Hamann}, J. and {Handley}, W. and {Hansen}, F.~K. and {Herranz}, D. and {Hildebrandt}, S.~R. and {Hivon}, E. and {Huang}, Z. and {Jaffe}, A.~H. and {Jones}, W.~C. and {Karakci}, A. and {Keih{\"a}nen}, E. and {Keskitalo}, R. and {Kiiveri}, K. and {Kim}, J. and {Kisner}, T.~S. and {Knox}, L. and {Krachmalnicoff}, N. and {Kunz}, M. and {Kurki-Suonio}, H. and {Lagache}, G. and {Lamarre}, J. -M. and {Lasenby}, A. and {Lattanzi}, M. and {Lawrence}, C.~R. and {Le Jeune}, M. and {Lemos}, P. and {Lesgourgues}, J. and {Levrier}, F. and {Lewis}, A. and {Liguori}, M. and {Lilje}, P.~B. and {Lilley}, M. and {Lindholm}, V. and {L{\'o}pez-Caniego}, M. and {Lubin}, P.~M. and {Ma}, Y. -Z. and {Mac{\'\i}as-P{\'e}rez}, J.~F. and {Maggio}, G. and {Maino}, D. and {Mandolesi}, N. and {Mangilli}, A. and {Marcos-Caballero}, A. and {Maris}, M. and {Martin}, P.~G. and {Martinelli}, M. and {Mart{\'\i}nez-Gonz{\'a}lez}, E. and {Matarrese}, S. and {Mauri}, N. and {McEwen}, J.~D. and {Meinhold}, P.~R. and {Melchiorri}, A. and {Mennella}, A. and {Migliaccio}, M. and {Millea}, M. and {Mitra}, S. and {Miville-Desch{\^e}nes}, M. -A. and {Molinari}, D. and {Montier}, L. and {Morgante}, G. and {Moss}, A. and {Natoli}, P. and {N{\o}rgaard-Nielsen}, H.~U. and {Pagano}, L. and {Paoletti}, D. and {Partridge}, B. and {Patanchon}, G. and {Peiris}, H.~V. and {Perrotta}, F. and {Pettorino}, V. and {Piacentini}, F. and {Polastri}, L. and {Polenta}, G. and {Puget}, J. -L. and {Rachen}, J.~P. and {Reinecke}, M. and {Remazeilles}, M. and {Renzi}, A. and {Rocha}, G. and {Rosset}, C. and {Roudier}, G. and {Rubi{\~n}o-Mart{\'\i}n}, J.~A. and {Ruiz-Granados}, B. and {Salvati}, L. and {Sandri}, M. and {Savelainen}, M. and {Scott}, D. and {Shellard}, E.~P.~S. and {Sirignano}, C. and {Sirri}, G. and {Spencer}, L.~D. and {Sunyaev}, R. and {Suur-Uski}, A. -S. and {Tauber}, J.~A. and {Tavagnacco}, D. and {Tenti}, M. and {Toffolatti}, L. and {Tomasi}, M. and {Trombetti}, T. and {Valenziano}, L. and {Valiviita}, J. and {Van Tent}, B. and {Vibert}, L. and {Vielva}, P. and {Villa}, F. and {Vittorio}, N. and {Wandelt}, B.~D. and {Wehus}, I.~K. and {White}, M. and {White}, S.~D.~M. and {Zacchei}, A. and {Zonca}, A.},
        title = "{Planck 2018 results. VI. Cosmological parameters}",
      journal = {\aap},
         year = 2020,
        month = sep,
       volume = {641},
          eid = {A6},
        pages = {A6},
          doi = {10.1051/0004-6361/201833910},
archivePrefix = {arXiv},
       eprint = {1807.06209},
 primaryClass = {astro-ph.CO},
       adsurl = {https://ui.adsabs.harvard.edu/abs/2020A&A...641A...6P}
}

@ARTICLE{iliev2006cosmological,
       author = {{Iliev}, Ilian T. and {Ciardi}, Benedetta and {Alvarez}, Marcelo A. and {Maselli}, Antonella and {Ferrara}, Andrea and {Gnedin}, Nickolay Y. and {Mellema}, Garrelt and {Nakamoto}, Taishi and {Norman}, Michael L. and {Razoumov}, Alexei O. and {Rijkhorst}, Erik-Jan and {Ritzerveld}, Jelle and {Shapiro}, Paul R. and {Susa}, Hajime and {Umemura}, Masayuki and {Whalen}, Daniel J.},
        title = "{Cosmological radiative transfer codes comparison project - I. The static density field tests}",
      journal = {\mnras},
         year = 2006,
        month = sep,
       volume = {371},
       number = {3},
        pages = {1057-1086},
          doi = {10.1111/j.1365-2966.2006.10775.x},
archivePrefix = {arXiv},
       eprint = {astro-ph/0603199},
 primaryClass = {astro-ph},
       adsurl = {https://ui.adsabs.harvard.edu/abs/2006MNRAS.371.1057I}
}

@ARTICLE{mellema2006simulating,
       author = {{Mellema}, Garrelt and {Iliev}, Ilian T. and {Pen}, Ue-Li and {Shapiro}, Paul R.},
        title = "{Simulating cosmic reionization at large scales - II. The 21-cm emission features and statistical signals}",
      journal = {\mnras},
         year = 2006,
        month = oct,
       volume = {372},
       number = {2},
        pages = {679-692},
          doi = {10.1111/j.1365-2966.2006.10919.x},
archivePrefix = {arXiv},
       eprint = {astro-ph/0603518},
 primaryClass = {astro-ph},
       adsurl = {https://ui.adsabs.harvard.edu/abs/2006MNRAS.372..679M}
}

@ARTICLE{giri202421,
       author = {{Giri}, Sambit K. and {Bianco}, Michele and {Schaeffer}, Timoth{\'e}e and {Iliev}, Ilian T. and {Mellema}, Garrelt and {Schneider}, Aurel},
        title = "{The 21-cm signal during the end stages of reionization}",
      journal = {\mnras},
         year = 2024,
        month = sep,
       volume = {533},
       number = {2},
        pages = {2364-2378},
          doi = {10.1093/mnras/stae1999},
archivePrefix = {arXiv},
       eprint = {2403.04838},
 primaryClass = {astro-ph.CO},
       adsurl = {https://ui.adsabs.harvard.edu/abs/2024MNRAS.533.2364G}
}

@ARTICLE{Mellema:2005ht,
       author = {{Mellema}, Garrelt and {Iliev}, Ilian T. and {Alvarez}, Marcelo A. and {Shapiro}, Paul R.},
        title = "{C$^{2}$-ray: A new method for photon-conserving transport of ionizing radiation}",
      journal = {\na},
         year = 2006,
        month = mar,
       volume = {11},
       number = {5},
        pages = {374-395},
          doi = {10.1016/j.newast.2005.09.004},
archivePrefix = {arXiv},
       eprint = {astro-ph/0508416},
 primaryClass = {astro-ph},
       adsurl = {https://ui.adsabs.harvard.edu/abs/2006NewA...11..374M}
}

@ARTICLE{Hirling:2023cjx,
       author = {{Hirling}, Patrick and {Bianco}, Michele and {Giri}, Sambit K. and {Iliev}, Ilian T. and {Mellema}, Garrelt and {Kneib}, Jean-Paul},
        title = "{pyC$^{2}$Ray: A flexible and GPU-accelerated radiative transfer framework for simulating the cosmic epoch of reionization}",
      journal = {Astronomy and Computing},
         year = 2024,
        month = jul,
       volume = {48},
          eid = {100861},
        pages = {100861},
          doi = {10.1016/j.ascom.2024.100861},
archivePrefix = {arXiv},
       eprint = {2311.01492},
 primaryClass = {astro-ph.CO},
       adsurl = {https://ui.adsabs.harvard.edu/abs/2024A&C....4800861H}
}

@ARTICLE{2004ApJ...613....1F,
       author = {{Furlanetto}, Steven R. and {Zaldarriaga}, Matias and {Hernquist}, Lars},
        title = "{The Growth of H II Regions During Reionization}",
      journal = {\apj},
         year = 2004,
        month = sep,
       volume = {613},
       number = {1},
        pages = {1-15},
          doi = {10.1086/423025},
archivePrefix = {arXiv},
       eprint = {astro-ph/0403697},
 primaryClass = {astro-ph},
       adsurl = {https://ui.adsabs.harvard.edu/abs/2004ApJ...613....1F}
}

@ARTICLE{2014MNRAS.440.1662S,
       author = {{Sobacchi}, Emanuele and {Mesinger}, Andrei},
        title = "{Inhomogeneous recombinations during cosmic reionization}",
      journal = {\mnras},
         year = 2014,
        month = may,
       volume = {440},
       number = {2},
        pages = {1662-1673},
          doi = {10.1093/mnras/stu377},
archivePrefix = {arXiv},
       eprint = {1402.2298},
 primaryClass = {astro-ph.CO},
       adsurl = {https://ui.adsabs.harvard.edu/abs/2014MNRAS.440.1662S}
}

@ARTICLE{2022MNRAS.514.1302D,
       author = {{Davies}, Frederick B. and {Furlanetto}, Steven R.},
        title = "{Improved treatments of the ionizing photon mean free path in seminumerical simulations of reionization}",
      journal = {\mnras},
         year = 2022,
        month = jul,
       volume = {514},
       number = {1},
        pages = {1302-1314},
          doi = {10.1093/mnras/stac1005},
archivePrefix = {arXiv},
       eprint = {2103.09821},
 primaryClass = {astro-ph.CO},
       adsurl = {https://ui.adsabs.harvard.edu/abs/2022MNRAS.514.1302D}
}

@ARTICLE{2021MNRAS.501....1G,
       author = {{Greig}, Bradley and {Mesinger}, Andrei and {Koopmans}, L{\'e}on V.~E. and {Ciardi}, Benedetta and {Mellema}, Garrelt and {Zaroubi}, Saleem and {Giri}, Sambit K. and {Ghara}, Raghunath and {Ghosh}, Abhik and {Iliev}, Ilian T. and {Mertens}, Florent G. and {Mondal}, Rajesh and {Offringa}, Andr{\'e} R. and {Pandey}, Vishambhar N.},
        title = "{Interpreting LOFAR 21-cm signal upper limits at z {\ensuremath{\approx}} 9.1 in the context of high-z galaxy and reionization observations}",
      journal = {\mnras},
         year = 2021,
        month = jan,
       volume = {501},
       number = {1},
        pages = {1-13},
          doi = {10.1093/mnras/staa3593},
archivePrefix = {arXiv},
       eprint = {2006.03203},
 primaryClass = {astro-ph.CO},
       adsurl = {https://ui.adsabs.harvard.edu/abs/2021MNRAS.501....1G}
}

@ARTICLE{2015MNRAS.449.4246G,
       author = {{Greig}, Bradley and {Mesinger}, Andrei},
        title = "{21CMMC: an MCMC analysis tool enabling astrophysical parameter studies of the cosmic 21 cm signal}",
      journal = {\mnras},
         year = 2015,
        month = jun,
       volume = {449},
       number = {4},
        pages = {4246-4263},
          doi = {10.1093/mnras/stv571},
archivePrefix = {arXiv},
       eprint = {1501.06576},
 primaryClass = {astro-ph.CO},
       adsurl = {https://ui.adsabs.harvard.edu/abs/2015MNRAS.449.4246G}
}

@ARTICLE{10.1111/j.1365-2966.2004.08604.x,
       author = {{Bharadwaj}, Somnath and {Ali}, Sk. Saiyad},
        title = "{On using visibility correlations to probe the HI distribution from the dark ages to the present epoch - I. Formalism and the expected signal}",
      journal = {\mnras},
         year = 2005,
        month = feb,
       volume = {356},
       number = {4},
        pages = {1519-1528},
          doi = {10.1111/j.1365-2966.2004.08604.x},
archivePrefix = {arXiv},
       eprint = {astro-ph/0406676},
 primaryClass = {astro-ph},
       adsurl = {https://ui.adsabs.harvard.edu/abs/2005MNRAS.356.1519B}
}

@ARTICLE{kingma2017,
       author = {{Kingma}, Diederik P. and {Ba}, Jimmy},
        title = "{Adam: A Method for Stochastic Optimization}",
      journal = {arXiv e-prints},
         year = 2014,
        month = dec,
          eid = {arXiv:1412.6980},
        pages = {arXiv:1412.6980},
          doi = {10.48550/arXiv.1412.6980},
archivePrefix = {arXiv},
       eprint = {1412.6980},
 primaryClass = {cs.LG},
       adsurl = {https://ui.adsabs.harvard.edu/abs/2014arXiv1412.6980K}
}

@INPROCEEDINGS{NEURIPS2019_9015,
       author = {{Paszke}, Adam and {Gross}, Sam and {Massa}, Francisco and {Lerer}, Adam and {Bradbury}, James and {Chanan}, Gregory and {Killeen}, Trevor and {Lin}, Zeming and {Gimelshein}, Natalia and {Antiga}, Luca and {Desmaison}, Alban and {Kopf}, Andreas and {Yang}, Edward and {DeVito}, Zach and {Raison}, Martin and {Tejani}, Alykhan and {Chilamkurthy}, Sasank and {Steiner}, Benoit and {Fang}, Lu and {Bai}, Junjie and {Chintala}, Soumith},
        title = "{PyTorch: An Imperative Style, High-Performance Deep Learning Library}",
    booktitle = {Advances in Neural Information Processing Systems},
         year = 2019,
       volume = {32},
        pages = {8024-8035},
archivePrefix = {arXiv},
       eprint = {1912.01703},
 primaryClass = {cs.LG},
       adsurl = {https://ui.adsabs.harvard.edu/abs/2019arXiv191201703P}
}

@ARTICLE{2013PASP..125..306F,
       author = {{Foreman-Mackey}, Daniel and {Hogg}, David W. and {Lang}, Dustin and {Goodman}, Jonathan},
        title = "{emcee: The MCMC Hammer}",
      journal = {\pasp},
         year = 2013,
        month = mar,
       volume = {125},
       number = {925},
        pages = {306},
          doi = {10.1086/670067},
archivePrefix = {arXiv},
       eprint = {1202.3665},
 primaryClass = {astro-ph.IM},
       adsurl = {https://ui.adsabs.harvard.edu/abs/2013PASP..125..306F}
}

@ARTICLE{Zhou_2022,
       author = {{Zhou}, Yihao and {La Plante}, Paul},
        title = "{Understanding the Impact of Semi-numeric Reionization Models when Using CNNs}",
      journal = {\pasp},
         year = 2022,
        month = apr,
       volume = {134},
       number = {1034},
          eid = {044001},
        pages = {044001},
          doi = {10.1088/1538-3873/ac5f5d},
archivePrefix = {arXiv},
       eprint = {2112.03443},
 primaryClass = {astro-ph.CO},
       adsurl = {https://ui.adsabs.harvard.edu/abs/2022PASP..134d4001Z}
}

@ARTICLE{Solt_2026,
       author = {{Solt}, Jasper and {Pober}, Jonathan C. and {Bach}, Stephen H.},
        title = "{Mitigating Simulator Dependence in AI Parameter Inference for the Epoch of Reionization: The Importance of Simulation Diversity}",
      journal = {arXiv e-prints},
         year = 2026,
        month = jan,
          eid = {arXiv:2601.05229},
        pages = {arXiv:2601.05229},
          doi = {10.48550/arXiv.2601.05229},
archivePrefix = {arXiv},
       eprint = {2601.05229},
 primaryClass = {astro-ph.CO},
       adsurl = {https://ui.adsabs.harvard.edu/abs/2026arXiv260105229S}
}

@ARTICLE{2012Natur.487...70V,
       author = {{Visbal}, Eli and {Barkana}, Rennan and {Fialkov}, Anastasia and {Tseliakhovich}, Dmitriy and {Hirata}, Christopher M.},
        title = "{The signature of the first stars in atomic hydrogen at redshift 20}",
      journal = {\nat},
         year = 2012,
        month = jul,
       volume = {487},
       number = {7405},
        pages = {70-73},
          doi = {10.1038/nature11177},
archivePrefix = {arXiv},
       eprint = {1201.1005},
 primaryClass = {astro-ph.CO},
       adsurl = {https://ui.adsabs.harvard.edu/abs/2012Natur.487...70V}
}

@ARTICLE{2012MNRAS.424.1335F,
       author = {{Fialkov}, Anastasia and {Barkana}, Rennan and {Tseliakhovich}, Dmitriy and {Hirata}, Christopher M.},
        title = "{Impact of the relative motion between the dark matter and baryons on the first stars: semi-analytical modelling}",
      journal = {\mnras},
         year = 2012,
        month = aug,
       volume = {424},
       number = {2},
        pages = {1335-1345},
          doi = {10.1111/j.1365-2966.2012.21318.x},
archivePrefix = {arXiv},
       eprint = {1110.2111},
 primaryClass = {astro-ph.CO},
       adsurl = {https://ui.adsabs.harvard.edu/abs/2012MNRAS.424.1335F}
}

@ARTICLE{2013MNRAS.432.2909F,
       author = {{Fialkov}, Anastasia and {Barkana}, Rennan and {Visbal}, Eli and {Tseliakhovich}, Dmitriy and {Hirata}, Christopher M.},
        title = "{The 21-cm signature of the first stars during the Lyman-Werner feedback era}",
      journal = {\mnras},
         year = 2013,
        month = jul,
       volume = {432},
       number = {4},
        pages = {2909-2916},
          doi = {10.1093/mnras/stt650},
archivePrefix = {arXiv},
       eprint = {1212.0513},
 primaryClass = {astro-ph.CO},
       adsurl = {https://ui.adsabs.harvard.edu/abs/2013MNRAS.432.2909F}
}

@ARTICLE{2014MNRAS.437L..36F,
       author = {{Fialkov}, Anastasia and {Barkana}, Rennan and {Pinhas}, Arazi and {Visbal}, Eli},
        title = "{Complete history of the observable 21 cm signal from the first stars during the pre-reionization era}",
      journal = {\mnras},
         year = 2014,
        month = jan,
       volume = {437},
       number = {1},
        pages = {L36-L40},
          doi = {10.1093/mnrasl/slt135},
archivePrefix = {arXiv},
       eprint = {1306.2354},
 primaryClass = {astro-ph.CO},
       adsurl = {https://ui.adsabs.harvard.edu/abs/2014MNRAS.437L..36F}
}

@ARTICLE{2004ApJ...609..474B,
       author = {{Barkana}, Rennan and {Loeb}, Abraham},
        title = "{Unusually Large Fluctuations in the Statistics of Galaxy Formation at High Redshift}",
      journal = {\apj},
         year = 2004,
        month = jul,
       volume = {609},
       number = {2},
        pages = {474-481},
          doi = {10.1086/421079},
archivePrefix = {arXiv},
       eprint = {astro-ph/0310338},
 primaryClass = {astro-ph},
       adsurl = {https://ui.adsabs.harvard.edu/abs/2004ApJ...609..474B}
}

@ARTICLE{2011MNRAS.418..906T,
       author = {{Tseliakhovich}, Dmitriy and {Barkana}, Rennan and {Hirata}, Christopher M.},
        title = "{Suppression and spatial variation of early galaxies and minihaloes}",
      journal = {\mnras},
         year = 2011,
        month = dec,
       volume = {418},
       number = {2},
        pages = {906-915},
          doi = {10.1111/j.1365-2966.2011.19541.x},
archivePrefix = {arXiv},
       eprint = {1012.2574},
 primaryClass = {astro-ph.CO},
       adsurl = {https://ui.adsabs.harvard.edu/abs/2011MNRAS.418..906T}
}

@ARTICLE{2026MNRAS.549ag878L,
       author = {{Liu}, Yuchen and {de Lera Acedo}, Eloy and {Sims}, Peter H.},
        title = "{Bayesian model comparison and validation with Gaussian Process Regression for interferometric 21-cm signal recovery}",
      journal = {\mnras},
         year = 2026,
        month = jun,
       volume = {549},
       number = {1},
          eid = {stag878},
        pages = {stag878},
          doi = {10.1093/mnras/stag878},
archivePrefix = {arXiv},
       eprint = {2511.10499},
 primaryClass = {astro-ph.CO},
       adsurl = {https://ui.adsabs.harvard.edu/abs/2026MNRAS.549ag878L}
}

@ARTICLE{2024MNRAS.533.2876O,
       author = {{O'Hara}, Oscar S.~D. and {Dulwich}, Fred and {de Lera Acedo}, Eloy and {Dhandha}, Jiten and {Gessey-Jones}, Thomas and {Anstey}, Dominic and {Fialkov}, Anastasia},
        title = "{Understanding spectral artefacts in SKA-Low 21-cm cosmology experiments: the impact of cable reflections}",
      journal = {\mnras},
         year = 2024,
        month = sep,
       volume = {533},
       number = {3},
        pages = {2876-2892},
          doi = {10.1093/mnras/stae1952},
archivePrefix = {arXiv},
       eprint = {2402.04008},
 primaryClass = {astro-ph.CO},
       adsurl = {https://ui.adsabs.harvard.edu/abs/2024MNRAS.533.2876O}
}

@ARTICLE{pochinda_inference_2024,
       author = {{Pochinda}, Simon and {Gessey-Jones}, Thomas and {Bevins}, Harry T.~J. and {Fialkov}, Anastasia and {Heimersheim}, Stefan and {Abril-Cabezas}, Irene and {de Lera Acedo}, Eloy and {Singh}, Saurabh and {Sikder}, Sudipta and {Barkana}, Rennan},
        title = "{Constraining the properties of Population III galaxies with multiwavelength observations}",
      journal = {\mnras},
         year = 2024,
        month = jun,
       volume = {531},
       number = {1},
        pages = {1113-1132},
          doi = {10.1093/mnras/stae1185},
archivePrefix = {arXiv},
       eprint = {2312.08095},
 primaryClass = {astro-ph.CO},
       adsurl = {https://ui.adsabs.harvard.edu/abs/2024MNRAS.531.1113P}
}

@ARTICLE{2015MNRAS.450L..61H,
       author = {{Handley}, W.~J. and {Hobson}, M.~P. and {Lasenby}, A.~N.},
        title = "{polychord: nested sampling for cosmology.}",
      journal = {\mnras},
         year = 2015,
        month = jun,
       volume = {450},
        pages = {L61-L65},
          doi = {10.1093/mnrasl/slv047},
archivePrefix = {arXiv},
       eprint = {1502.01856},
 primaryClass = {astro-ph.CO},
       adsurl = {https://ui.adsabs.harvard.edu/abs/2015MNRAS.450L..61H}
}

@ARTICLE{2015MNRAS.453.4384H,
       author = {{Handley}, W.~J. and {Hobson}, M.~P. and {Lasenby}, A.~N.},
        title = "{POLYCHORD: next-generation nested sampling}",
      journal = {\mnras},
         year = 2015,
        month = nov,
       volume = {453},
       number = {4},
        pages = {4384-4398},
          doi = {10.1093/mnras/stv1911},
archivePrefix = {arXiv},
       eprint = {1506.00171},
 primaryClass = {astro-ph.IM},
       adsurl = {https://ui.adsabs.harvard.edu/abs/2015MNRAS.453.4384H}
}

@ARTICLE{10.1214/06-BA127,
       author = {{Skilling}, John},
        title = "{Nested sampling for general Bayesian computation}",
      journal = {Bayesian Analysis},
         year = 2006,
        month = dec,
       volume = {1},
       number = {4},
        pages = {833-859},
          doi = {10.1214/06-BA127},
       adsurl = {https://ui.adsabs.harvard.edu/abs/2006BayAn...1..833S}
}

@ARTICLE{2022NRvMP...2...39A,
       author = {{Ashton}, Greg and {Bernstein}, Noam and {Buchner}, Johannes and {Chen}, Xi and {Cs{\'a}nyi}, G{\'a}bor and {Fowlie}, Andrew and {Feroz}, Farhan and {Griffiths}, Matthew and {Handley}, Will and {Habeck}, Michael and {Higson}, Edward and {Hobson}, Michael and {Lasenby}, Anthony and {Parkinson}, David and {P{\'a}rtay}, Livia B. and {Pitkin}, Matthew and {Schneider}, Doris and {Speagle}, Joshua S. and {South}, Leah and {Veitch}, John and {Wacker}, Philipp and {Wales}, David J. and {Yallup}, David},
        title = "{Nested sampling for physical scientists}",
      journal = {Nature Reviews Methods Primers},
         year = 2022,
        month = may,
       volume = {2},
          eid = {39},
        pages = {39},
          doi = {10.1038/s43586-022-00121-x},
archivePrefix = {arXiv},
       eprint = {2205.15570},
 primaryClass = {stat.CO},
       adsurl = {https://ui.adsabs.harvard.edu/abs/2022NRvMP...2...39A}
}

@ARTICLE{2013ApJS..208...19H,
       author = {{Hinshaw}, G. and {Larson}, D. and {Komatsu}, E. and {Spergel}, D.~N. and {Bennett}, C.~L. and {Dunkley}, J. and {Nolta}, M.~R. and {Halpern}, M. and {Hill}, R.~S. and {Odegard}, N. and {Page}, L. and {Smith}, K.~M. and {Weiland}, J.~L. and {Gold}, B. and {Jarosik}, N. and {Kogut}, A. and {Limon}, M. and {Meyer}, S.~S. and {Tucker}, G.~S. and {Wollack}, E. and {Wright}, E.~L.},
        title = "{Nine-year Wilkinson Microwave Anisotropy Probe (WMAP) Observations: Cosmological Parameter Results}",
      journal = {\apjs},
         year = 2013,
        month = oct,
       volume = {208},
       number = {2},
          eid = {19},
        pages = {19},
          doi = {10.1088/0067-0049/208/2/19},
archivePrefix = {arXiv},
       eprint = {1212.5226},
 primaryClass = {astro-ph.CO},
       adsurl = {https://ui.adsabs.harvard.edu/abs/2013ApJS..208...19H}
}

@ARTICLE{2020MNRAS.493.4728G,
       author = {{Ghara}, R. and {Giri}, S.~K. and {Mellema}, G. and {Ciardi}, B. and
         {Zaroubi}, S. and {Iliev}, I.~T. and {Koopmans}, L.~V.~E. and
         {Chapman}, E. and {Gazagnes}, S. and {Gehlot}, B.~K. and {Ghosh}, A. and
         {Jeli{\'c}}, V. and {Mertens}, F.~G. and {Mondal}, R. and {Schaye}, J. and
         {Silva}, M.~B. and {Asad}, K.~M.~B. and {Kooistra}, R. and
         {Mevius}, M. and {Offringa}, A.~R. and {Pandey}, V.~N. and
         {Yatawatta}, S.},
        title = "{Constraining the intergalactic medium at z {\ensuremath{\approx}} 9.1 using LOFAR Epoch of Reionization observations}",
      journal = {\mnras},
         year = 2020,
        month = apr,
       volume = {493},
       number = {4},
        pages = {4728-4747},
          doi = {10.1093/mnras/staa487},
archivePrefix = {arXiv},
       eprint = {2002.07195},
 primaryClass = {astro-ph.CO},
       adsurl = {https://ui.adsabs.harvard.edu/abs/2020MNRAS.493.4728G}
}

@ARTICLE{Harnois12,
       author = {{Harnois-D{\'e}raps}, J. and {Pen}, U.-L. and {Iliev}, I.~T. and {Merz}, H. and {Emberson}, J.~D. and {Desjacques}, V.},
        title = "{High-performance P$^{3}$M N-body code: CUBEP$^{3}$M}",
      journal = {\mnras},
         year = 2013,
        month = nov,
       volume = {436},
       number = {1},
        pages = {540-559},
          doi = {10.1093/mnras/stt1591},
archivePrefix = {arXiv},
       eprint = {1208.5098},
 primaryClass = {astro-ph.CO},
       adsurl = {https://ui.adsabs.harvard.edu/abs/2013MNRAS.436..540H}
}

@ARTICLE{Watson2013TheAges,
       author = {{Watson}, William A. and {Iliev}, Ilian T. and {D'Aloisio}, Anson and {Knebe}, Alexander and {Shapiro}, Paul R. and {Yepes}, Gustavo},
        title = "{The halo mass function through the cosmic ages}",
      journal = {\mnras},
         year = 2013,
        month = aug,
       volume = {433},
       number = {2},
        pages = {1230-1245},
          doi = {10.1093/mnras/stt791},
archivePrefix = {arXiv},
       eprint = {1212.0095},
 primaryClass = {astro-ph.CO},
       adsurl = {https://ui.adsabs.harvard.edu/abs/2013MNRAS.433.1230W}
}

@ARTICLE{giri2019neutral,
       author = {{Giri}, Sambit K. and {Mellema}, Garrelt and {Aldheimer}, Thomas and
         {Dixon}, Keri L. and {Iliev}, Ilian T.},
        title = "{Neutral island statistics during reionization from 21-cm tomography}",
      journal = {\mnras},
         year = 2019,
        month = oct,
       volume = {489},
       number = {2},
        pages = {1590-1605},
          doi = {10.1093/mnras/stz2224},
archivePrefix = {arXiv},
       eprint = {1903.01294},
 primaryClass = {astro-ph.GA},
       adsurl = {https://ui.adsabs.harvard.edu/abs/2019MNRAS.489.1590G}
}

@ARTICLE{2019JCAP...02..058G,
       author = {{Giri}, Sambit K. and {D'Aloisio}, Anson and {Mellema}, Garrelt and
         {Komatsu}, Eiichiro and {Ghara}, Raghunath and {Majumdar}, Suman},
        title = "{Position-dependent power spectra of the 21-cm signal from the epoch of reionization}",
      journal = {\jcap},
         year = 2019,
        month = feb,
       volume = {2019},
       number = {2},
          eid = {058},
        pages = {058},
          doi = {10.1088/1475-7516/2019/02/058},
archivePrefix = {arXiv},
       eprint = {1811.09633},
 primaryClass = {astro-ph.CO},
       adsurl = {https://ui.adsabs.harvard.edu/abs/2019JCAP...02..058G}
}

@ARTICLE{2017PhRvD..95b3513H,
       author = {{Heinrich}, Chen He and {Miranda}, Vinicius and {Hu}, Wayne},
        title = "{Complete reionization constraints from Planck 2015 polarization}",
      journal = {\prd},
         year = 2017,
        month = jan,
       volume = {95},
       number = {2},
          eid = {023513},
        pages = {023513},
          doi = {10.1103/PhysRevD.95.023513},
archivePrefix = {arXiv},
       eprint = {1609.04788},
 primaryClass = {astro-ph.CO},
       adsurl = {https://ui.adsabs.harvard.edu/abs/2017PhRvD..95b3513H}
}

@ARTICLE{2002PhRvD..66j3511L,
       author = {{Lewis}, Antony and {Bridle}, Sarah},
        title = "{Cosmological parameters from CMB and other data: A Monte Carlo approach}",
      journal = {\prd},
         year = 2002,
        month = nov,
       volume = {66},
       number = {10},
          eid = {103511},
        pages = {103511},
          doi = {10.1103/PhysRevD.66.103511},
archivePrefix = {arXiv},
       eprint = {astro-ph/0205436},
 primaryClass = {astro-ph},
       adsurl = {https://ui.adsabs.harvard.edu/abs/2002PhRvD..66j3511L}
}

@ARTICLE{Pietschke2025,
       author = {{Pietschke}, Yannic and {Heneka}, Caroline and {Schlenker}, Tom and {Ore}, Ayodele and {Schosser}, Benedikt},
        title = "{Direct reconstruction of the Reionization history from 21cm 2D Power Spectra}",
      journal = {\jcap},
         year = 2025,
        month = oct,
       volume = {2025},
       number = {10},
          eid = {039},
        pages = {039},
          doi = {10.1088/1475-7516/2025/10/039},
       adsurl = {https://ui.adsabs.harvard.edu/abs/2025JCAP...10..039P}
}

@ARTICLE{Hutter2021,
       author = {{Hutter}, Anne and {Dayal}, Pratika and {Yepes}, Gustavo and {Gottl{\"o}ber}, Stefan and {Legrand}, Laurent and {Ucci}, Graziano},
        title = "{Astraeus I: the interplay between galaxy formation and reionization}",
      journal = {\mnras},
         year = 2021,
        month = may,
       volume = {503},
       number = {3},
        pages = {3698-3723},
          doi = {10.1093/mnras/stab602},
archivePrefix = {arXiv},
       eprint = {2004.08401},
 primaryClass = {astro-ph.GA},
       adsurl = {https://ui.adsabs.harvard.edu/abs/2021MNRAS.503.3698H}
}

@ARTICLE{Hutter2025,
       author = {{Hutter}, Anne and {Cueto}, Elie R. and {Dayal}, Pratika and {Gottl{\"o}ber}, Stefan and {Trebitsch}, Maxime and {Yepes}, Gustavo},
        title = "{ASTRAEUS: X. Indications of a top-heavy initial mass function in highly star-forming galaxies from JWST observations at z > 10}",
      journal = {\aap},
         year = 2025,
        month = feb,
       volume = {694},
          eid = {A254},
        pages = {A254},
          doi = {10.1051/0004-6361/202452460},
archivePrefix = {arXiv},
       eprint = {2410.00730},
 primaryClass = {astro-ph.GA},
       adsurl = {https://ui.adsabs.harvard.edu/abs/2025A&A...694A.254H}
}

@INPROCEEDINGS{durkan2019neural,
       author = {{Durkan}, Conor and {Bekasov}, Artur and {Murray}, Iain and {Papamakarios}, George},
        title = "{Neural Spline Flows}",
    booktitle = {Advances in Neural Information Processing Systems},
         year = 2019,
       volume = {32},
archivePrefix = {arXiv},
       eprint = {1906.04032},
 primaryClass = {stat.ML},
       adsurl = {https://ui.adsabs.harvard.edu/abs/2019arXiv190604032D}
}

@ARTICLE{2025arXiv250816329C,
       author = {{Cooper}, Nadia and {Norregaard}, Carina and {Meriot}, Romain and {Pritchard}, Jonathan R.},
        title = "{Simulation based inference of the ionization history from the 2D 21 cm power spectrum}",
      journal = {arXiv e-prints},
         year = 2025,
        month = aug,
          eid = {arXiv:2508.16329},
        pages = {arXiv:2508.16329},
          doi = {10.48550/arXiv.2508.16329},
archivePrefix = {arXiv},
       eprint = {2508.16329},
 primaryClass = {astro-ph.CO},
       adsurl = {https://ui.adsabs.harvard.edu/abs/2025arXiv250816329C}
}

@ARTICLE{murray202021cmfast,
       author = {{Murray}, Steven G. and {Greig}, Bradley and {Mesinger}, Andrei and {Mu{\~n}oz}, Julian B. and {Qin}, Yuxiang and {Park}, Jaehong and {Watkinson}, Catherine A.},
        title = "{21cmFAST v3: A Python-integrated C code for generating 3D realizations of the cosmic 21cm signal.}",
      journal = {The Journal of Open Source Software},
         year = 2020,
        month = oct,
       volume = {5},
       number = {54},
          eid = {2582},
        pages = {2582},
          doi = {10.21105/joss.02582},
archivePrefix = {arXiv},
       eprint = {2010.15121},
 primaryClass = {astro-ph.IM},
       adsurl = {https://ui.adsabs.harvard.edu/abs/2020JOSS....5.2582M}
}

@ARTICLE{qin2020tale,
       author = {{Qin}, Yuxiang and {Mesinger}, Andrei and {Park}, Jaehong and {Greig}, Bradley and {Mu{\~n}oz}, Julian B.},
        title = "{A tale of two sites - I. Inferring the properties of minihalo-hosted galaxies from current observations}",
      journal = {\mnras},
         year = 2020,
        month = jun,
       volume = {495},
       number = {1},
        pages = {123-140},
          doi = {10.1093/mnras/staa1131},
archivePrefix = {arXiv},
       eprint = {2003.04442},
 primaryClass = {astro-ph.CO},
       adsurl = {https://ui.adsabs.harvard.edu/abs/2020MNRAS.495..123Q}
}

@ARTICLE{munoz2022impact,
       author = {{Mu{\~n}oz}, Julian B. and {Qin}, Yuxiang and {Mesinger}, Andrei and {Murray}, Steven G. and {Greig}, Bradley and {Mason}, Charlotte},
        title = "{The impact of the first galaxies on cosmic dawn and reionization}",
      journal = {\mnras},
         year = 2022,
        month = apr,
       volume = {511},
       number = {3},
        pages = {3657-3681},
          doi = {10.1093/mnras/stac185},
archivePrefix = {arXiv},
       eprint = {2110.13919},
 primaryClass = {astro-ph.CO},
       adsurl = {https://ui.adsabs.harvard.edu/abs/2022MNRAS.511.3657M}
}

@ARTICLE{mcquinn2006cosmological,
       author = {{McQuinn}, Matthew and {Zahn}, Oliver and {Zaldarriaga}, Matias and {Hernquist}, Lars and {Furlanetto}, Steven R.},
        title = "{Cosmological Parameter Estimation Using 21 cm Radiation from the Epoch of Reionization}",
      journal = {\apj},
         year = 2006,
        month = dec,
       volume = {653},
       number = {2},
        pages = {815-834},
          doi = {10.1086/505167},
archivePrefix = {arXiv},
       eprint = {astro-ph/0512263},
 primaryClass = {astro-ph},
       adsurl = {https://ui.adsabs.harvard.edu/abs/2006ApJ...653..815M}
}

@ARTICLE{ltu-ili,
       author = {{Ho}, Matthew and {Bartlett}, Deaglan J. and {Chartier}, Nicolas and {Cuesta-Lazaro}, Carolina and {Ding}, Simon and {Lapel}, Axel and {Lemos}, Pablo and {Lovell}, Christopher C. and {Makinen}, T. Lucas and {Modi}, Chirag and {Pandya}, Viraj and {Pandey}, Shivam and {Perez}, Lucia A. and {Wandelt}, Benjamin and {Bryan}, Greg L.},
        title = "{LtU-ILI: An All-in-One Framework for Implicit Inference in Astrophysics and Cosmology}",
      journal = {The Open Journal of Astrophysics},
         year = 2024,
        month = jul,
       volume = {7},
          eid = {54},
        pages = {54},
          doi = {10.33232/001c.120559},
archivePrefix = {arXiv},
       eprint = {2402.05137},
 primaryClass = {astro-ph.IM},
       adsurl = {https://ui.adsabs.harvard.edu/abs/2024OJAp....7E..54H}
}

@ARTICLE{tejero-cantero2020sbi,
       author = {{Tejero-Cantero}, Alvaro and {Boelts}, Jan and {Deistler}, Michael and {Lueckmann}, Jan-Matthis and {Durkan}, Conor and {Gon{\c{c}}alves}, Pedro J. and {Greenberg}, David S. and {Macke}, Jakob H.},
        title = "{sbi: A toolkit for simulation-based inference}",
      journal = {The Journal of Open Source Software},
         year = 2020,
        month = aug,
       volume = {5},
       number = {52},
          eid = {2505},
        pages = {2505},
          doi = {10.21105/joss.02505},
archivePrefix = {arXiv},
       eprint = {2007.09114},
 primaryClass = {cs.LG},
       adsurl = {https://ui.adsabs.harvard.edu/abs/2020JOSS....5.2505T}
}

@ARTICLE{BoeltsDeistler_sbi_2025,
       author = {{Boelts}, Jan and {Deistler}, Michael and {Gloeckler}, Manuel and {Tejero-Cantero}, {\'A}lvaro and {Lueckmann}, Jan-Matthis and {Moss}, Guy and {Steinbach}, Peter and {Moreau}, Thomas and {Muratore}, Fabio and {Linhart}, Julia and {Durkan}, Conor and {Vetter}, Julius and {Miller}, Benjamin Kurt and {Herold}, Maternus and {Ziaeemehr}, Abolfazl and {Pals}, Matthijs and {Gruner}, Theo and {Bischoff}, Sebastian and {Krouglova}, Nastya and {Gao}, Richard and {Lappalainen}, Janne K. and {Mucs{\'a}nyi}, B{\'a}lint and {Pei}, Felix and {Schulz}, Auguste and {Stefanidi}, Zinovia and {Rodrigues}, Pedro and {Schr{\"o}der}, Cornelius and {Abu Zaid}, Faried and {Beck}, Jonas and {Kapoor}, Jaivardhan and {Greenberg}, David S. and {Gon{\c{c}}alves}, Pedro J. and {Macke}, Jakob H.},
        title = "{sbi reloaded: a toolkit for simulation-based inference workflows}",
      journal = {The Journal of Open Source Software},
         year = 2025,
        month = apr,
       volume = {10},
       number = {108},
          eid = {7754},
        pages = {7754},
          doi = {10.21105/joss.07754},
archivePrefix = {arXiv},
       eprint = {2411.17337},
 primaryClass = {cs.LG},
       adsurl = {https://ui.adsabs.harvard.edu/abs/2025JOSS...10.7754B}
}

@ARTICLE{papamakarios2019neural,
       author = {{Papamakarios}, George},
        title = "{Neural Density Estimation and Likelihood-free Inference}",
      journal = {arXiv e-prints},
         year = 2019,
        month = oct,
          eid = {arXiv:1910.13233},
        pages = {arXiv:1910.13233},
          doi = {10.48550/arXiv.1910.13233},
archivePrefix = {arXiv},
       eprint = {1910.13233},
 primaryClass = {stat.ML},
       adsurl = {https://ui.adsabs.harvard.edu/abs/2019arXiv191013233P}
}

@INPROCEEDINGS{lueckmann2017flexible,
       author = {{Lueckmann}, Jan-Matthis and {Gon{\c{c}}alves}, Pedro J. and {Bassetto}, Giacomo and {{\"O}cal}, Kaan and {Nonnenmacher}, Marcel and {Macke}, Jakob H.},
        title = "{Flexible statistical inference for mechanistic models of neural dynamics}",
    booktitle = {Advances in Neural Information Processing Systems},
         year = 2017,
       volume = {30},
archivePrefix = {arXiv},
       eprint = {1711.01861},
 primaryClass = {stat.ML},
       adsurl = {https://ui.adsabs.harvard.edu/abs/2017arXiv171101861L}
}

@ARTICLE{cranmer2020frontier,
       author = {{Cranmer}, Kyle and {Brehmer}, Johann and {Louppe}, Gilles},
        title = "{The frontier of simulation-based inference}",
      journal = {Proceedings of the National Academy of Science},
         year = 2020,
        month = dec,
       volume = {117},
       number = {48},
        pages = {30055-30062},
          doi = {10.1073/pnas.1912789117},
archivePrefix = {arXiv},
       eprint = {1911.01429},
 primaryClass = {stat.ML},
       adsurl = {https://ui.adsabs.harvard.edu/abs/2020PNAS..11730055C}
}

@ARTICLE{talts2018validating,
       author = {{Talts}, Sean and {Betancourt}, Michael and {Simpson}, Daniel and {Vehtari}, Aki and {Gelman}, Andrew},
        title = "{Validating Bayesian Inference Algorithms with Simulation-Based Calibration}",
      journal = {arXiv e-prints},
         year = 2018,
        month = apr,
          eid = {arXiv:1804.06788},
        pages = {arXiv:1804.06788},
          doi = {10.48550/arXiv.1804.06788},
archivePrefix = {arXiv},
       eprint = {1804.06788},
 primaryClass = {stat.ME},
       adsurl = {https://ui.adsabs.harvard.edu/abs/2018arXiv180406788T}
}

@ARTICLE{2025A&A...698A..80M,
       author = {{Meriot}, R. and {Semelin}, B. and {Cornu}, D.},
        title = "{Comparison of Bayesian inference methods using the LORELI II database of hydro-radiative simulations of the 21-cm signal}",
      journal = {\aap},
         year = 2025,
        month = jun,
       volume = {698},
          eid = {A80},
        pages = {A80},
          doi = {10.1051/0004-6361/202452901},
archivePrefix = {arXiv},
       eprint = {2411.03093},
 primaryClass = {astro-ph.CO},
       adsurl = {https://ui.adsabs.harvard.edu/abs/2025A&A...698A..80M}
}

@ARTICLE{2024A&A...683A..24M,
       author = {{Meriot}, R. and {Semelin}, B.},
        title = "{The LORELI database: 21 cm signal inference with 3D radiative hydrodynamics simulations}",
      journal = {\aap},
         year = 2024,
        month = mar,
       volume = {683},
          eid = {A24},
        pages = {A24},
          doi = {10.1051/0004-6361/202347591},
archivePrefix = {arXiv},
       eprint = {2310.02684},
 primaryClass = {astro-ph.CO},
       adsurl = {https://ui.adsabs.harvard.edu/abs/2024A&A...683A..24M}
}

@ARTICLE{Mondal2017,
       author = {{Mondal}, Rajesh and {Bharadwaj}, Somnath and {Majumdar}, Suman},
        title = "{Statistics of the epoch of reionization (EoR) 21-cm signal - II. The evolution of the power-spectrum error-covariance}",
      journal = {\mnras},
         year = 2017,
        month = jan,
       volume = {464},
       number = {3},
        pages = {2992-3004},
          doi = {10.1093/mnras/stw2599},
archivePrefix = {arXiv},
       eprint = {1606.03874},
 primaryClass = {astro-ph.CO},
       adsurl = {https://ui.adsabs.harvard.edu/abs/2017MNRAS.464.2992M}
}

@ARTICLE{Mondal2015,
       author = {{Mondal}, R. and {Bharadwaj}, S. and {Majumdar}, S. and {Bera}, A. and {Acharyya}, A.},
        title = "{The effect of non-Gaussianity on error predictions for the Epoch of Reionization (EoR) 21-cm power spectrum.}",
      journal = {\mnras},
         year = 2015,
        month = apr,
       volume = {449},
        pages = {L41-L45},
          doi = {10.1093/mnrasl/slv015},
archivePrefix = {arXiv},
       eprint = {1409.4420},
 primaryClass = {astro-ph.CO},
       adsurl = {https://ui.adsabs.harvard.edu/abs/2015MNRAS.449L..41M}
}

@INPROCEEDINGS{liu2008isolation,
       author = {{Liu}, Fei Tony and {Ting}, Kai Ming and {Zhou}, Zhi-Hua},
        title = "{Isolation Forest}",
    booktitle = {2008 Eighth IEEE International Conference on Data Mining},
         year = 2008,
        month = dec,
        pages = {413-422},
          doi = {10.1109/ICDM.2008.17}
}

@ARTICLE{choudhury2018,
       author = {{Choudhury}, Tirthankar Roy and {Paranjape}, Aseem},
        title = "{Photon number conservation and the large-scale 21 cm power spectrum in seminumerical models of reionization}",
      journal = {\mnras},
         year = 2018,
        month = dec,
       volume = {481},
       number = {3},
        pages = {3821-3837},
          doi = {10.1093/mnras/sty2551},
archivePrefix = {arXiv},
       eprint = {1807.00836},
 primaryClass = {astro-ph.CO},
       adsurl = {https://ui.adsabs.harvard.edu/abs/2018MNRAS.481.3821C}
}

@ARTICLE{maity2023,
       author = {{Maity}, Barun and {Choudhury}, Tirthankar Roy},
        title = "{Efficient exploration of reionization parameters for the upcoming 21 cm observations using a photon-conserving seminumerical model SCRIPT}",
      journal = {\mnras},
         year = 2023,
        month = may,
       volume = {521},
       number = {3},
        pages = {4140-4155},
          doi = {10.1093/mnras/stad791},
archivePrefix = {arXiv},
       eprint = {2211.12909},
 primaryClass = {astro-ph.CO},
       adsurl = {https://ui.adsabs.harvard.edu/abs/2023MNRAS.521.4140M}
}

@ARTICLE{choudhury2024,
       author = {{Choudhury}, T. Roy and {Paranjape}, A. and {Maity}, B.},
        title = "{A GPR-based emulator for semi-numerical reionization code SCRIPT: parameter inference from 21 cm data}",
      journal = {\jcap},
         year = 2024,
        month = mar,
       volume = {2024},
       number = {3},
          eid = {027},
        pages = {027},
          doi = {10.1088/1475-7516/2024/03/027},
archivePrefix = {arXiv},
       eprint = {2311.04556},
 primaryClass = {astro-ph.IM},
       adsurl = {https://ui.adsabs.harvard.edu/abs/2024JCAP...03..027C}
}

@ARTICLE{music_2011,
       author = {{Hahn}, Oliver and {Abel}, Tom},
        title = "{Multi-scale initial conditions for cosmological simulations}",
      journal = {\mnras},
         year = 2011,
        month = aug,
       volume = {415},
       number = {3},
        pages = {2101-2121},
          doi = {10.1111/j.1365-2966.2011.18820.x},
archivePrefix = {arXiv},
       eprint = {1103.6031},
 primaryClass = {astro-ph.CO},
       adsurl = {https://ui.adsabs.harvard.edu/abs/2011MNRAS.415.2101H}
}

@ARTICLE{cobaya2021,
       author = {{Torrado}, Jes{\'u}s and {Lewis}, Antony},
        title = "{Cobaya: code for Bayesian analysis of hierarchical physical models}",
      journal = {\jcap},
         year = 2021,
        month = may,
       volume = {2021},
       number = {5},
          eid = {057},
        pages = {057},
          doi = {10.1088/1475-7516/2021/05/057},
archivePrefix = {arXiv},
       eprint = {2005.05290},
 primaryClass = {astro-ph.IM},
       adsurl = {https://ui.adsabs.harvard.edu/abs/2021JCAP...05..057T}
}

@ARTICLE{Gelman_Rubin1992,
       author = {{Gelman}, Andrew and {Rubin}, Donald B.},
        title = "{Inference from Iterative Simulation Using Multiple Sequences}",
      journal = {Statistical Science},
         year = 1992,
        month = jan,
       volume = {7},
       number = {4},
        pages = {457-472},
          doi = {10.1214/ss/1177011136},
       adsurl = {https://ui.adsabs.harvard.edu/abs/1992StaSc...7..457G}
}

@ARTICLE{bianco2025deep,
       author = {{Bianco}, Michele and {Giri}, Sambit K. and {Sharma}, Rohit and {Chen}, Tianyue and {Krishna}, Shreyam Parth and {Finlay}, Chris and {Nistane}, Viraj and {Denzel}, Philipp and {De Santis}, Massimo and {Ghorbel}, Hatem},
        title = "{Deep learning approach for identification of H II regions during reionization in 21-cm observations - III. Image recovery}",
      journal = {\mnras},
         year = 2025,
        month = jul,
       volume = {541},
       number = {1},
        pages = {234-250},
          doi = {10.1093/mnras/staf973},
archivePrefix = {arXiv},
       eprint = {2408.16814},
 primaryClass = {astro-ph.CO},
       adsurl = {https://ui.adsabs.harvard.edu/abs/2025MNRAS.541..234B}
}

@ARTICLE{Bianco2021,
       author = {{Bianco}, Michele and {Giri}, Sambit K. and {Iliev}, Ilian T. and {Mellema}, Garrelt},
        title = "{Deep learning approach for identification of H II regions during reionization in 21-cm observations}",
      journal = {\mnras},
         year = 2021,
        month = aug,
       volume = {505},
       number = {3},
        pages = {3982-3997},
          doi = {10.1093/mnras/stab1518},
archivePrefix = {arXiv},
       eprint = {2102.06713},
 primaryClass = {astro-ph.IM},
       adsurl = {https://ui.adsabs.harvard.edu/abs/2021MNRAS.505.3982B}
}

@ARTICLE{wang2021regularizing,
       author = {{Wang}, Yulin and {Huang}, Gao and {Song}, Shiji and {Pan}, Xuran and {Xia}, Yitong and {Wu}, Cheng},
        title = "{Regularizing Deep Networks with Semantic Data Augmentation}",
      journal = {IEEE Transactions on Pattern Analysis and Machine Intelligence},
         year = 2022,
        month = jul,
       volume = {44},
       number = {7},
        pages = {3733-3748},
          doi = {10.1109/TPAMI.2021.3052951},
archivePrefix = {arXiv},
       eprint = {2007.10538},
 primaryClass = {cs.CV},
       adsurl = {https://ui.adsabs.harvard.edu/abs/2020arXiv200710538W}
}

@ARTICLE{Feroz:2008xx,
       author = {{Feroz}, F. and {Hobson}, M.~P. and {Bridges}, M.},
        title = "{MULTINEST: an efficient and robust Bayesian inference tool for cosmology and particle physics}",
      journal = {\mnras},
         year = 2009,
        month = oct,
       volume = {398},
       number = {4},
        pages = {1601-1614},
          doi = {10.1111/j.1365-2966.2009.14548.x},
archivePrefix = {arXiv},
       eprint = {0809.3437},
 primaryClass = {astro-ph},
       adsurl = {https://ui.adsabs.harvard.edu/abs/2009MNRAS.398.1601F}
}

@ARTICLE{Hermans+2019,
       author = {{Hermans}, Joeri and {Begy}, Volodimir and {Louppe}, Gilles},
        title = "{Likelihood-free MCMC with Amortized Approximate Ratio Estimators}",
      journal = {arXiv e-prints},
         year = 2019,
        month = mar,
          eid = {arXiv:1903.04057},
        pages = {arXiv:1903.04057},
          doi = {10.48550/arXiv.1903.04057},
archivePrefix = {arXiv},
       eprint = {1903.04057},
 primaryClass = {stat.ML},
       adsurl = {https://ui.adsabs.harvard.edu/abs/2019arXiv190304057H}
}

@ARTICLE{Cranmer+2015,
       author = {{Cranmer}, Kyle and {Pavez}, Juan and {Louppe}, Gilles},
        title = "{Approximating Likelihood Ratios with Calibrated Discriminative Classifiers}",
      journal = {arXiv e-prints},
         year = 2015,
        month = jun,
          eid = {arXiv:1506.02169},
        pages = {arXiv:1506.02169},
          doi = {10.48550/arXiv.1506.02169},
archivePrefix = {arXiv},
       eprint = {1506.02169},
 primaryClass = {stat.AP},
       adsurl = {https://ui.adsabs.harvard.edu/abs/2015arXiv150602169C}
}

@ARTICLE{Durkan+2020,
       author = {{Durkan}, Conor and {Murray}, Iain and {Papamakarios}, George},
        title = "{On Contrastive Learning for Likelihood-free Inference}",
      journal = {arXiv e-prints},
         year = 2020,
        month = feb,
          eid = {arXiv:2002.03712},
        pages = {arXiv:2002.03712},
          doi = {10.48550/arXiv.2002.03712},
archivePrefix = {arXiv},
       eprint = {2002.03712},
 primaryClass = {stat.ML},
       adsurl = {https://ui.adsabs.harvard.edu/abs/2020arXiv200203712D}
}

@ARTICLE{Miller+2022,
       author = {{Miller}, Benjamin Kurt and {Weniger}, Christoph and {Forr{\'e}}, Patrick},
        title = "{Contrastive Neural Ratio Estimation for Simulation-based Inference}",
      journal = {arXiv e-prints},
         year = 2022,
        month = oct,
          eid = {arXiv:2210.06170},
        pages = {arXiv:2210.06170},
          doi = {10.48550/arXiv.2210.06170},
archivePrefix = {arXiv},
       eprint = {2210.06170},
 primaryClass = {stat.ML},
       adsurl = {https://ui.adsabs.harvard.edu/abs/2022arXiv221006170M}
}

@ARTICLE{Thomas+2016,
       author = {{Thomas}, Owen and {Dutta}, Ritabrata and {Corander}, Jukka and {Kaski}, Samuel and {Gutmann}, Michael U.},
        title = "{Likelihood-free inference by ratio estimation}",
      journal = {arXiv e-prints},
         year = 2016,
        month = nov,
          eid = {arXiv:1611.10242},
        pages = {arXiv:1611.10242},
          doi = {10.48550/arXiv.1611.10242},
archivePrefix = {arXiv},
       eprint = {1611.10242},
 primaryClass = {stat.ML},
       adsurl = {https://ui.adsabs.harvard.edu/abs/2016arXiv161110242T}
}

@BOOK{silverman+1986,
       author = {{Silverman}, B.~W.},
        title = "{Density Estimation for Statistics and Data Analysis}",
         year = 1986,
    publisher = {Chapman and Hall},
       adsurl = {https://ui.adsabs.harvard.edu/abs/1986desd.book.....S}
}

@ARTICLE{Buchner+2016,
       author = {{Buchner}, Johannes},
        title = "{A statistical test for Nested Sampling algorithms}",
      journal = {Statistics and Computing},
         year = 2016,
        month = jan,
       volume = {26},
       number = {1-2},
        pages = {383-392},
          doi = {10.1007/s11222-014-9512-y},
archivePrefix = {arXiv},
       eprint = {1407.5459},
 primaryClass = {stat.CO},
       adsurl = {https://ui.adsabs.harvard.edu/abs/2016S&C....26..383B}
}

@ARTICLE{Buchner+2019,
       author = {{Buchner}, Johannes},
        title = "{Collaborative Nested Sampling: Big Data versus Complex Physical Models}",
      journal = {\pasp},
         year = 2019,
        month = oct,
       volume = {131},
       number = {1004},
        pages = {108005},
          doi = {10.1088/1538-3873/aae7fc},
archivePrefix = {arXiv},
       eprint = {1707.04476},
 primaryClass = {stat.CO},
       adsurl = {https://ui.adsabs.harvard.edu/abs/2019PASP..131j8005B}
}

@ARTICLE{songaila2010evolution,
       author = {{Songaila}, Antoinette and {Cowie}, Lennox L.},
        title = "{The Evolution of Lyman Limit Absorption Systems to Redshift Six}",
      journal = {\apj},
         year = 2010,
        month = oct,
       volume = {721},
       number = {2},
        pages = {1448-1466},
          doi = {10.1088/0004-637X/721/2/1448},
archivePrefix = {arXiv},
       eprint = {1007.3262},
 primaryClass = {astro-ph.CO},
       adsurl = {https://ui.adsabs.harvard.edu/abs/2010ApJ...721.1448S}
}

@ARTICLE{Buchner+2021,
       author = {{Buchner}, Johannes},
        title = "{UltraNest - a robust, general purpose Bayesian inference engine}",
      journal = {The Journal of Open Source Software},
         year = 2021,
        month = apr,
       volume = {6},
       number = {60},
          eid = {3001},
        pages = {3001},
          doi = {10.21105/joss.03001},
archivePrefix = {arXiv},
       eprint = {2101.09604},
 primaryClass = {stat.CO},
       adsurl = {https://ui.adsabs.harvard.edu/abs/2021JOSS....6.3001B}
}

@ARTICLE{2025arXiv251113370Z,
       author = {{Zhang}, Jiajun and {Shan}, Huanyuan},
        title = "{Constraining the Neutral Hydrogen Fraction from SKA Simulated Observation using a Double-Gaussian Decomposition Technique}",
      journal = {arXiv e-prints},
         year = 2025,
        month = nov,
          eid = {arXiv:2511.13370},
        pages = {arXiv:2511.13370},
          doi = {10.48550/arXiv.2511.13370},
archivePrefix = {arXiv},
       eprint = {2511.13370},
 primaryClass = {astro-ph.CO},
       adsurl = {https://ui.adsabs.harvard.edu/abs/2025arXiv251113370Z}
}

@ARTICLE{giri2026impact,
       author = {{Giri}, Sambit K. and {Mertens}, Florent},
        title = "{Impact of 21-cm foreground mitigation strategies on reionization power spectrum constraints}",
      journal = {\mnras},
         year = 2026,
        month = aug,
       volume = {550},
       number = {2},
          eid = {stag1258},
        pages = {stag1258},
          doi = {10.1093/mnras/stag1258},
archivePrefix = {arXiv},
       eprint = {2606.14682},
 primaryClass = {astro-ph.CO},
       adsurl = {https://ui.adsabs.harvard.edu/abs/2026MNRAS.550g1258G}
}

@ARTICLE{2019NatAs...3.1030A,
       author = {{An}, Tao and {Wu}, Xiang-Ping and {Hong}, Xiaoyu},
        title = "{SKA data take centre stage in China}",
      journal = {Nature Astronomy},
         year = 2019,
        month = nov,
       volume = {3},
        pages = {1030-1030},
          doi = {10.1038/s41550-019-0943-4},
       adsurl = {https://ui.adsabs.harvard.edu/abs/2019NatAs...3.1030A}
}

@ARTICLE{2022SCPMA..6529501A,
       author = {{An}, Tao and {Wu}, Xiaocong and {Lao}, Baoqiang and {Guo}, Shaoguang and {Xu}, Zhijun and {Lv}, Weijia and {Zhang}, Yingkang and {Zhang}, Zhongli},
        title = "{Status and progress of China SKA Regional Centre prototype}",
      journal = {Science China Physics, Mechanics, and Astronomy},
         year = 2022,
        month = dec,
       volume = {65},
       number = {12},
          eid = {129501},
        pages = {129501},
          doi = {10.1007/s11433-022-1981-8},
archivePrefix = {arXiv},
       eprint = {2206.13022},
 primaryClass = {astro-ph.IM},
       adsurl = {https://ui.adsabs.harvard.edu/abs/2022SCPMA..6529501A}
}

@ARTICLE{2022MNRAS.513.5109G,
       author = {{Georgiev}, Ivelin and {Mellema}, Garrelt and {Giri}, Sambit K. and {Mondal}, Rajesh},
        title = "{The large-scale 21-cm power spectrum from reionization}",
      journal = {\mnras},
         year = 2022,
        month = jul,
       volume = {513},
       number = {4},
        pages = {5109-5124},
          doi = {10.1093/mnras/stac1230},
archivePrefix = {arXiv},
       eprint = {2110.13190},
 primaryClass = {astro-ph.CO},
       adsurl = {https://ui.adsabs.harvard.edu/abs/2022MNRAS.513.5109G}
}

@ARTICLE{1978MNRAS.183..341W,
       author = {{White}, S.~D.~M. and {Rees}, M.~J.},
        title = "{Core condensation in heavy halos: a two-stage theory for galaxy formation and clustering.}",
      journal = {\mnras},
         year = 1978,
        month = may,
       volume = {183},
        pages = {341-358},
          doi = {10.1093/mnras/183.3.341},
       adsurl = {https://ui.adsabs.harvard.edu/abs/1978MNRAS.183..341W}
}

@ARTICLE{2001PhR...349..125B,
       author = {{Barkana}, Rennan and {Loeb}, Abraham},
        title = "{In the beginning: the first sources of light and the reionization of the universe}",
      journal = {\physrep},
         year = 2001,
        month = jul,
       volume = {349},
       number = {2},
        pages = {125-238},
          doi = {10.1016/S0370-1573(01)00019-9},
archivePrefix = {arXiv},
       eprint = {astro-ph/0010468},
 primaryClass = {astro-ph},
       adsurl = {https://ui.adsabs.harvard.edu/abs/2001PhR...349..125B}
}

@ARTICLE{2018PhR...780....1D,
       author = {{Dayal}, Pratika and {Ferrara}, Andrea},
        title = "{Early galaxy formation and its large-scale effects}",
      journal = {\physrep},
         year = 2018,
        month = dec,
       volume = {780},
        pages = {1-64},
          doi = {10.1016/j.physrep.2018.10.002},
archivePrefix = {arXiv},
       eprint = {1809.09136},
 primaryClass = {astro-ph.GA},
       adsurl = {https://ui.adsabs.harvard.edu/abs/2018PhR...780....1D}
}

@ARTICLE{2009MNRAS.397.1926W,
       author = {{Wyithe}, J. Stuart B. and {Loeb}, Abraham},
        title = "{The 21-cm power spectrum after reionization}",
      journal = {\mnras},
         year = 2009,
        month = aug,
       volume = {397},
       number = {4},
        pages = {1926-1934},
          doi = {10.1111/j.1365-2966.2009.15019.x},
archivePrefix = {arXiv},
       eprint = {0808.2323},
 primaryClass = {astro-ph},
       adsurl = {https://ui.adsabs.harvard.edu/abs/2009MNRAS.397.1926W}
}

@ARTICLE{10.1093/mnras/stw3026,
       author = {{Greig}, Bradley and {Mesinger}, Andrei},
        title = "{The global history of reionization}",
      journal = {\mnras},
         year = 2017,
        month = mar,
       volume = {465},
       number = {4},
        pages = {4838-4852},
          doi = {10.1093/mnras/stw3026},
archivePrefix = {arXiv},
       eprint = {1605.05374},
 primaryClass = {astro-ph.CO},
       adsurl = {https://ui.adsabs.harvard.edu/abs/2017MNRAS.465.4838G}
}

@ARTICLE{10.1093/mnras/stac825,
       author = {{Greig}, Bradley and {Mesinger}, Andrei and {Davies}, Frederick B. and {Wang}, Feige and {Yang}, Jinyi and {Hennawi}, Joseph F.},
        title = "{IGM damping wing constraints on reionization from covariance reconstruction of two z {\ensuremath{\gtrsim}} 7 QSOs}",
      journal = {\mnras},
         year = 2022,
        month = jun,
       volume = {512},
       number = {4},
        pages = {5390-5403},
          doi = {10.1093/mnras/stac825},
archivePrefix = {arXiv},
       eprint = {2112.04091},
 primaryClass = {astro-ph.CO},
       adsurl = {https://ui.adsabs.harvard.edu/abs/2022MNRAS.512.5390G}
}

@ARTICLE{refId0,
       author = {{Mason}, Charlotte A. and {Chen}, Zuyi and {Stark}, Daniel P. and {Lu}, Ting-Yi and {Topping}, Michael and {Tang}, Mengtao},
        title = "{Constraints on the z {\ensuremath{\sim}} 6-13 intergalactic medium from JWST spectroscopy of Lyman-alpha damping wings in galaxies}",
      journal = {\aap},
         year = 2026,
        month = jan,
       volume = {705},
          eid = {A114},
        pages = {A114},
          doi = {10.1051/0004-6361/202553820},
archivePrefix = {arXiv},
       eprint = {2501.11702},
 primaryClass = {astro-ph.GA},
       adsurl = {https://ui.adsabs.harvard.edu/abs/2026A&A...705A.114M}
}

@ARTICLE{2007MNRAS.377.1043M,
       author = {{McQuinn}, Matthew and {Lidz}, Adam and {Zahn}, Oliver and {Dutta}, Suvendra and {Hernquist}, Lars and {Zaldarriaga}, Matias},
        title = "{The morphology of HII regions during reionization}",
      journal = {\mnras},
         year = 2007,
        month = may,
       volume = {377},
       number = {3},
        pages = {1043-1063},
          doi = {10.1111/j.1365-2966.2007.11489.x},
archivePrefix = {arXiv},
       eprint = {astro-ph/0610094},
 primaryClass = {astro-ph},
       adsurl = {https://ui.adsabs.harvard.edu/abs/2007MNRAS.377.1043M}
}
\bibliographystyle{mnras}
\appendix
\section{Numerical values of the scores}
The tables in this appendix list the numerical values of the different scores for each team for the different data sets. The listed scores are the Z-scores for the different redshift intervals $Z(z_i)$, the Mahalanobis distance $D_\mathrm{M}$ and the SDC3b-score, $S_\mathrm{SDC3b}$.

\begin{table*}
\caption{PS1 metrics}
\label{tab:leader1}
\begin{tabular}{llllll}
\hline
Team&$Z(z_1)$&$Z(z_2)$&$Z(z_3)$&$D_M$&$S_{\rm SDC3b}$\\
\hline
21 To Infinities&1.155&0.152&1.575&1.952&0.2445\\
Akashganga&1.929&3.654&4.043&7.679&0.0146\\
Cantabrigians&5.827&3.96&2.469&8.362&0.9995\\
COTSS-21&4.63&0.889&1.461&10.733&0.1495\\
EoR PIE MC&2.023&0.253&1.155&15.063&0.0004\\
EoR PIE&3.635&2.225&0.61&10.19&0.4275\\
HIMALAYA&8.943&2.354&0.656&9.845&0.0012\\
HIstorians ANN&0.289&3.623&3.884&5.319&0.1169\\
HIstorians GPR&1.08&25.53&3.884&25.846&0.0\\
HIstorians LCDM Bayesian&0.942&0.246&1.095&1.465&0.058\\
Imperial-Nottingham&4.159&1.675&2.811&16.536&0.9448\\
KUSANAGI&0.063&0.279&0.282&0.402&0.2714\\
Loreli A&59.841&19.623&6.955&68.826&0.0\\
Loreli B&2.513&1.749&1.879&3.593&0.3606\\
ReionYuga&4.354&2.712&16.129&16.894&0.0012\\
RISE&1.876&1.031&0.229&3.395&0.2214\\
Modern SEarCH&16.97&24.498&13.056&32.68&0.0\\
Traditional SEarCH&9.322&13.658&13.325&14.964&0.0\\
Shuimu-Tianlai A&1.147&0.958&0.379&1.198&0.5713\\
Shuimu-Tianlai B&0.543&0.002&0.44&2.225&0.5512\\
Shuimu-Tianlai C&0.142&0.494&0.6&1.109&0.6465\\
ToSKA - Explicit likelihood&1.317&1.737&1.828&2.995&0.092\\
ToSKA - Model Selection&8.788&4.075&1.034&10.244&0.6407\\
ToSKA - SBI general&3.652&3.657&0.664&4.36&0.9272\\
YEYE&2.131&4.159&5.566&11.281&0.0\\

\hline
\end{tabular}
\end{table*}

\begin{table*}
\caption{PS2 metrics}
\begin{tabular}{llllll}
\hline
% Team&$Z(z_1)$&$Z(z_2)$&$Z(z_3)$&Mahalanobis&SDC3b score\\
Team&$Z(z_1)$&$Z(z_2)$&$Z(z_3)$&$D_M$&$S_{\rm SDC3b}$\\
\hline
21 To Infinities&2.858&2.366&18.605&18.987&0.0\\
Akashganga&2.359&0.756&0.887&5.406&0.9999\\
Cantabrigians&0.092&0.934&0.27&2.431&1.0\\
COTSS-21&1.395&1.358&1.71&2.511&0.3895\\
EoR PIE MC&1.149&0.812&2.506&7.793&0.0001\\
EoR PIE&0.706&0.538&0.016&6.523&0.3485\\
HIMALAYA&0.384&0.712&1.881&2.974&0.0495\\
HIstorians ANN&1.315&1.686&$>>1$&$>>1$&0.0\\
HIstorians GPR&1.315&1.686&$>>1$&$>>1$&0.0\\
HIstorians LCDM Bayesian&3.274&1.634&3.754&5.242&0.0\\
Imperial-Nottingham&2.487&1.147&6.531&19.449&0.5066\\
KUSANAGI&0.677&0.404&0.345&0.861&0.093\\
Loreli A&81.273&17.637&2.548&84.401&0.0\\
Loreli B&0.442&2.517&3.502&4.335&0.0254\\
ReionYuga&11.833&17.518&22.7&30.913&0.0\\
RISE&2.21&1.857&1.277&2.264&0.0734\\
Modern SEarCH&1.782&3.378&2.267&4.457&0.6811\\
Traditional SEarCH&1.272&2.728&4.787&5.659&0.7485\\
Shuimu-Tianlai A&0.069&0.424&0.761&0.929&0.3745\\
Shuimu-Tianlai B&1.384&1.5&0.464&3.316&0.0336\\
Shuimu-Tianlai C&1.116&1.109&0.454&2.952&0.067\\
ToSKA - Explicit likelihood&1.402&1.733&2.43&2.683&0.65\\
ToSKA - Model Selection&1.432&1.833&2.88&3.377&0.6604\\
ToSKA - SBI general&1.078&1.458&2.134&4.004&0.3737\\
YEYE&0.162&0.329&0.841&1.239&0.4828\\
\hline
\label{tab:leader2}
\end{tabular}
\end{table*}

\begin{table*}
\caption{PS3/IM1 metrics}
\begin{tabular}{llllll}
\hline
Team&$Z(z_1)$&$Z(z_2)$&$Z(z_3)$&$D_M$&$S_{\rm SDC3b}$\\
\hline
HIMALAYA PS3&8.087&2.542&0.443&11.459&0.001\\
HIMALAYA IM1&7.462&2.733&0.052&8.865&0.0014\\
Modern SEarCH PS3&1.627&2.397&1.757&3.412&0.0477\\
Modern SEarCH IM1&21.465&3.338&13.714&24.497&0.0\\
Traditional SEarCH PS3&7.24&10.12&10.629&11.381&0.0\\
Traditional SEarCH IM1&21.465&3.338&13.714&24.497&0.0\\
YEYE PS3&3.463&5.065&5.437&5.523&0.0\\
YEYE IM1&5.272&0.396&0.756&5.341&0.0538\\
\hline
\label{tab:leader3}
\end{tabular}
\end{table*}

\section{List of computational facility partners}\label{sec:partners}
\subsection{ASTRON / SURF (Amsterdam, Netherlands)}
The ASTRON/SURF facility provided access to Spider\footnote{\url{https://doc.spider.surfsara.nl}}, a versatile high-throughput data-processing platform designed for processing large, structured data sets. Spider is a combined CPU and GPU system, located at the SURF data centre in Amsterdam, the Netherlands.

\subsection{China SRC (Shanghai, China)} 
The China Square Kilometre Array (SKA) Regional Centre prototype (cnSRC) \citep{2019NatAs...3.1030A,2022SCPMA..6529501A} is a cutting-edge computing platform that employs a novel hybrid architecture.
This advanced setup integrated general-purpose Intel x86 CPUs, GPUs, and ARM processors, offering unparalleled flexibility and efficiency for data processing tasks.
The prototype comprises 35 Intel x86 CPU nodes with a total of 2240 cores (up to 128 cores per node), 12 ARM CPU nodes with 1152 cores (96 cores per node), and 4 GPU nodes equipped with 16 Nvidia V100 and 8 A100 GPUs,
collectively delivering approximately 800 TFlops of computing power. The CPU nodes contribute an additional 196 TFlops of computational capacity \footnote{\url{https://shaoska-user-guide.readthedocs.io/}}.
In terms of storage, cnSRC offers an approximately 9\,PB of distributed storage capacity, supported by a high-speed internal network enabling data exchange at rates of 100-200\,Gb/s between compute and storage nodes. Certain nodes are capable of accommodating up to 4\,TB of memory, with a maximum of 36\,GB per core, ensuring efficient processing of large-scale datasets.
Furthermore, cnSRC is connected to a 2\,Gbps transcontinental internet link, facilitating seamless international collaboration, and maintains a 200 Mbps connection with other SKA regional centre nodes to support global research initiatives.

%This work used resources of the China SKA Regional Centre prototype funded by the Ministry of Science and Technology of the People's Republic of China and Chinese Academy of Sciences, National SKA Program of China (Grant No. 2022SKA0130103, 2022SKA0120102). Shaoguang Guo is supported by the international partnership program of the Chinese Academy of Sciences (Grant No. 018GJHZ2024025GC).

\subsection{Galicia Supercomputing Centre/ CESGA (Santiago de Compostela, Spain)}
The Galician Supercomputing Centre (CESGA) facility provided access to FinisTerrae-III \footnote{\url{https://cesga-docs.gitlab.io/ft3-user-guide/overview.html}} supercomputer, located in Santiago de Compostela, Spain. FinisTerrae-III is an HPC system integrating Intel x86 CPUs, AMD CPUs and Nvidia GPUs, with over 384 nodes, totalling more than 22,000 x86 cores and 157 GPUs with Infiniband interconnect and 6 PB of high performance storage Lustre Filesystem.

\subsection{GENCI / IDRIS (Orsay, France)}
GENCI (\textit{Grand Equipement National de Calcul Intensif}) granted CPU and NVIDIA V100 GPU computing resources to two teams on the Jean Zay supercomputer hosted by IDRIS (Institute for Development and Resources in Intensive Scientific Computing).

IDRIS is the national supercomputer centre of the CNRS (French National Centre for Scientific Research) for High Performance Computing and Artificial Intelligence. It currently hosts Jean Zay, a supercomputer comprising 720 Intel CPU nodes and 843 GPU nodes equipped with V100, A100 or H100 NVIDIA GPUs for a total of 3704 GPUs. Following three successive extensions, the cumulative peak performance of Jean Zay reached 126 Pflop/s starting in July 2024.

\subsection{INAF (Italy)}
The Italian National Institute for Astrophysics (INAF), through its “Unità Scientifica Centrale 8” (USC8), provided computational support via the distributed HPC facility “Pleiadi.” This infrastructure consists of 192 nodes, totaling over 7,000 Intel Xeon E5-2697 V4 cores, interconnected via an Omni-Path HFI Silicon 100 Series network at 100 Gbit, with external connectivity at 1 Gb/s, and managed through the SLURM workload scheduler.  

INAF allocated up to 8 CPU nodes (128 GB RAM each) at the Institute of Radioastronomy in Bologna and up to 6 CPU nodes (128 GB RAM each, each with an available GPU) at the Astrophysical Observatory of Catania. 

\subsection {JPSRC (Tokyo, Japan)}
The Japanese SKA Regional Centre (JPSRC) prototype is an inter-university cooperation led by the SKA1 Japan Promotion Group (SKAJ) of the National Astronomical Observatory of Japan (NAOJ). JPSRC contains heterogeneous workstations with 152 CPU cores, 6 GPUs, 1 NEC SX-Aurora TSUBASA vector engine, as well as 1.86\,TB DRAM memory, 3.71\,TB M.2/SSD, and 338\,TB HDD storage. The workstations were federated by OpenStack as virtual machines, but JPSRC solely provided one of the workstations as a stand-alone machine, separating the server used for SDC3a from those used to deploy SRCNet v0.1. 

The provided workstation was equipped with an AMD EPYC 7713P 64 core CPU, 2 x NVIDIA Quadro A6000 48GB, 1.024\,TB DRAM memory, and 162 TB HDD, and the network speed was typically 1\,Gbps. The operating system was Linux Ubuntu 22.04, and an environment for docker and singularity containerizations was supported. 

\subsection {SPSRC/IAA-CSIC (Granada, Spain)}
The Spanish prototype of an SRC (a.k.a. espSRC), being developed at the Instituto de Astrofísica de Andalucía (IAA-CSIC) in  Spain,  is one of the 17 national initiatives contributing to the SRCNet development.  The espSRC OpenStack cloud gathers 240 CPU cores and 2.5\,TB of memory across five compute hypervisors, plus 600+ TB of SSD usable storage capacity managed by Ceph. The servers are interconnected by a 100\,Gbps network and the cluster is connected to RedIRIS (the Spanish National Research Network ) with a 10\,Gbps link. 

Hosted teams were provided with a virtual machine with 32 cores and 128\,GB memory. 100\,GB root disk using local SAS SSD and up to 5\,TB of additional block storage were assigned. The operating system and the software in the virtual machine were fully customizable by the team through pseudo-sudo access. Accessibility to the machine was flexible, provided by ssh, remote desktop and a clientless remote desktop gateway.

\subsection {Swiss National Computing Centre/ CSCS (Lugano, Switzerland)} 
The Swiss National Supercomputing Centre (CSCS) is a key partner in deploying and operating the Swiss SRCNet node, contributing its expertise and computational resources to support scientific research. For this project, one team was granted access to Piz Daint, CSCS’s former flagship supercomputer, which served as a testbed for Swiss SRCNet node science activities.  

At its peak, the now decommissioned Piz Daint was one of the most powerful supercomputers in Europe, comprising over 5,000 NVIDIA P100 GPGPUs. The team was allocated 15,000 node hours on this system, utilizing nodes equipped with one NVIDIA P100 GPU, 12 Intel Xeon CPU cores, and 65 GB of RAM. To support large-scale computations and data-intensive workflows, the project leveraged over 15 TB of long-term storage along with up to 10 PB of scratch space for high-speed temporary data access.

\subsection {UC-LCA (Coimbra, Portugal)}
The Laboratory for Advanced Computing of the University of Coimbra (UC-LCA) provided access to its Navigator cluster, which includes CPU and GPU nodes. The resources included 20\,TB of shared storage space (including image cube data) and 100,000 core-hours for the team assigned to UC-LCA. The team could use up to 1032 cores of Intel Xeon E5-2697v2 or up to 400 cores of Intel Xeon Gold 6148. Each node with these processors has 96 GB RAM and the interconnect is infiniband FDR/EDR. 

\subsection {UKSRC, IRIS-CAM (Cambridge, UK)}	
The UK SKA regional centre (UKSRC) resource at the University of Cambridge comprises a multi node HPC/GPU cluster running a Slurm batch scheduler. An OpenStack-hosted Platform-as-a-Service Azimuth applications portal is also available on request. Each team was allocated 100,000 core hours, 1000 GPU hours and 20\,TB of storage.

\subsection{SWESRC, Onsala Space Observatory/Chalmers e-Commons (Gothenburg, Sweden)}

Onsala Space Observatory hosts the Swedish node of the SRCNet (sweSRC) at Chalmers e-Commons in Gothenburg, Sweden. At sweSRC, we operate an OpenStack cloud that comprises a total of 1400 CPU cores, six GPUs of different types, 27 TB of RAM and a total of 400 TB of flash storage that is provisioned via a Ceph Cluster.

In the context of the SDC3b challenge, sweSRC provided one team with a virtual machine (VM) that had 64 CPUs, 384 GB of RAM and 2 TB of Ceph-provided HDD storage. The team had full control of the VM via pseudo-sudo access and login was performed via regular ssh access.

\clearpage

% here are the affilitations
\section*{Affiliatons}
1 SKA Observatory, Jodrell Bank, Lower Whitington, Macclesfield, SK11 9FT, UK\\
2 National Radio Astronomy Observatory, 520 Edgemont Road, Charlottesville, VA 22903, USA\\
3 The NSF-Simons AI Institute for Cosmic Origins, USA, 201 E. 24th Street, POB 4.102, Austin, Texas 78712-1229\\
4 Institute for Particle Physics and Astrophysics (ETHZ), Wolfgang-Pauli-Str 27, 8093 Zurich, Switzerland\\
5 Department of Astronomy and Oskar Klein Centre, AlbaNova, Stockholm University, SE-10691 Stockholm, Sweden\\
6 Max-Planck-Institut für Astrophysik, Garching 85748, Germany\\
7 Berkeley Center for Cosmological Physics, University of California, Berkeley, CA 94720, United States\\
8 Department of Electronics and Communication Engineering, 
Institute of Aeronautical Engineering, Hyderabad\\
9 Univ Lyon, Univ Lyon1, Ens de Lyon, CNRS, CRAL UMR5574, F-69230, Saint-Genis-Laval, France\\
10 Scuola Normale Superiore, Piazza dei Cavalieri 7, 56126 Pisa, Italy\\
11 Theoretical and Scientific Data Science, Scuola Internazionale Superiore di Studi Avanzati (SISSA), Via Bonomea 265, 34136 Trieste, Italy\\
12 Institute of Physics, Laboratory of Astrophysics, École Polytechnique Fédérale de Lausanne (EPFL), Switzerland\\
13 National Centre for Radio Astrophysics, TIFR, Pune University Campus, Ganeshkhind, Pune 411007, India\\
14 Inter-University Centre for Astronomy and Astrophysics, Pune University Campus, Ganeshkhind, Pune 411007, India\\
15 Department of Physics and Astronomy, University of the Western Cape, Robert Sobukwe Road, Bellville, 7535, South Africa\\
16 National Astronomical Observatories, Chinese Academy of Sciences, Beijing 100101, China\\
17 Centre for Strings, Gravitation and Cosmology, Department of Physics, Indian Institute of Technology, Madras, Chennai 600036, India\\
18 , International Centre for Theoretical Sciences, TIFR, Bengaluru  560089, India\\
19 Blackett Laboratory, Imperial College London, Prince Consort Road, London, SW7 2AZ, UK\\
20 Haute Ecole Arc Ingénierie, University of Applied Sciences and Arts Western Switzerland (HES-SO), Saint-Imier, Switzerland\\
21 Department of Physics, Indian Institute of Technology Kharagpur, Kharagpur 721302, India\\
22 Institut d’Astrophysique Spatiale, CNRS, Universit´e Paris-Saclay, 91405 Orsay, France\\
23 Institut für Theoretische Physik, Universität Heidelberg, Philosophenweg 16, 69120 Heidelberg, Germany\\
24 Department of Physics, National Institute of Technology Calicut, Calicut 673601, Kerala, India\\
25 Cavendish Astrophysics, University of Cambridge, JJ Thomson Avenue, Cambridge CB3 0HE, UK\\
26 Kavli Institute for Cosmology, University of Cambridge, Madingley Road, Cambridge CB3 0HA, UK\\
27 Department of Astronomy, Astrophysics and Space Engineering, Indian Institute of Technology Indore, Indore 453552, India\\
28 Max-Planck-Institut für Astronomie, Königstuhl 17, 69117 Heidelberg, Germany\\
29 IFPU, Institute for Fundamental Physics of the Universe, Via Beirut 2, I-34151 Trieste, Italy\\
30 SISSA - International School for Advanced Studies, Via Bonomea 265, I-34136 Trieste, Italy; INAF - Osservatorio Astronomico di Trieste, Via G. B. Tiepolo 11, I-34143 Trieste, Italy\\
31 Department of Astronomy, Tsinghua University, Beijing 100084, China\\
32 Max Planck Institute for Radioastronomy, Auf dem Hügel 69, Bonn, 53121,Germany\\
33 Ishigakijima Astronomical Observatory, National Astronomical Observatory of Japan, National Institutes of Natural Sciences, 1024-1 Arakawa, Ishigaki, Okinawa, 907-0024, Japan\\
34 Research Center for the Early Universe, Graduate School of Science, The University of Tokyo, 7-3-1 Hongo, Bunkyo, Tokyo 113-0033, Japan\\
35 Astrophysics Research Center of the Open University (ARCO), The Open University of Israel, 1 University Road P.O. Box 808, Ra'anana 4353701, Israel\\
36 Department of Physics, Indian Institute of Technology, Madras, Chennai 600036, India\\
37 Département de Physique Théorique, Université de Genève, 24 quai Ernest Ansermet, 1211 Genève 4, Switzerland\\
38 Department of Physics and Trottier Space Institute, McGill University, Qu´ebec, H3A 2T8, Canada\\
39 LUX, Observatoire de Paris, Université PSL, Sorbonne Université, CNRS, 75014 Paris, France\\
40 Department of Computer Science, University of Nevada, Las Vegas, NV 89154, USA\\
41 School of Physics and Astronomy, Sun Yat-sen University, Zhuhai 519082, China\\
42 Liaoning Key Laboratory of Cosmology and Astrophysics, College of Sciences, Northeastern University, Shenyang 110819, China\\
43 Department of Physics, University of Chicago, Chicago, IL 60637, USA\\
44 Shanghai Astronomical Observatory, Chinese Academy of Sciences, Shanghai 200030, China\\
45 Department of Physics and Astronomy, The Johns Hopkins University, Baltimore, MD 21218, USA\\
46 State Key Laboratory of Radio Astronomy and Technology, National Astronomical Observatories, CAS, Beijing 100101, China\\
47 School of Astronomy and Space Science, University of Chinese Academy of Sciences, Beijing 100049, China\\
48 School of Physics and Astronomy, Tel-Aviv University, Tel-Aviv 69978, Israel\\
49 Department of Physics, Imperial College London, Blackett Laboratory, Prince Consort Road, London SW7 2AZ, UK; I-X Centre for AI in Science, Translation and Innovation Hub (I-HUB), Imperial College London, White City Campus, 84 Wood Lane, London W12 0BZ, UK\\
50 Department of Physics, Graduate School of Science, The University of Tokyo, 7-3-1 Hongo, Bunkyo, Tokyo 133-0033, Japan\\
51 David A. Dunlap Department of Astronomy and Astrophysics, University of Toronto, 50 St. George Street, Toronto, Canada\\
52 School of Physics and Astronomy, The University of Nottingham, University Park, Nottingham, NG7 2RD, UK\\
53 Centre for Artificial Intelligence, Zurich University of Applied Sciences (ZHAW), Technikumstrasse 71, 8400 Winterthur, Switzerland\\
54 Institute of Astronomy, University of Cambridge, Madingley Road, CB3 0HA, UK\\
55 School of Physics and Astronomy, Sun Yat-Sen University, No.2 Daxue Road, Zhuhai 519082, China\\
56 Raman Research Institute, C. V. Raman Avenue, Sadashivanagar, Bengaluru 560080, India\\
57 School of Physical Sciences, National Institute of Science Education and Research, Bhubaneswar 752050, Odisha, India\\
58 School of Science, Jiangxi University of Science and Technology, Ganzhou 341000, China\\
59 Global College, Shanghai Jiao Tong University, Shanghai 200240, China\\
60 Department of Physics and Astronomy, Chalmers University of Technology, Onsala Space Observatory, SE-43992 Onsala, Sweden\\
61 ASTRON, Netherlands Institute for Radio Astronomy, Oude Hoogeveensedijk 4, 7991 PD Dwingeloo, The Netherlands\\
62 Institute for Astronomy, University of Vienna, Türkenschanzstrasse 17, A-1180 Vienna, Austria\\
63 Graduate School of Science and Technology, Kumamoto University, 39-1 Kumamotoshi, Kumamoto 860-8555, Japan\\
64 Kapteyn Astronomical Institute, University of Groningen, P.O.Box 800, 9700AV Groningen, the Netherlands\\
65 School of Physics and Electronic Science, Guizhou Normal University, Guiyang 550001, China\\
66 Department of Physics and Astronomy “Ettore Majorana”, University of Catania, Via Santa Sofia 64, 95123 Catania, Italy\\
67 Research School of Astronomy and Astrophysics, Australian National University, Canberra ACT 2611, Australia\\
68 Department of Physics, Stellenbosch University, Matieland, 7602, South Africa\\
69 Department of Physics, Indian Institute of Science, Bangalore 560012, India\\
70 Institute for Data Science, University of Applied Sciences and Arts Northwestern Switzerland (FHNW), Bahnhofstrasse 6, Windisch, 5210, Switzerland\\
71 Department of Physics, Liaoning Normal University, Dalian 116029, China\\
72 South-Western Institute for Astronomy Research, Key Laboratory of Survey Science of Yunnan Province, Yunnan University, Kunming, Yunnan 650500, People's Republic of China\\
73 Division of Particle and Astrophysical Science, Nagoya University, Furo-cho, Chikusa-ku, Nagoya 464-8602, Japan\\
74 The Research Center for Statistical Machine Learning, The Institute of Statistical Mathematics, 10-3 Midori-cho, Tachikawa, Tokyo 190-8562, Japan\\
75 Physics Department, Astrophysics Group, Imperial College London, Prince Consort Road, London SW7 2AZ, UK\\
76 School of Physics and Technology, Wuhan University, Wuhan 430072, China\\
77 National Institute for Theoretical and Computational Sciences (NITheCS), Merensky Building, Merriman Street, Stellenbosch, 7600, South Africa\\
78 School of Physics, Ningxia University, Yinchuan 750021, China\\
79 School of Physics and Astronomy, Beijing Normal University, Beijing 100875, China\\
80 School of Artificial Intelligence, Beijing Normal University, Beijing 100875, China\\
81 Institute for Advanced Research, Nagoya University, Furo-cho Chikusa-ku, Nagoya 464-8601, Japan\\
82 Graduate School of Science, Division of Particle and Astrophysical Science, Nagoya University, Furocho, Chikusa-ku, Nagoya, Aichi 464-8602, Japan.\\
83 College of Big Data and Information Engineering, Guizhou University, Guiyang 550025, China\\
84 State Key Laboratory of Public Big Data, Guizhou University, Guiyang 550025, China\\

\end{document}